\documentclass{article}

\usepackage[left=3cm, right=3cm, top=2cm, bottom = 2cm]{geometry}
\usepackage[xindy]{glossaries}
\usepackage[hidelinks]{hyperref} 

\usepackage{xspace}

\newcommand{\Genom}{G\textsuperscript{en}oM\xspace}
\newcommand{\Vanda}{\textsc{Vanda}\xspace}

\newcommand{\Gwen}{\textsc{Gwendolen}\xspace}

\newcommand{\eventually}{\ensuremath{\lozenge}}
\newcommand{\always}{\ensuremath{\Box}}
\newcommand{\lbelief}[2]{\ensuremath{\mathrm{{\mathcal B}}_{#1} \, #2}}

\newacronym{ltl}{LTL}{Linear-time Temporal Logic}

\newacronym{ctl}{CTL}{Computation Tree Logic}

\newacronym{ptl}{PTL}{Probabilistic Temporal Logic}

\newacronym{pctl}{PCTL}{Probabilistic Computation Tree Logic}

\newacronym{pctl*}{PCTL*}{Probabilistic Computation Tree Logic*}

\newacronym{tctl}{TCTL}{Timed Computation Tree Logic}

\newacronym{stl}{STL}{Signal Temporal Logic}

\newacronym{ltl-x}{LTL-X}{a version of Linear Time Logic without the `next' operator}

\newacronym{mtl}{MTL}{Metric Temporal Logic}

\newacronym{iactlk-x}{IACTLK-X}{a fragment of the temporal-epistemic logic CTLK, without the `next' operator}

\newacronym{mucalc}{$\mu$-calculus}{a strict superset of other temporal logics}

\newacronym{fotl}{FOTL}{First-Order Temporal Logic}

\newacronym{bdi}{BDI}{Belief-Desire-Intention}

\newacronym{prs}{PRS}{Procedural Reasoning System}

\newacronym{are}{ARE}{Autonomous Reasoning Engine}

\newacronym{dmars}{dMARS}{distributed Multi-Agent Reasoning System}

\newacronym{fsm}{FSM}{Finite-State Machine}

\newacronym{pfsm}{PFSM}{Probabilistic Finite-State Machine}

\newacronym{apfsm}{APFSM}{\textit{Autonomous} Probabilistic Finite-State Machine}

\newacronym{pha}{PHA}{Probabilistic Hybrid Automata}

\newacronym{fsa}{FSA}{Finite-State Automata}

\newacronym{ta}{TA}{Timed Automata}

\newacronym{gspn}{GSPN}{Generalised Stochastic Petri Net}

\newacronym{mopn}{MOPN}{Marked Ordinary Petri Net}

\newacronym{ba}{BA}{B\"{u}chi Automata}

\newacronym{dtmc}{DTMC}{Discrete-Time Markov Chain}

\newacronym{pars}{PARS}{Process Algebra for Robot Schemas}

\newacronym{fsp}{FSP}{Finite State Processes}

\newacronym{piadl}{$\pi$ADL}{$\pi$ calculus combined with ADL}

\newacronym{csp}{CSP}{Communicating Sequential Processes}

\newacronym{ccs}{CCS}{Calculus of Communicating Systems}

\newacronym{wsccs}{WSCCS}{Weighted Synchronous CCS}

\newacronym{csp|b}{CSP$\|$B}{an integrated formalism combining CSP and B}

\newacronym{fdr}{FDR}{the Failures-Divergences Refinement checker}

\newacronym{jpf}{JPF}{Java PathFinder}

\newacronym{ajpf}{AJPF}{Agent Java PathFinder}

\newacronym{mcmas}{MCMAS}{Model Checker for Multi-Agent Systems}

\newacronym{ltlmop}{LTLMoP}{Linear Temporal Logic MissiOn Planning}

\newacronym{smt}{SMT}{Satisfiability Modulo Theories}

\newacronym{adl}{ADL}{Architecture Description Language}

\newacronym{aadl}{AADL}{Architecture Analysis and Design Language}

\newacronym{lts}{LTS}{Labelled-Transition System}

\newacronym{uml}{UML}{Unified Modelling Language}

\newacronym{sysml}{SySML}{Systems Modelling Language}

\newacronym{robotml}{RobotML}{Robot Modelling Language}

\newacronym{dsl}{DSL}{Domain Specific Language}

\newacronym{sct}{SCT}{Supervisory Control Theory}

\newacronym{ide}{IDE}{Integrated Development Environment}

\newacronym{mde}{MDE}{Model-Driven Engineering}

\newacronym{ai}{AI}{Artificial Intelligence}

\newacronym{fdir}{FDIR}{Fault Detection, Isolation and Recovery}

\newacronym{ifm}{iFM}{Integrated Formal Methods}

\newacronym{ros}{ROS}{Robot Operating System}

\newacronym{amcl}{AMCL}{Adaptive Monte Carlo Localisation}

\newacronym{dl}{\text{\upshape\textsf{d{\kern-0.05em}L}}}{differential dynamic logic}

\newacronym{vm}{VM}{Virtual Machine}

\newacronym{bip}{BIP}{Behaviour Interaction Priority}

\newacronym{fol}{FOL}{First-Order Logic}

\newacronym{owl}{OWL}{Web Ontology Language}

\newacronym{ria}{RIA}{Restore Invariant Approach}

\newacronym{ocl}{OCL}{Object Constraint Language}

\newacronym{sil}{SIL}{Safety Integrity Level}

\newacronym{rcl}{RCL}{ROS Contract Language}

\newacronym{rv}{RV}{Runtime Verification}

\newacronym{rml}{RML}{Runtime Monitoring Language}

\newacronym{slam}{SLAM}{Simultaneous Localisation and Mapping}

\newacronym{cfs}{cFS}{Core Flight System}

\usepackage{appendix}
\usepackage{placeins}
\usepackage[utf8]{inputenc}
\usepackage[T1]{fontenc}
\usepackage{lmodern}
\usepackage[UKenglish]{isodate}

\usepackage{amsmath,amsfonts}
\usepackage{algorithmic}
\usepackage{graphicx}
\usepackage{textcomp}
\usepackage{xcolor}
\usepackage{dafny}
\usepackage{rotating}
\usepackage{multirow}
\usepackage{hhline}
\usepackage{tabularx}
\usepackage{caption}
\usepackage{subcaption}

\RequirePackage{doi}

\def\lland{\,\land\,}

\usepackage{tikz}
\usetikzlibrary{shapes.geometric, arrows}
\tikzstyle{process} = [rectangle, minimum width=3em, minimum height=2em, text centered, draw=black]
\tikzstyle{arrow} = [thick,->,>=stealth]
\usepackage{tablefootnote}
\usetikzlibrary{shapes,backgrounds,arrows,calc,positioning}

\usepackage{xspace}
\usepackage{adjustbox}
\usepackage{paralist}
\usepackage{tabularx}

\usepackage{zed-csp}
\usepackage{gensymb}

\usepackage[nounderscore]{syntax}
\usepackage{programs}
\usepackage{listings}

\lstdefinelanguage{RCL}
{
emph={node, inputs, outputs, topics, assume, guarantee},
otherkeywords={matches},
morekeywords={:, forall, in, out, and, or},
emphstyle={\color{blue}},
keywordstyle={\color{red}},
sensitive=false,
morecomment=[l]{//},
morecomment=[s]{/*}{*/},
morestring=[b]",
}

\newcommand{\inlineRCL}[1]{\lstinline[language=RCL, basicstyle=\ttfamily]{#1}}
\newcommand{\dummycommand}{\end{lstlisting}}

\usepackage{multirow}
\usepackage{hhline}
\usepackage{tabularx}
\usepackage{authblk}
\usepackage{scalerel}

\usepackage{orcidlink}

\begin{document}

\renewcommand{\thelstlisting}{\arabic{lstlisting}}

\title{A Compositional Approach\\ to\\ Verifying Modular Robotic Systems}

\author[1]{Matt Luckcuck \orcidlink{0000-0002-6444-9312} \thanks{This work began in 2018 while all authors were employed by the University of Liverpool, UK.
The work was supported by UK Research and Innovation, and EPSRC Hubs for Robotics and AI in Hazardous Environments: EP/R026092 (FAIR-SPACE), EP/R026173 (ORCA), and EP/R026084 (RAIN); and through grant EP/V026801 (TAS Verifiability Node).
Cardoso's and Fisher's work was supported by the Royal Academy of Engineering under the Chairs in Emerging Technologies scheme. some of Luckcuck and Farrell's work was performed while employed by Maynooth University, Ireland; and The University of Manchester, UK. Some of Ferrando's work was performed while employed by the University of Genova, Italy. 

To ensure open access, the authors have applied a Creative Commons Attribution (CC BY) licence to any Author Accepted Manuscript version arising.

The formal models, data, and source code supporting the findings reported in this paper are openly available from the Zenodo repository at: \url{https://doi.org/10.5281/zenodo.6941344}.
}} 
\author[2]{Marie Farrell \orcidlink{0000-0001-7708-3877}}
\author[3]{Angelo Ferrando \orcidlink{0000-0002-8711-4670}} 
\author[4]{Rafael C. Cardoso \orcidlink{0000-0001-6666-6954}} 
\author[2]{Louise A. Dennis \orcidlink{0000-0003-1426-1896}}
\author[2]{Michael Fisher \orcidlink{0000-0002-0875-3862} }

\affil[1]{School of Computer Science, University of Nottingham, United Kingdom}
\affil[2]{Department of Computer Science, University of Manchester, United Kingdom}
\affil[3]{Department of Physical, Computer and Mathematical Sciences, University of Modena and Reggio Emilia, Italy}
\affil[4]{Department of Computing Science, University of Aberdeen, United Kingdom}

\date{\today}

\maketitle

\begin{abstract}
\noindent Robotic systems used in safety-critical scenarios often rely on modular software architectures, and increasingly include autonomous components. Verifying that these modular robotic systems behave as expected requires approaches that can cope with, and preferably take advantage of, this inherent modularity. This paper describes a compositional approach to specifying the nodes in robotic systems built using the \gls{ros}, where each node is specified using \gls{fol} assume-guarantee contracts that link the specification to the \gls{ros} implementation. We introduce inference rules that facilitate the composition of these node-level contracts to derive system-level properties. We also present a novel Domain-Specific Language, the \gls{rcl}, which captures a node's \gls{fol} specification and links this contract to its implementation. \gls{rcl} contracts can be automatically translated, by our tool \textsc{Vanda}, into executable monitors; which we use to verify the contracts at runtime. We illustrate our approach through the specification and verification of an autonomous rover engaged in the remote inspection of a nuclear site, and finish with smaller examples that illustrate other useful features of our framework.

\end{abstract}

\glsresetall

\section{Introduction}
\label{sec:intro}

Robotic systems are increasingly deployed in industrial, often safety-critical, scenarios such as monitoring offshore structures~\cite{Hastie:18}, nuclear inspection and decommissioning~\cite{Bogue:11,Workington18}, and space exploration~\cite{Wilcox:92,FloresAbad:14}. Engineering the software to control a robotic system is a complex task, often supported by modular software frameworks, such as the \gls{ros}~\cite{quigley2009ros} or \Genom{}~\cite{foughali:tel-02080063,Fleury1997}. It is crucial to ensure that the software controlling a robot behaves correctly, particularly as modern robotic systems become more autonomous, more complex, and are used in dynamic environments that they share with humans. The generality and flexibility of robotic software frameworks also means that guaranteeing their correct behaviour is challenging~\cite{Halder2017}. 


The state-of-the-art for verification of autonomous and robotic systems includes a variety of formal methods that can be used for specification and verification~\cite{Luckcuck2019} -- non-formal methods such as field tests and simulation-based testing are also common.
Formal methods that we see being used include: model-checking~\cite{clarke1999model}, which exhaustively explores the state space to establish that a property holds; runtime verification~\cite{DBLP:journals/jlp/LeuckerS09}, which monitors system behaviour at runtime; and theorem-provers~\cite{bertot2013interactive}, demonstrating by mathematical proof that the system behaves correctly.
Different components of a robotic system may be better suited to different verification techniques, but linking the outputs of multiple techniques remains a challenge. Previously, we have argued that robotics is a domain in which \textit{integrating} (formal and non-formal) verification methods is both a necessity to be dealt with and an opportunity to be grasped~\cite{Farrell2018}.

\emph{Assume-Guarantee} reasoning~\cite{Jones83} (or the specification of \emph{pre-} and \emph{post-}conditions) is a well established compositional verification technique. A pair of pre- and post-conditions form a \textit{contract}~\cite{Meyer92}. 
Specifying a system using Assume-Guarantee or \emph{pre-} and \emph{post-}conditions enables it to be decomposed into modules, so that each module can be verified separately against its associated contract. However, care must be taken when specifying the contracts, because they still require validation against the actual requirements of the system.


This paper presents our approach to verifying robotic systems that are developed using \gls{ros}. This particular robotic software framework was chosen because of its prevalence in the literature. A \gls{ros} system is composed of nodes that communicate using message passing via buffered communication channels. The nodes coordinate to control the robot's overall behaviour. Typically, each node will be specialised to perform a different function, with different nodes (or collections of nodes) often requiring distinct verification techniques. For example, machine learning components will likely be verified via testing, whereas a planner might be mathematically modelled and reasoned about using  formal verification. Our approach uses the encapsulation provided by \gls{ros} nodes to provide compositionality. Fundamentally, we address the research question: 
\begin{quote} 
\textit{Can we use a compositional and heterogeneous approach to verify ROS-based systems?}
\end{quote}

\noindent Our approach begins by (manually) abstracting the graph of nodes in the \gls{ros} program (generated by \gls{ros} Graph or similar) into a more manageable model of the system, a model that focusses on the system's most critical components. Although some nodes maintain a one-to-one correspondence with the \gls{ros} software, abstraction can involve dropping some nodes from the system model or combining related nodes into, what we call, a \textit{compound node}. 

Once the \gls{ros} software is abstracted to a manageable system model, we specify each of the nodes in this model using Assume-Guarantee contracts, written in our contract specification language that is based on \gls{fol}. \gls{fol} was chosen for contract specifications because it is both expressive and widely understood, flattening our approach's learning curve.

We also provide a calculus that can be used to combine the module contracts and to derive system-level properties that correspond to the system's requirements. The calculus uses temporal operators from \gls{fotl}~\cite{Practical11} to represent the connections between the contracts. 

Once the contracts are verified against the system's requirements, we can take two further steps in parallel. The contracts are used to \textit{guide} the verification of each node, using heterogeneous (formal or non-formal) verification approaches chosen to suit each node. Providing a formal link between \gls{fol} and every verification approach is not in scope for this paper. Using the contracts as a guide enables the use of formal and informal links between the contracts and verification. We leave the verifiers to choose the most suitable verification method for each node, because they are best placed to make this choice. Meanwhile, our tool, \textsc{Vanda}, automatically synthesises runtime monitors from the contracts. The overall approach builds on two pieces of previous work:
\smallskip
\begin{compactitem}
\item the initial presentation of the contract calculus in~\cite{Cardoso20e}, which we significantly update in this paper; and,
\item our application of heterogeneous verification approaches to a simulation of the Mars Curiosity rover~\cite{Cardoso2020}, which we also update with additional steps.
\end{compactitem}
\smallskip
Our work provides four contributions:
\smallskip
\begin{compactenum}
\item A compositional approach to specifying the pre- and post-conditions of robotic systems constructed using \gls{ros}, which is supported by;
\item a \gls{dsl}, called the \gls{rcl}, that links a contract to the \gls{ros} implementation; 
\item a calculus containing inference rules for combining the contracts so that we can derive system-level (safety and mission) properties; and,
\item a tool-chain that synthesises runtime monitors from the system's \gls{rcl} contracts. 
\end{compactenum}
Our approach enables the introduction of a formal specification to an existing \gls{ros} system. As such, we use the intended behaviour of the nodes in a system as the starting point for specifying its contracts, and combine them using the calculus. Our \gls{dsl} and prototype parsing and monitor-generation tool (\textsc{Vanda}) support users in writing grammatically correct contracts. 

In summary, the contracts: are structured using \gls{rcl}; are reasoned about using our calculus; guide the heterogeneous verification of the nodes; and are then used to generate runtime monitors that provide a safety net, ensuring that the contracts have been verified correctly and the system is obeying its requirements. 
We validate our approach by applying it to a rover robot performing a remote inspection task inside a nuclear storage facility (\S\ref{sec:exampleSystem}). This example system was developed independently of our work, and we use our approach to introduce contracts and formal verification. We also describe the specification and verification of individual nodes, showing how our approach can guide verification using a variety of different formalisms.



The remainder of this paper is structured as follows. 
In \S\ref{sec:related}, we discuss related work. 
\S\ref{sec:specifyingRobots} describes our compositional approach to verifying modular robotic systems using \gls{fol} contracts, including: the calculus for combining node specifications, a description of how \gls{rcl} supports contract specification, and the automatic synthesis of runtime monitors from \gls{rcl}.
\S\ref{sec:exampleSystem} presents the specification and verification of our Case Study, a remote inspection rover.
In \S\ref{sec:discussion} we discuss some interesting characteristics of our framework, such as the use of a system's modularity, how our approach fits into the robotic software development process and, in~\S\ref{sec:towards}, we describe how our approach can be applied to non-\gls{ros} systems.
Finally, \S\ref{sec:conclusion} concludes the paper and presents avenues for future work.

\section{Related Work}
\label{sec:related}

This section discusses approaches in the literature that are related to our work. We have grouped these into Compositional Verification and Reliable Software Engineering (\S\ref{sec:CVandRSE}), which covers work using assume/guarantee contracts; and Robotics (\S\ref{sec:robotics}), which covers work relating to specifying robotic systems specifically. Some of the cited work could fit into both of these categories, but we have added them to the most relevant category based on their main contributions and publication venue.

\subsection{Compositional Verification and Reliable Software Engineering} 
\label{sec:CVandRSE}

Our approach encourages the development of systems as a composition of sub-systems, as does the work on the Pacti \cite{rouquette2023early} tool for assume-guarantee contracts. Pacti supports polyhedral constraints in which the terms are expressed as linear inequalities with real coefficients. Pacti also supports several contract operations (for example refinement and composition) to reason about the relationship between contracts.
One key feature of Pacti is that the contract operations are agnostic of the specific algebra used, though the work in~\cite{rouquette2023early} only implements polyhedral constraints so far, with other formalisms ``such as LTL and nonlinear constraints'' suggested as future work.
In contrast to the language used by Pacti, we use FOL which is more expressive and actually capable of expressing Pacti contracts. However, Pacti's contract operations are more developed than ours, as it includes refinement between contracts and a \textit{quotient} operator that can discover the specification of a missing subsystem that will combine with the existing specification to meet a top-level system specification. Our work focusses on specifying contracts that guide the verification of the system and enable the automatic synthesis of \gls{rv} monitors, both of which are absent from the work on Pacti.

Compositional verification is applied to the SIENA event-notification middleware~\cite{caporuscio2004compositional} where a global system property is decomposed into local properties that only hold on sub-parts of the system. They use compositional model-checking and the system models (labelled transition systems) are translated to Promela, using SPIN for verification. The various properties are related using simulation, a notion that in~\cite{caporuscio2004compositional} seems similar to formal refinement. Our work provides a broader approach; we are not restricted to model-checking and we enable heterogeneous verification. That said, their way of relating global and sub-properties is interesting and will likely inspire future directions for our work.

Compositional Assume-Guarantee reasoning has been used to define contracts for system modules~\cite{TheFrenchReport2012}, with rule-defined contract composition. Their rules share the same aim as our work and are extended by~\cite{Li2017}, using a variant of \gls{stl} to describe behaviour and contracts. They used the rules in~\cite{TheFrenchReport2012} to produce whole system assumptions and guarantees. They target closed-loop control systems, whereas our approach targets robotic systems that are written in general-purpose programming languages {and leverages heterogeneous specification}. Related compositional approaches include OCRA~\cite{cimatti2013ocra} and AGREE~\cite{cofer2012compositional}, though neither explicitly incorporates heterogeneous verification.

CoCoSpec~\cite{champion2016cocospec} is a language that provides assume/guarantee contracts for reactive systems. CoCoSpec extends the Lustre specification language and uses the Kind2 model-checker for compositional verification~\cite{champion2016kind}. This approach is specialised for synchronous communications, which differ from the event-based communications that we target, and their contract semantics is more restrictive than ours. {Specifically, CoCoSpec with Kind2 uses logical implication requiring the user to show that the lower-level node specifications imply a predefined top-level node specification; whereas we derive the top-level contract from the combination of the lower-level contracts.} Further, it is not clear how their support for compositional verification can be extended to support heterogeneous components such as those in our example. 

In previous work by some of the authors of this paper, a combination of  NASA's Formal Requirements Elicitation Tool (FRET), CoCoSpec, and Event-B were used to verify an inspection rover example system~\cite{bourbouh2021integrating}. The work used a Simulink model of the rover's architecture to define individual components. {The combination of techniques was needed to adequately verify the rover system. For example, the CoCoSpec model could not be verified against the planner requirements for grids larger than $4 \times 4$ so Event-B's proof-based approach was used to verify these properties for any size of grid. However, doing so forced us to verify an abstracted planner model in place of the Simulink model, potentially sacrificing some accuracy for tractability.} In contrast, our work focusses on systems developed in ROS and uses FOL contracts to reason about system-level properties.

We take inspiration from Broy's approach to systems engineering~\cite{Broy18} which presents three kinds of artefacts: (1) system-level requirements, (2) functional system specification, and (3) logical subsystem architecture. These are represented as logical predicates in the form of assertions, with relationships defined between them that extend to assume/commitment contracts. The treatment of these contracts is purely logical, and we present a similar technique that, instead of assertions, uses Assume-Guarantee contracts and is specialised to the software engineering of robotic systems. 

Ruchkin et al. describe the Integration Property Language (IPL)~\cite{ruchkin2018ipl}, { which also uses \gls{fol} with temporal operations to specify \textit{integration properties} (properties involving multiple models and formalisms) for heterogeneous models, specifically aimed at cyber-physical systems. They target \textit{architectural views} (abstracted, behaviourless component models that are annotated with types and properties) which handle the difficult task of integrating heterogeneous models. IPL specifications can be verified using \gls{smt} solvers and model checkers. Our approach avoids the difficulty of integrating the node-level models by combining the nodes, more abstract, \gls{fol} contracts and verifying the heterogeneous models of each node in isolation. }

Publish-subscribe architectures, like ROS, are popular in many domains. Baresi, et al. present the Loupe model-checker for publish-subscribe architectures \cite{baresi2010loupe}. They essentially embed the communications infrastructure within the verification checker to reduce the state space for verification. Such approaches are certainly relevant for our work, but, they do not explicitly focus on ROS or support a compositional approach to verification of individual system modules.

\subsection{Robotics}
\label{sec:robotics}

Many approaches for building safe robotic systems focus on \gls{ros}, which is a well established middleware that supports interoperability and modularity in the development of robotic software. A safety-critical working group\footnote{\url{https://github.com/ros-safety} Accessed: 03/11/2023} for ROS2 has been developing tools, libraries, and documentation to support the safe engineering of safety-critical \gls{ros} systems. For example, they provide a \emph{contracts} package\footnote{\url{https://github.com/ros-safety/contracts_lite} Accessed: 03/17/2023}, where a contract is a combination of pre-/post-conditions and assertions over the implementation of C++ functions. The advantage of this approach is that the contracts work directly in the implementation's source code. However, limiting the library to C++ is a disadvantage because \gls{ros} nodes may also be written in Python, Java, etc. In comparison, our approach is more general and not limited to the verification of individual functions, but also allows the verification of the system as a whole through the use of our inference rules. Further, their ad-hoc contract language is less expressive than \gls{fol}. An interesting line of future work may include updating \textsc{Vanda}, our prototype tool, to produce contracts compatible with the ROS2 contracts package for applications fully implemented in C++.

A similar approach is shown in~\cite{DBLP:conf/ecai/BasuGLNBIS08} for \Genom\ instead of \gls{ros}, which uses the Behaviour-Interaction-Priority (BIP) framework for incremental composition of heterogeneous components. They offer synthesis of functional-level controllers by synchronising dependencies between controllers. They verify safety properties and detect deadlock conditions using model checking and ``observers'' (runtime monitors). 
Another compositional approach in~\cite{Spellini2019} uses Assume-Guarantee contracts to decompose the control software of multi-robot systems and targets \gls{ros}. The individual components or robots are decomposed into sub-problems and then recomposed using contracts to provide system-level validation. Finally, the resulting synthesised controller is integrated into \gls{ros}. 

Drona~\cite{10.1007/978-3-030-03421-4_8} is a toolchain for programming safety-critical robots, with support for ROS. Their \gls{dsl}, called $P$, is based on state machines. It offers compositional Assume-Guarantee testing and a runtime assurance system to check that the assumptions made at design-time hold at runtime. In contrast, our \gls{dsl} is used purely to specify contracts for verification, we do not directly interfere with the system's implementation. Additionally, our runtime monitors are automatically synthesised from the contracts, while theirs require additional specification.

The Declarative Robot Safety (DeRoS)~\cite{Adam16} is a \gls{dsl} with a declarative syntax for specifying safety-related constraints in \gls{ros}, which lowers the barrier to using this approach -- similarly to our use of a textual version of \gls{fol}. 
In DeRoS, each contract is a refinement of the system-level contract; it is not clear if they use a similar technique to our inference rules to ensure that the contracts are composed correctly at design time. 
Unfortunately, DeRoS is not publicly available, or it could have been an alternative way to synthesise our monitors. 

Another \gls{dsl}, PROMISE~\cite{10.1145/3377812.3382143}, is designed to describe mission specifications for multi-robot systems. PROMISE has been integrated in Eclipse as a plugin to provide a graphical interface for users which allows the automatic generation of behaviour from the mission specification, sending missions to the robots, and runtime management of missions. 

Other related work provides automatic static verification of system-wide properties for message-passing in \gls{ros} applications~\cite{9341085}. The specification of the safety properties is written in a \gls{dsl}. This is then translated into \gls{fotl} to be used in Electrum~\cite{10.1145/2950290.2950318} which provides an automatic \emph{Analyser}. Their approach is embedded in HAROS~\cite{7759661}, a framework for quality assessment of \gls{ros} software that offers a visualisation interface for safety issues. Instead, our work focuses on compositional verification of nodes/modules in \gls{ros}, which are supported by inference rules to automatically generate system-wide properties. Further, we are not limited to static verification, our approach also provides runtime verification by  synthesising monitors from the contracts.

SOTER~\cite{8809550} is a programming framework to support the development of robotic systems by capturing runtime safety assurance principles. Their high-level \gls{dsl} can be used to implement reactive systems, make use of systematic testing techniques, and support runtime assurances. The case study and experiments presented use \gls{ros}. Our approach differs in that we are not limited to runtime assurances, we also encourage a variety of offline verification techniques.

RoboSC is a \gls{dsl}, and accompanying tool, that enables the specification of \gls{ros} nodes as \gls{fsa} and then synthesises supervisory controller nodes~\cite{Wesselink2023}. The events in the \gls{fsa} are the \gls{ros} communication topics that trigger state changes in the automaton. The supervisory controller nodes aim to enforce a node's (user-described) requirements. RoboSC always adds the overhead of communication to the \gls{ros} middleware, which increases the average time of a controller cycle by approx. 957\% (adding an average of 33.5 microseconds). 
Our approach can be used to enforce requirements, or to simply observe and log deviations from the specification; the latter option avoids some of the overheads of the additional nodes needed for \gls{rv}.

RoboChart~\cite{10.1007/s10270-018-00710-z} is a \gls{dsl} based on the \gls{uml}, which supports verification and automated reasoning of robotic systems using model checking and theorem proving. Its notation is based on state-machines, with a restricted set of constructs. 
\gls{ide} support is available through an Eclipse plugin called RoboTool that automatically generates C++ code for state machines and controllers, but does not yet offer automatic code deployment in \gls{ros}. 

As robotic systems become more complex, supporting heterogeneous verification becomes even more important~\cite{Farrell2018}. Crucially, the modular structure of \gls{ros} systems facilitates the use of heterogeneous verification methods. For example, recent work used various verification techniques on an autonomous space debris removal grasping system that was developed in \gls{ros}~\cite{farrell2021formal}. Another example is Antlab~\cite{10.1145/3126513}, a multi-robot task server for declarative multi-robot programming based on \gls{ros}. 
Our calculus could be used to derive system-level contracts for systems like these.

\subsection{Summary}

Many approaches partially address the challenge that we tackle, and there are various foundational approaches to compositional verification that are not restricted to any particular domain. We take inspiration from some of these, but devise a calculus that is specifically tailored for \gls{ros} systems. Other approaches are limited to specific tools and so neither support nor harness the power of incorporating a suite of heterogeneous verification approaches in the way that our approach does. To ensure traceability and consistency, we also provide a way of automatically generating runtime monitors so that we can support both static and dynamic verification for systems that operate in the real world.

\section{Specifying Verifiable Robotic Systems}

\label{sec:specifyingRobots}
\label{sec:spec}

As mentioned in \S\ref{sec:intro}, our work enables the introduction of a formal specification to existing \gls{ros} programs. 
Recall that a \gls{ros} system is composed of \textit{nodes}. Each node may subscribe to receive messages from, or publish messages to, a \textit{topic} (a buffered communication channel). Each topic is described by the message type(s) that it can accept. \gls{ros} contains several built-in message types such as \texttt{string}, \texttt{bool}, and \texttt{int8}; and custom types can also be added. 

Our verification approach begins with a manual analysis of the \gls{ros} program, to abstract it into a more manageable system model. We then specify a contract for each of the system's nodes in \textit{typed} \gls{fol}, which was chosen to lower the barrier to learning to write contracts. 
The contracts are then combined and reasoned about using our calculus, which uses the $\bigcirc$ (``next'') and $\lozenge$ (``eventually'') temporal operators from \gls{fotl}.
We use the contracts to guide the modelling and verification of individual nodes, while our tool, \textsc{Vanda}, automatically synthesises monitors from the contracts to provide a safety-net that checks that their guarantees are obeyed at runtime.
\smallskip

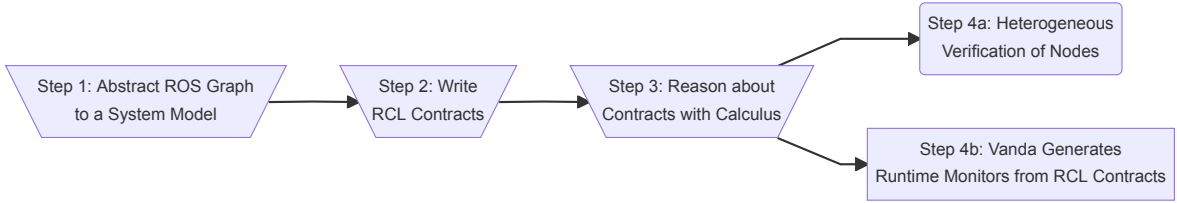
\begin{figure}[t]
\centering
\scalebox{0.65}{
\begin{tikzpicture}[font=\sffamily]

\node (Step1) [trapezium, draw, align=center, text width = 3cm, shape border rotate=180]{Step 1: Abstract ROS Graph into System Model};

\node (Step2) [trapezium, draw, align=center, text width =2cm, shape border rotate=180, right of=Step1, node distance=5cm]{Step 2: Write RCL Contracts};



\node (Step3) [trapezium, draw, align=center, text width = 3cm, shape border rotate=180, right of=Step2, node distance=5cm]{Step 3: Reason about Contracts with Calculus};

\node (Step4a) [draw, rectangle, rounded corners, text width =3cm, right of = Step3, yshift = 2cm, node distance=4cm] {Step 4a: Heterogeneous Verification of Nodes};

\node (Step4b) [process, text width =3cm, right of = Step3, yshift = -2cm, node distance=4cm] {Step 4b: Vanda Generates Runtime Monitors from RCL Contracts};

\draw [arrow, ultra thick] (Step1) -- (Step2);
\draw [arrow, ultra thick] (Step2) -- (Step3);
\draw [arrow, ultra thick] (Step3) -- (Step4a);
\draw [arrow, ultra thick] (Step3) -- (Step4b);
\end{tikzpicture}
}
\caption{A flowchart of the steps in our approach. The trapezium-shaped steps are manual processes, the rectangle with rounded corners (Step 4a) is a process that may be manual or automatic (depending on the verification tools available), and the rectangle with right-angled corners (Step 4b) is an automatic process. Note that Steps 4a and 4b can happen in parallel. If needed, previous steps may be revisited.
\label{fig:process}}
\end{figure}

\newpage
Our approach is split into the following five steps (shown in Fig.~\ref{fig:process}):
\begin{itemize}
\item[\textbf{Step 1 (Manual):}] Abstract the \gls{ros} program into a more manageable system model, containing the nodes that are critical to the program's correct behaviour. 
 Depending on the \gls{ros} program, this step may involve some form of abstraction, for example by combining multiple nodes into a \textit{compound node}, or omitting nodes that are known to be reliable.

\item[\textbf{Step 2 (Manual):}]  Write the \gls{rcl} contract for each node. The contract contains the \gls{fol} \emph{assume} and \emph{guarantee} conditions, and describes the \gls{ros} topics that the node uses (its inputs and outputs). Our \gls{rcl} tool, \textsc{Vanda}, parses the contracts, identifying where the format of the language has not been adhered to, and can synthesise runtime monitors (see Step 4b). 

\item[\textbf{Step 3 (Manual):}] Use our calculus to reason about the combination of the node contracts. This generates a list of system-level properties 
that can be used to verify the system's requirements. The calculus also identifies malformed contracts, which we debug until the inference rules are valid. This includes checking/verifying that any simplifying assumptions made to successfully apply the proof rules are met.

\item[\textbf{Step 4a (Semi-Automatic):}] Verify the nodes, using a suite of heterogeneous verification approaches. This step uses the contracts to guide the verification,  which is especially important where there is not a formal link between the verification approach and \gls{fol}. 
 
\item[\textbf{Step 4b (Automatic):}] Automatically synthesise runtime monitors for nodes from their \gls{rcl} contracts. {These monitors verify the contracts' guarantees at runtime. The assumptions are accounted for during the compositional reasoning step, where the calculus verifies that a node's assumptions are satisfied by the guarantees of the node(s) that provide its inputs.} 
\end{itemize}

\noindent This paper focuses on \gls{ros} systems, but Steps 1 and 2 above could be applied to a system where the nodes/modules are classes or methods. 
We discuss how this might work in \S\ref{sec:towards}.

The steps are numbered sequentially, but they are \textit{not} intended to be only followed linearly;
Steps 4a and 4b can also be performed in parallel. Additionally, previous steps may be revisited, if needed. For example, as identified above, Step 3 could highlight a malformed contract that would be debugged and potentially rewritten, which revisits Step 2. Similarly, Step 4a could reveal that a specification is too restrictive to be verified (for example), which could trigger a re-write (revisiting Step 2) or the restrictive part of the specification could be left to be monitored in Step 4b. Step 1 can also be revisited if we later find that an abstraction in the system model is troublesome, for example we may need to split a compound node back into its constituent parts.

It is important to note the difference in purpose between Steps 3 and 4a/4b. In Step 3, the calculus is used to combine the contracts and reveal the system-level property (or properties) that is produced by the combination of their guarantees. The verification in Step 4a statically verifies that the nodes implement their contracts, which then implies the system-level property holds. And the monitors in Step 4b check the guarantees hold at runtime; if all the monitors do not conclude $false$\footnote{Because our runtime monitors are checking for violation of a property, they return $false$ when no violation is detected.}, then the system-level property holds. If one monitor concludes $false$, then we know {that the monitored guarantee has been violated. In many cases, this also identifies the software node responsible for not satisfying the contract. However, a guarantee violation may also be the consequence of an unmet assumption. In our approach, this risk is mitigated by the compositional reasoning in Step 3, since the calculus checks that the assumptions of connected nodes follow from the guarantees of the preceding nodes. Thus, for most internal software interactions, an assumption violation corresponds to a guarantee violation of another monitored node. Assumptions of environmental inputs require separate consideration, and explicit assumption monitors could provide further diagnostic information in such cases.}

In the remainder of this section we describe: the process of abstracting a \gls{ros} system into a system model, \S\ref{sec:diagram}; 
how our contracts are described and composed, \S\ref{sec:fol}; 
 our approach to writing the contracts, \S\ref{sec:dsl}; our calculus for combining the contracts, \S\ref{sec:combination}; how the contracts can guide heterogeneous verification, \S\ref{sec:guidingHV}; and our runtime monitoring approach, \S\ref{sec:rv}.

\subsection{System Model}
\label{sec:diagram}

To start writing the contracts we need a description of system's nodes. Because we are using \gls{ros}, we can make use of the \emph{rqt\_graph} library\footnote{\emph{rqt\_graph} library: \url{http://wiki.ros.org/rqt_graph} Accessed: 03/11/2023}, which automatically generates a graph (called a \gls{ros} graph) for the system that contains all of its nodes and the communication links between them.  
\gls{ros} graphs can also display topics and actions (used to execute long-running tasks), but using only the nodes and communication links is enough for our purposes. 
If the graph is simple enough, then we can use it as the system model. However, the graph is often very large, with many nodes from well-tested libraries, so we might choose to abstract the ROS graph into a  more manageable system model. 

This section describes a heuristic approach to generating a more compact system model based on the \gls{ros} graph.
\begin{enumerate}
    \item \emph{Generate the ROS graph:} use the \emph{rqt\_graph} library to generate a graph of the system. This can be used ``as is'' (skip to Step 3 in this methodology) or can be further abstracted as detailed in the next steps.
    
    \item \emph{Remove nodes:} remove nodes that match the following conditions:
    \begin{itemize}
        \item nodes from libraries that have been demonstrated to be reliable in most cases (e.g., the \emph{move base} library\footnote{Move Base ROS Library: \url{http://wiki.ros.org/move_base} Accessed: 03/11/2023} for path planning in ROS) through community experimentation and testing;
        \item and nodes that are simple or have no impact on the nodes or properties that are being verified.        
    \end{itemize}
    
    \item \emph{Combine related nodes into compound nodes:} some nodes may be simple parts of a larger group or sub-system. These nodes can be merged into a \emph{compound} node, making sure that it retains all the information needed for verification and that it matches the implementation of the original nodes.
    
    \item \emph{Add external nodes:}  
    some nodes may be external to ROS, such as autonomous components (e.g., rational agents) and image processing (e.g., machine learning), and therefore do not appear on a ROS graph. These nodes are added to the system diagram, alongside a description of how they interact and communicate with the other nodes.
\end{enumerate}

\noindent This heuristic approach indicates how a more tractable system model can be distilled from a complicated ROS graph. The most important aspect is that the abstracted system model must still resemble the implementation of the \gls{ros} nodes. We provide an example application of this heuristic in \S\ref{sec:case1step1}.

\subsection{Background Concepts for the First-Order Logic Contracts} 
\label{sec:fol}

Our contracts use the standard definition of \gls{fol} with quantifiers ($\forall$, $\exists$) and logical connectives ($\land$, $\lor$, $\lnot$, $\Rightarrow$, $\iff$) over logical propositions including basic set theory~\cite{huth2004logic}.
For a given component/node, $C$, its contract comprises  $\mathcal{A}_C(\overline{i_C})$ (assumption/pre-condition) and $\mathcal{G}_C(\overline{o_C})$ (guarantee/post-condition), where $\overline{i_C}$ and $\overline{o_C}$ are each a vector of variables representing the node's inputs/outputs, respectively. (Note that, when discussing one specific component, we often omit the $C$ subscript.)

Complex robotic systems often produce and consume \textit{streams} of data. Our approach to stream semantics is based on well-established work in the area of stream logic programming~\cite{Ueda85,Gregory87}. In our approach, a stream is a list of data: $[ e | Tail ]$, where $e$ denotes the first element in the list and $Tail$ denotes the remaining elements. When receiving data, a component takes $e$ from the stream, processes it, and recurses over $Tail$.

Each contract states that if a node consumes input data $\overline{i}$ from its input stream ($\mathsf{InStream}([\overline{i}\mid S])$) and $\mathcal{A}(\overline{i})$ holds (i.e., $\overline{i}$ satisfies the assumption/pre-condition for correct operation of the node) then eventually the node will place some data, $\overline{o}$ on its output stream that satisfies the node's guarantee.  So if its current output stream was $T$, i.e. $\mathsf{OutStream}(T)$, before the execution of the node\footnote{$T$ is the sequence of outputs so far.} functionality on $\overline{i}$, then afterwards the output stream will be $\mathsf{OutStream}([\overline{o}\mid T])$ and $\mathcal{G}(\overline{o})$ will hold.
While our contract assumptions and guarantees are expressed using FOL, we represent the meaning of the contract by a small extension using the ``eventually'' operator, `$\lozenge$' from \gls{ltl}~\cite{Pnueli77temporal}.  Thus, a contract guarantees:
$$
\begin{array}{c}
  ( \mathsf{InStream}([\overline{i}\mid S])\lland
    \mathcal{A}(\overline{i})\lland
    \mathsf{OutStream}(T) )\\
  \Rightarrow\\
  \lozenge ( \mathsf{InStream}(S) \lland
    \mathcal{G}(\overline{o}) \lland
    \mathsf{OutStream}([\overline{o}\mid T]) )
\end{array}
  $$
for any streams $S$ and $T$.

Using the temporal logic $\lozenge$ operator enables us to abstract away from internal computation/activity. Since the internal computation will never be instantaneous, nor do we have precise timing constraints, the execution of one component is described as \emph{eventually} completing. Hence the use of the ``sometime in the future'' temporal operator `$\lozenge$'. Later in the development process, this very general temporal constraint might be refined to more precise real-time computational properties.

As in Stream Logic Programming, we will have rules to deal with end cases, such as when the remaining input list/vector is empty. Note that in such cases we might choose to terminate the processing, to suspend and wait until the list or vector is non-empty, to perform some exception handling, or undertake any other required computation. 
Note also that we make the simplifying assumption that each contract consumes one element from the input stream and generates (at most) one element for the output stream. In practice, this can be generalised to contracts consuming multiple input elements either by combining the input elements into a new (compound) element or by having several sub-contracts to handle the different input elements (as has been shown within Stream Logic Programming languages). However, for simplicity of description we keep our assumption that one input element is consumed at a time.

Also, a component can consume, or produce, multiple streams. For example, we might have a component consuming items from several streams to generate a combined output (on one stream). Whether we take an item from each stream simultaneously or just take an item from one of the streams (or any combination of these approaches) will depend on the component, and the component verification should account for this where appropriate. Note that we deliberately say nothing about the global behaviour of concurrent streams, for example whether one is generated more quickly than another, instead focusing on the first element on each relevant stream.
These concurrency aspects might well be explored in future work but, for this initial investigation, we concentrate on the straightforward case for clarity of explanation.

Although we use \gls{fol} to specify contracts, we require a machine-readable syntax for capturing and generating monitors for contracts. For this, we introduce \gls{rcl} in the next subsection.

\begin{figure*}[t]
    \centering
    \includegraphics[width=\textwidth]{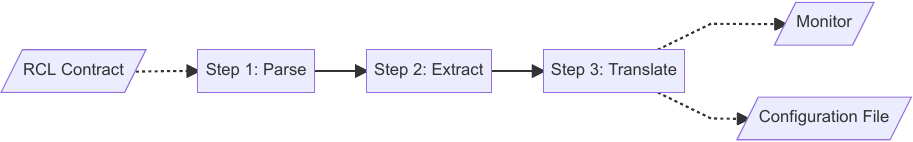}
    \caption{A flow chart showing \textsc{Vanda}'s monitor synthesis workflow. The rectangles represent steps in the workflow, and the solid lines show the flow of control between the steps. The parallelograms represent input or output files, and the dotted lines show the file being input to or output from a step. \label{fig:vandaFlow} }
\end{figure*}

\begin{table}[t]
\centering
\begin{tabularx}{\textwidth}{rX}
    start :& contract\_clause+ \\
    contract\_clause : & node\_clause | context\_clause \\
    context\_clause : & ``context'' ``\{'' (type\_declaration \\ ~& \hspace{1em}| constant\_declaration)+ ``\}'' \\
    type\_declaration : & STRING ``:'' type\_declaration\_part ``;" \\
    constant\_declaration : & STRING ``='' type\_declaration\_part ``;''\\
    type\_declaration\_part : & ? a function declaration, set, sequence, or tuple ? \\
    node\_clause : & ``node'' STRING ``\{'' inputs \hspace{0.2em} outputs \hspace{0.2em} topic\_list (assume)* \hspace{0.2em} (guarantee)+  ``\}''\\
    inputs : & ``inputs'' ``('' (io\_var (``,'' io\_var)*)?  ``)'' \\
    outputs :  & ``outputs'' ``('' (io\_var (``,'' io\_var)*)? ``)'' \\
    io\_var : & STRING ``:'' STRING \\
    topic\_list : & ``topics'' ``('' (topic (``,'' topic)*)? ``)'' \\
    topic : & TYPE \hspace{0.2em} STRING \\ & \hspace{1em} (``matches'' ``('' topic\_match\_name ``)'' )? \\
    topic\_match\_name : & io\_pointer? STRING \\
    io\_pointer :  & ``in.'' | ``out.'' \\
    assume : & (``assume'') ``('' formula ``)'' \\ 
    guarantee : & (``guarantee'') ``('' formula ``)'' \\
    formula : & ? Textual Definition of \gls{fol} ? \\
    TYPE : & ? Types from \gls{ros} or custom type ? \\
    STRING : & ? Character String?     
\end{tabularx}
 \caption{A simplified description of \gls{rcl}'s syntax in Extended Backus-Naur form (EBNF). A ``|'' shows alternatives, quotation marks shows literals, a ``?'' shows zero or one instances, a ``*'' shows zero or more instances, a ``+'' shows one or more instances. Textual descriptions of a rule are enclosed in ``?''. }
    \label{tab:rclGrammar}
\end{table}

\subsection{Specifying Nodes in the ROS Contract Language}
\label{sec:dsl} 

\gls{rcl} captures a node's \gls{fol} contract; plus a description of the node's inputs and outputs, and the \gls{ros} topics to which they correspond. The contracts describe the behaviour of a node at a higher level of abstraction than the \gls{ros} implementation or a task/mission specification. {The \gls{rcl} contracts are agnostic of the communication type between \gls{ros} nodes (topics or services, for example) but the link in the \gls{rcl} contracts between inputs/outputs and a corresponding \gls{ros} topic enables the synthesis of \gls{rv} monitors. This work uses a version of our chosen \gls{rv} framework (see \S\ref{sec:rv}) that only monitors topics, the stream abstraction we describe in \S\ref{sec:fol} does not prevent an extension that can capture \gls{ros} services and actions. We discuss this possible extension, supported by a recent update to the \gls{rv} framework, further in \S\ref{sec:rv}.
} 

We define three rules to map the values from a \gls{ros} message to the matching variable in an \gls{rcl} contract. The rule that is used depends on the type of the \gls{rcl} variable that the \gls{ros} message matches. These rules are suitable for most mappings but may not cover every situation, so bespoke rules can be added to capture more complicated mappings. We discuss the application of these mapping rules for our case study in \S~\ref{sec:case1step2}. 

The simplest rule is the \textbf{Scalar} rule, which is applied when the \gls{rcl} variable is a scalar type. The rule maps a single variable in a \gls{ros} topic to a single \gls{rcl} variable of the same type. For example, if the \gls{ros} topic and \gls{rcl} variable are both integers, we can map one directly to the other.

The \textbf{Collection} rule applies when the \gls{rcl} variable is a collection type (a set, sequence, or tuple) containing constants. If the \gls{ros} message is a single value, then it should map to one of the elements in the collection. We use this to map \gls{ros} strings into \gls{rcl}, for example if the \gls{ros} message is a string such as ``stop'' or ``go'' \textbf{Collection} can map this into a constant in \gls{rcl} if the contract defines the variable's type as $\{stop, go\}$. Because the contracts assume that the variable is drawn from the given collection, it is left for the developer to decide how to handle situations where the value of the \gls{ros} message is not present in the collection. 

The \textbf{Function} rule applies when the \gls{rcl} variable is a function. The rule maps the values of the variables in the \gls{ros} message to the function's parameters. If there is a mismatch between the number of variables in the \gls{ros} message and the number of parameters in the \gls{rcl} function, then we map the variables from the \gls{ros} message that match the types of the function parameters in the order that they appear in the message. For example, if the \gls{ros} message contains two variables \texttt{int16, int16} and the \gls{rcl} function is $\mathbb{N} \cross \mathbb{N} \rightarrow \mathbb{B}$, then \textbf{Function} applies the \textbf{Scalar} rule to map the first \texttt{int16} variable to the first $\mathbb{N}$ parameter and the second \texttt{int16} variable to the first $\mathbb{N}$ parameter. However, if the \gls{ros} message has three variables \texttt{string, int16, int16} then \textbf{Function} will ignore the \texttt{string} variable and apply \textbf{Scalar} to map the two \textit{int16} variables as before.

Each assumption and guarantee is declared separately, in a plain-text version of \gls{fol}, which uses keywords such as \texttt{forall} and \texttt{in} to represent logical operators. Table~\ref{tab:rclGrammar} shows a simplified Extended Backus-Naur form (EBNF) grammar for an \gls{rcl} contract.
The ``TYPE'' rule defines the built-in \gls{ros} message types and allows custom types to be defined.
Listing~\ref{list:CS1AgentRCL} (\S\ref{sec:case1step2}) shows one of the \gls{rcl} contracts that we used during this work.

\textsc{Vanda}\footnote{\textsc{Vanda}, which means `Oath' in Quenya, is available at: \url{https://github.com/autonomy-and-verification/ros-contract-language/tree/v0.3.1-ras}. Accessed: 16/01/2026} parses \gls{rcl} files, synthesises executable monitors (\S\ref{sec:rv}), and can produce \LaTeX{} versions of the contracts. It is written in Python3 and uses the Lark parsing library\footnote{Lark parsing library: \url{https://github.com/lark-parser/lark} Accessed: 03/11/2023}. 
Fig.~\ref{fig:vandaFlow} shows the steps taken (and files involved) in the synthesis of a monitor from an \gls{rcl} contract. This process comprises the following three steps. (1) \textbf{parse}: parse a contract file and, if the contract is well-formed, produce a parse tree. (2) \textbf{extract}: pre-process the parse tree from Step 1 to extract the node name, topics and, guarantees into a \texttt{Contract} object. 
(3) \textbf{translate}: a \textit{contract translator} uses the \texttt{Contract} object from Step 2 to produce the configuration file and monitor structure. The guarantees are translated by a \textit{\gls{fol} translator} class, which is a Lark \texttt{Interpreter} that converts the \gls{fol} parse tree into the monitoring language (\S\ref{sec:rv}). 

{Note that the current implementation of Vanda extracts the node name, topics, and guarantees for monitor synthesis. Assumptions remain part of the RCL contract and are central to the compositional reasoning in Step 3, where the calculus checks if a node's assumptions are entailed by the guarantees of the nodes that provide its inputs. Consequently, many assumptions do not need to be monitored separately because they are either trivially true, supplied by the environment, or already represented as guarantees of other nodes. Monitoring only guarantees also reduces runtime overhead, since the generated ROSMonitoring monitors need to evaluate fewer conditions at runtime.}

\textsc{Vanda} is built so as to make it easy to change the input and output formats. The input language is defined in a Lark grammar, so that we can update \gls{rcl} if needed. The translation is implemented by two files (the contract translator and \gls{fol} translator mentioned previously), so updating {\Vanda} to produce monitors in a different formalism should be relatively straightforward.

So far, we have described how contracts for individual nodes are expressed in \gls{fol} and discussed how these are encoded in \gls{rcl}. Next we present our calculus for reasoning about how specifications for individual nodes are combined. 

\subsection{Calculus for Combining Node Specifications} 
\label{sec:combination}

Nodes in a (modular)  system can be linked as long as their input types and  requirements match. The basic way to describe
these structures is to first have the contract capture all input and output variables and then describe how they are
combined using suitable inference rules. We compose the contracts of individual nodes in a number of ways, the simplest being sequential composition:\linebreak

\begin{minipage}{0.7\textwidth}
\textbf{R1:}
$$
\begin{array}{l}
\forall \overline{i_1},\overline{o_1}\cdot \mathcal{A}_1(\overline{i_1})\ \Rightarrow\ \lozenge \mathcal{G}_1(\overline{o_1})\\
\hspace{70pt}\vdots \\
\forall\overline{i_n},\overline{o_n}\cdot \mathcal{A}_n(\overline{i_n})\ \Rightarrow\ \lozenge \mathcal{G}_n(\overline{o_n})\\
\overline{o_1} = \overline{i_2} \land \ldots \land \overline{o_{n-1}} = \overline{i_n}\\
\vdash\ \bigg( \forall\overline{o_1},\overline{i_2}\cdot \mathcal{G}_1(\overline{o_1})\ \Rightarrow\ \mathcal{A}_2(\overline{i_2})\\
\hspace{68pt}\land \hspace{100pt} \\
\hspace{70pt}\vdots \hspace{100pt} \\
\hspace{68pt}\land \hspace{100pt} \\
\ \ \ \ \ \forall\overline{o_{n-1}},\overline{i_n}\cdot \mathcal{G}_{n-1}(\overline{o_{n-1}})\ \Rightarrow\ \mathcal{A}_n(\overline{i_n}) \bigg) \\ 
\hline 
\bigg( \forall\overline{i_1},\overline{o_n}\cdot \mathcal{A}_1(\overline{i_1})\ \Rightarrow\ \lozenge\mathcal{G}_n(\overline{o_n}) \bigg)
\end{array}
$$
\end{minipage}
\qquad
\begin{minipage}{0.3\textwidth}
\scalebox{0.7}{
\hspace{-20pt}
\begin{tikzpicture}[node distance=1em]
\node (N1) [process]{\parbox[t][][t]{1cm}{\centering{$1$}}};
\node (A1)[left of = N1, xshift = -2em, yshift = 2em]{\color{blue}$\mathcal{A}_1(\overline{i_1})$};
\node (G1)[right of = N1, xshift = 2em, yshift = 2em]{\color{blue}$\mathcal{G}_1(\overline{o_1})$};

\node (N2) [below of=N1, yshift=-4em]{\parbox[t][][t]{1cm}{\centering{$\bullet$}}};
\node (A2)[below of = N2]{$\bullet$};
\node (G2)[below of = A2]{$\bullet$};

\node (N3) [process, below of=G2, yshift = -4em]{\parbox[t][][t]{1cm}\centering{$n$}};
\node (A3)[left of = N3, xshift = -2em, yshift = 2em]{\color{blue}$\mathcal{A}_{n}(\overline{i_{n}})$};
\node (G3)[right of = N3, xshift = 2em, yshift = 2em]{\color{blue}$\mathcal{G}_{n}(\overline{o_{n}})$};

\draw [arrow] (N1) -- node[above]{}(N2);
\draw [arrow] (G2) -- node[above]{}(N3);
\end{tikzpicture}
}
\end{minipage}
\linebreak
\linebreak

\noindent Rule \textbf{R1} applies to nodes that are connected in a \textit{linear sequence}, where the output of one is the input of the next, etc. 
If the guarantee of the first implies the assumption of the second etc. then, if the assumption of the first node holds, we can conclude that the guarantees of the final node in the sequence will eventually hold. This rule provides a basic starting point, however, robotic systems are generally more complex than this, often including multiple, branching outputs. 
\linebreak

\begin{minipage}{0.9\textwidth}
\begin{minipage}{0.6\textwidth}
\textbf{R2:}
$$
\begin{array}{l}
\forall \overline{i_1},\overline{o_1}\cdot \mathcal{A}_1(\overline{i_1})\ \Rightarrow\ \lozenge \mathcal{G}_1(\overline{o_1})\\
\hspace{70pt}\vdots \\
\forall\overline{i_n},\overline{o_n}\cdot\ \mathcal{A}_n(\overline{i_n})\ \Rightarrow\ \lozenge \mathcal{G}_n(\overline{o_n})\\
\overline{o}_1 = \overline{i_2} \cup\ldots\cup \overline{i_n}\\
\vdash\ \bigg( \forall\overline{o_1},\overline{i_2}, \ldots , \overline{i_n} \cdot \ \mathcal{G}_1(\overline{o_1})\Rightarrow\ \mathcal{A}_2(\overline{i_2})
\\
\hspace{68pt}\land \hspace{100pt} \\
\hspace{70pt}\vdots \hspace{100pt} \\
\hspace{68pt}\land \hspace{40pt} \mathcal{A}_n(\overline{i_n}) \bigg)\\ 
\hline
\bigg( \forall\overline{i_1},\overline{o_2}, \ldots , \overline{o_n} \cdot \mathcal{A}_1(\overline{i_1})\\ \ \ \ \ \Rightarrow\ \lozenge\mathcal{G}_2(\overline{o_2}) \land \ldots \land \lozenge\mathcal{G}_n(\overline{o_n}) \bigg)
\end{array}
$$
\end{minipage}
\begin{minipage}{0.3\textwidth}
\scalebox{0.7}{
\hspace{-30pt}
\begin{tikzpicture}[node distance=1em]
\node (N1) [process,  yshift=-6em]{\parbox[t][][t]{1cm}{\centering{$1$}}};
\node (A1)[left of = N1, xshift = -2em, yshift = 2em]{\color{blue}$\mathcal{A}_1(\overline{i_1})$};
\node (G1)[right of = N1, xshift = 2em, yshift = 2em]{\color{blue}$\mathcal{G}_1(\overline{o_1})$};

\node (N2) [process, left of=N1, xshift=-5em, yshift = -6em]{\parbox[t][][t]{1cm}{\centering{$2$}}};
\node (A2)[left of = N2, xshift = -2em, yshift = -2em]{\color{blue}$\mathcal{A}_2(\overline{i_2})$};
\node (G2)[right of = N2, xshift = 2em, yshift = -2em]{\color{blue}$\mathcal{G}_2(\overline{o_2})$};


\node(dots)[right of = N2, xshift = 5em]{\parbox[t][][t]{1cm}{\centering{$\bullet\bullet\bullet$}}};

\node (NN) [process, right of=dots, xshift = 5em]{\parbox[t][][t]{1cm}\centering{$n$}};
\node (AN)[left of = NN, xshift = -2em, yshift = -2em]{\color{blue}$\mathcal{A}_n(\overline{i_n})$};
\node (GN)[right of = NN, xshift = 2em, yshift = -2em]{\color{blue}$\mathcal{G}_n(\overline{o_n})$};

\draw [arrow] (N1) -- node[above]{}(N2);
\draw [arrow] (N1) -- node[above]{}(NN);
\end{tikzpicture}
}
\end{minipage}
\end{minipage}
\linebreak
\linebreak

\noindent Rule \textbf{R2} is used when a node has \textit{branching outputs}. Here, the combined inputs of all of the `leaf' nodes is equal to the output of the `root' node, while the guarantee of the `root' node implies the assumptions of each `leaf' node, as shown above 
(where we assume that the union operator, $\cup$, can takes vectors as parameters and merge them in the same way as with sets). \smallskip

Rule \textbf{R3} deals with the converse architecture, where a node's input is the \textit{union of several outputs} from other nodes. As expected, this rule is essentially the dual of \textbf{R2}. Note that there is a simplifying assumption here: that the outputs persist once generated, according to the $\lozenge$ constraint. Under this assumption, all the required inputs (from nodes $1$ to $n-1$) will be available at the same time. In future work, we will consider weakening this assumption, thus adding further timing constraints. We assume that a user of this calculus ensures that this simplifying assumption is met in order to use the rules.\\

\begin{minipage}{0.9\textwidth}
\begin{minipage}{0.6\textwidth}
\textbf{R3:}
$$
\begin{array}{l}
\forall \overline{i_1},\overline{o_1}\cdot \mathcal{A}_1(\overline{i_1})\ \Rightarrow\ \lozenge \mathcal{G}_1(\overline{o_1})\\
\hspace{70pt}\vdots\\
\forall\overline{i_n},\overline{o_n}\cdot \mathcal{A}_n(\overline{i_n})\ \Rightarrow\ \lozenge \mathcal{G}_n(\overline{o_n})\\
\overline{i_n} = \overline{o_1}\cup \ldots \cup \overline{o_{n-1}}\\
\vdash\ \bigg( \forall\overline{i_n},\overline{o_1},\ldots, \overline{o_{n-1}}\cdot \mathcal{G}_1(\overline{o_1})
\\
\hspace{68pt}\land \hspace{100pt} \\
\hspace{70pt}\vdots \hspace{100pt} \\
\hspace{68pt}\land \hspace{100pt} \\
\hspace{40pt} \mathcal{G}_{n-1}(\overline{o_{n-1}})\ \Rightarrow \mathcal{A}_n(\overline{i_n}) \bigg) \\ 
\hline 
\bigg( \forall\overline{i_1},\ldots\ \overline{i_{n-1}},\overline{o_n}\cdot \mathcal{A}_1(\overline{i_1}) \land \ldots \land \mathcal{A}_{n-1}(\overline{i_{n-1}})\Rightarrow \lozenge\mathcal{G}_n(\overline{o_n}) \bigg)
\end{array}
$$
\end{minipage}
\begin{minipage}{0.3\textwidth}
\scalebox{0.7}{
\hspace{-30pt}
\begin{tikzpicture}[node distance=1em]
\node (NN) [process,  yshift=-6em]{\parbox[t][][t]{1cm}{\centering{$n$}}};
\node (ANN)[left of = NN, xshift = -2em, yshift = -2em]{\color{blue}$\mathcal{A}_n(\overline{i_n})$};
\node (GNN)[right of = NN, xshift = 2em, yshift = -2em]{\color{blue}$\mathcal{G}_n(\overline{o_n})$};

\node (N1) [process, left of=NN, xshift=-5em, yshift = 6em]{\parbox[t][][t]{1cm}{\centering{$1$}}};
\node (A1)[left of = N1, xshift = -2em, yshift = 2em]{\color{blue}$\mathcal{A}_1(\overline{i_1})$};
\node (G1)[right of = N1, xshift = 2em, yshift = 2em]{\color{blue}$\mathcal{G}_1(\overline{o_1})$};

\node(dots)[right of = N1, xshift = 5em]{\parbox[t][][t]{1cm}{\centering{$\bullet\bullet\bullet$}}};

\node (NN-1) [process, right of=dots, xshift = 5em]{\parbox[t][][t]{1cm}\centering{$n-1$}};
\node (AN-1)[left of = NN-1, xshift = -2em, yshift = 2em]{\color{blue}$\mathcal{A}_{n-1}(\overline{i_{n-1}})$};
\node (GN-1)[right of = NN-1, xshift = 2em, yshift = 2em]{\color{blue}$\mathcal{G}_{n-1}(\overline{o_{n-1}})$};

\draw [arrow] (N1) -- node[above]{}(NN);
\draw [arrow] (NN-1) -- node[above]{}(NN);
\end{tikzpicture}
}
\end{minipage}
\end{minipage}
\linebreak
\linebreak

These three simple inference rules (\textbf{R1}, \textbf{R2} and \textbf{R3}) constitute our basic calculus for reasoning about node-level \gls{fol} contracts in a robotic system. We do \emph{not} specify the fine-grained concurrency/streaming of processes/data but are just specifying the interface expectations of nodes in a robotic system. 

We use the $\lozenge$ operator from \gls{fotl} to represent that some arbitrary time has passed.  
Using the $\lozenge$ operator gives us the flexibility to describe and refine a range of temporal aspects. We assume that contracts are neither mutually dependent, nor circularly dependant. However, we intend to investigate this in future work and potentially leverage existing techniques like~\cite{elkader2015automated}.

We can extend \textbf{R1-3} with rule \textbf{R4}, below, which captures looping behaviour. We consider this rule to be an extension to our core calculus because it is directly derived from the first three rules.\\
 
\begin{minipage}{0.6\textwidth}
\textbf{R4:}
$$
\begin{array}{l}
\forall \overline{i_1},\overline{o_1}\cdot \mathcal{A}_1(\overline{i_1})\ \Rightarrow\ \lozenge \mathcal{G}_1(\overline{o_1})\\
\forall \overline{i_2},\overline{o_2}\cdot \mathcal{A}_2(\overline{i_2})\ \Rightarrow\ \lozenge \mathcal{G}_2(\overline{o_2}) \\
\forall\overline{i_3},\overline{o_3}\cdot\ \mathcal{A}_3(\overline{i_3})\ \Rightarrow\ \lozenge \mathcal{G}_3(\overline{o_3})\\
\forall\overline{i_4},\overline{o_4}\cdot\ \mathcal{A}_4(\overline{i_4})\ \Rightarrow\ \lozenge \mathcal{G}_4(\overline{o_4})\\
\overline{i_2} = \overline{o_1} \cup \overline{o_3} \land \overline{i_3} = \overline{o_2} \land \overline{o_3} \subseteq \overline{i_4}\\ 
\vdash\forall \overline{i_2}, \overline{o_3} \cdot \mathcal{A}_2(\overline{i_2}) \Rightarrow \lozenge \mathcal{G}_3(\overline{o_3})\\
\vdash\forall\overline{i_3},\overline{o_2},\overline{o_4} \cdot \mathcal{A}_3(\overline{i_3})\ \Rightarrow\ \lozenge \mathcal{G}_2(\overline{o_2}) \land \lozenge \mathcal{G}_4(\overline{o_4}) \\
\vdash\forall\overline{i_1},\overline{i_3},\overline{o_2} \cdot \mathcal{A}_1(\overline{i_1})\land  \mathcal{A}_3(\overline{i_3})\ \Rightarrow\ \lozenge \mathcal{G}_2(\overline{o_2}) \\\hline
\bigg(  \forall\overline{i_1},\overline{o_4} \cdot \mathcal{A}_1(\overline{i_1})\ \Rightarrow\ \lozenge\mathcal{G}_4(\overline{o_4}) \bigg)
\end{array}
$$
\end{minipage}
\qquad
\begin{minipage}{0.4\textwidth}
\scalebox{0.6}{
\begin{tikzpicture}[node distance=1em]
\node (N1) [process]{\parbox[t][][t]{1cm}{\centering{$1$}}};
\node (A1)[left of = N1, xshift = -2em, yshift = 2em]{\color{blue}$\mathcal{A}_1(\overline{i_1})$};
\node (G1)[right of = N1, xshift = 2em, yshift = 2em]{\color{blue}$\mathcal{G}_1(\overline{o_1})$};

\node (N2) [process, below of=N1, yshift=-5em]{\parbox[t][][t]{1cm}{\centering{$2$}}};
\node (A2)[left of = N2, xshift = -2em, yshift = 2em]{\color{blue}$\mathcal{A}_2(\overline{i_2})$};
\node (G2)[right of = N2, xshift = 2em, yshift = 2em]{\color{blue}$\mathcal{G}_2(\overline{o_2})$};

\node (N3) [process, below of=N2, yshift = -5em]{\parbox[t][][t]{1cm}\centering{$3$}};
\node (A3)[left of = N3, xshift = -2em, yshift = 2em]{\color{blue}$\mathcal{A}_{3}(\overline{i_{3}})$};
\node (G3)[right of = N3, xshift = 2em, yshift = 2em]{\color{blue}$\mathcal{G}_{3}(\overline{o_{3}})$};

\node (N4) [process, below of=N3, yshift = -5em]{\parbox[t][][t]{1cm}\centering{$4$}};
\node (A4)[left of = N4, xshift = -2em, yshift = 2em]{\color{blue}$\mathcal{A}_{4}(\overline{i_{4}})$};
\node (G4)[right of = N4, xshift = 2em, yshift = 2em]{\color{blue}$\mathcal{G}_{4}(\overline{o_{4}})$};

\draw [arrow] (N1) -- node[above]{}(N2);
\draw [arrow] (N2) to [out=-120,in=120](N3);
\draw [arrow] (N3) -- node[above]{}(N4);
\draw [arrow] (N3) to [out=60,in=-60] (N2);
\end{tikzpicture}
}
\vspace{20pt}
\end{minipage}
\noindent Observe that the three lines with the $\vdash$ symbol are a direct result of applying \textbf{R1}, \textbf{R2} and \textbf{R3}, respectively, to the nodes in the loop illustration above. The conclusion below the line is drawn from their combination and simplification. 
Although not part of our core calculus, we list \textbf{R4} here to explicitly capture the behaviour of loops and it can easily be extended to incorporate an arbitrary number of nodes between nodes 2 and 3 in the loop illustration above. Alternatively, we could augment \textbf{R4} to include fixed-point reasoning in first-order temporal logic~\cite{BB89mutl}. It is well-known that reasoning over loops can be complex, typically requiring the addition of loop invariants that are not always obvious or straightforward to define \cite{furia2014loop}. In the derivation above, we assume that all of the nodes are verified to meet their assumptions and guarantees and that they terminate, producing a correct result for each given input. When it comes to automating these derivation rules we may need to consider how to define appropriate, and ideally generic so that they can be easily instantiated, loop invariants but this is left as future work.

Our inference rules describe how node contracts should be interpreted when multiple nodes are combined in various ways.
These rules are simple and therefore transparently sound. For brevity, we omit the soundness proofs here because they do not add to the novelty of this paper.
It is a separate verification step to demonstrate that the individual node itself obeys its own contract and we discuss this in more detail in the next subsection. It is through this node verification process that users verify that whenever the assumptions for a particular node are true, then the guarantee should hold (after execution). In this way, our calculus, in practice, will avoid situations where trivially-true implication statements occur. Our calculus is thus to be used only with nodes {that have been verified to preserve their guarantees when their assumptions are satisfied by the environment.}

At present, our calculus is applied manually and we intend to investigate future ways to automate its application. This includes incorporating useful tactics for simplifying FOL formulas as needed. For this, we intend to incorporate a first-order theorem proving approach such as Vampire \cite{kovacs2013first}.

\subsection{Guiding Heterogeneous Verification}
\label{sec:guidingHV}

Our \gls{rcl} contracts provide a high-level specification of the system's requirements.  
These \textit{node-level} contracts are used to \textit{guide} the verification of individual nodes, because a formal link is often difficult or impossible. However, the \gls{fol} specification, and other information in the contract, can be used to inform the (formal or non-formal) verification process.

Each node can present its own challenges when verifying that it obeys its \gls{fol} contract, so the most suitable verification method should be chosen for each node. Some nodes may be amenable to formal verification, such as an agent that can be model checked for correctness properties (as in our previous work~\cite{Cardoso2020}). Other nodes may be based on neural networks, e.g. a vision classifier that might need a specific testing approach~\cite{huang_survey_2019}.
Some nodes might be more critical to the system's safety requirements than others, these nodes are likely to be the focus of the most robust formal verification (as we suggest in \cite{luckcuck_matt_2021_5012322, luckcuck_using_2021}). Our approach leaves whoever is verifying the node to choose the most suitable verification method(s), since they are best placed to make this decision. 

We use the contract's high-level specification as a guide, or a target for what properties a particular node should obey, supporting developing the system from abstract specifications. Specifying what a system should do is often the most time-consuming part of formal verification~\cite{rozier2016specification}, so reusing a specification throughout the development process makes that initial effort more worthwhile.  

When developing a system from an abstract specification, a developer can start by specifying a high-level contract for a node's basic behaviour, and then verify more concrete properties about the node's functionality using their chosen verification method.
For example, as in~\cite{bourbouh2021integrating}, a high-level contract for a planner might require that all of the plans it produces are obstacle-free, but the detailed planner verification might also verify that all points in the plan are valid (e.g. within the map), that the planner does not deadlock, or that the plan conforms to some measure of optimality. 

Crucially, our approach enables a system to be verified using a range of verification techniques without needing to decide beforehand which techniques will be used. 
As mentioned in \S\ref{sec:intro}, using a variety of heterogeneous verification techniques on one system can be beneficial, particularly in the robotics domain~\cite{Luckcuck2019, Farrell2018, Cardoso2020, bourbouh2021integrating, farrell2021formal}. 

The path from an \gls{rcl} contract to \textit{formal} verification is fairly clear, as we describe in \S\ref{sec:exampleSystem}. Some formal methods are also based on \gls{fol} and sets, so the guarantees in the contracts can be checked directly. Whereas, for other methods, more effort may be needed to capture the contract's properties in its specification language. In either case, the contract describes the properties that the chosen formal method must check.

The path from an \gls{rcl} contract to \textit{non-formal} verification is made easier because each contract provides an unambiguous specification of a node's requirements.
A suitably skilled test engineer should be able to, for example, create software tests, simulation scenarios, etc.; that check the node for the properties described in the guarantee. Having a formal specification provides validation that the `right' properties are being verified.

\subsection{A Runtime Verification Safety-Net}
\label{sec:rv}

The final part of our approach is to introduce a \emph{safety net} of \gls{rv} monitors, one for each \gls{rcl} contract. Each monitor compares a trace of events produced by the system, with a formal model of the intended behaviour. This \gls{rv} safety net ensures that the nodes conform to their contracts and, consequently, support the inferred system-level properties. 
 
We use the formalism-agnostic general-purpose framework, ROSMonitoring~\cite{DBLP:conf/taros/FerrandoC0AFM20}, which is built for \gls{rv} of \gls{ros} systems. The monitors used by ROSMonitoring follow a standard publish/subscribe pattern. Each monitor is a \gls{ros} node that subscribes to the topics needed to observe the behaviour that is relevant to its property, and publishes a message to inform the system that a violation was observed. We chose ROSMonitoring because its monitors can be easily distributed through the system to check the contracts for each node, and it can be used with various versions of \gls{ros}/ROS2.

\tikzset{every picture/.style={line width=0.75pt}} 

\begin{figure*}[t]
\centering

\scalebox{0.55}{

\begin{tikzpicture}[x=0.75pt,y=0.75pt,yscale=-1,xscale=1]

\draw   (90,84) -- (172,84) -- (172,124) -- (90,124) -- cycle ;
\draw  [fill={rgb, 255:red, 224; green, 214; blue, 123 }  ,fill opacity=1 ] (46.1,68) -- (23,68) -- (23,22) -- (56,22) -- (56,58.1) -- cycle -- (46.1,68) ; \draw   (56,58.1) -- (48.08,60.08) -- (46.1,68) ;
\draw    (55,50) .. controls (82.3,48.05) and (69.67,66.06) .. (88.49,82.72) ;
\draw [shift={(90,84)}, rotate = 218.99] [fill={rgb, 255:red, 0; green, 0; blue, 0 }  ][line width=0.75]  [draw opacity=0] (8.93,-4.29) -- (0,0) -- (8.93,4.29) -- cycle    ;


\draw  [fill={rgb, 255:red, 108; green, 154; blue, 209 }  ,fill opacity=1 ] (317.1,127) -- (294,127) -- (294,81) -- (327,81) -- (327,117.1) -- cycle -- (317.1,127) ; \draw   (327,117.1) -- (319.08,119.08) -- (317.1,127) ;
\draw  [fill={rgb, 255:red, 108; green, 154; blue, 209 }  ,fill opacity=1 ] (337.1,147) -- (314,147) -- (314,101) -- (347,101) -- (347,137.1) -- cycle -- (337.1,147) ; \draw   (347,137.1) -- (339.08,139.08) -- (337.1,147) ;
\draw  [fill={rgb, 255:red, 108; green, 154; blue, 209 }  ,fill opacity=1 ] (354.1,175) -- (331,175) -- (331,129) -- (364,129) -- (364,165.1) -- cycle -- (354.1,175) ; \draw   (364,165.1) -- (356.08,167.08) -- (354.1,175) ;
\draw    (172,104) .. controls (199.36,118.93) and (224.25,141.77) .. (302.32,139.04) ;
\draw [shift={(303.5,139)}, rotate = 537.8299999999999] [fill={rgb, 255:red, 0; green, 0; blue, 0 }  ][line width=0.75]  [draw opacity=0] (8.93,-4.29) -- (0,0) -- (8.93,4.29) -- cycle    ;

\draw  [fill={rgb, 255:red, 108; green, 154; blue, 209 }  ,fill opacity=1 ] (304.1,236) -- (281,236) -- (281,190) -- (314,190) -- (314,226.1) -- cycle -- (304.1,236) ; \draw   (314,226.1) -- (306.08,228.08) -- (304.1,236) ;
\draw    (172,104) .. controls (197.37,116.94) and (178.69,193.23) .. (279.47,190.05) ;
\draw [shift={(281,190)}, rotate = 537.77] [fill={rgb, 255:red, 0; green, 0; blue, 0 }  ][line width=0.75]  [draw opacity=0] (8.93,-4.29) -- (0,0) -- (8.93,4.29) -- cycle    ;

\draw  [dash pattern={on 4.5pt off 4.5pt}] (230,34.5) -- (400,34.5) -- (400,263) -- (230,263) -- cycle ;

\draw    (315.5,201) .. controls (364.01,214.86) and (381.16,123.85) .. (444.56,121.06) ;
\draw [shift={(446.5,121)}, rotate = 539.12] [fill={rgb, 255:red, 0; green, 0; blue, 0 }  ][line width=0.75]  [draw opacity=0] (8.93,-4.29) -- (0,0) -- (8.93,4.29) -- cycle    ;

\draw  [fill={rgb, 255:red, 194; green, 108; blue, 214 }  ,fill opacity=1 ] (469.1,133) -- (446,133) -- (446,87) -- (479,87) -- (479,123.1) -- cycle -- (469.1,133) ; \draw   (479,123.1) -- (471.08,125.08) -- (469.1,133) ;
\draw   (467,192.5) .. controls (467,179.52) and (490.95,169) .. (520.5,169) .. controls (550.05,169) and (574,179.52) .. (574,192.5) .. controls (574,205.48) and (550.05,216) .. (520.5,216) .. controls (490.95,216) and (467,205.48) .. (467,192.5) -- cycle ;
\draw    (316.53,215.81) .. controls (382.37,241.21) and (409.05,167.76) .. (465.29,191.74) ;
\draw [shift={(467,192.5)}, rotate = 204.56] [fill={rgb, 255:red, 0; green, 0; blue, 0 }  ][line width=0.75]  [draw opacity=0] (8.93,-4.29) -- (0,0) -- (8.93,4.29) -- cycle    ;
\draw [shift={(314.5,215)}, rotate = 22.38] [fill={rgb, 255:red, 0; green, 0; blue, 0 }  ][line width=0.75]  [draw opacity=0] (8.93,-4.29) -- (0,0) -- (8.93,4.29) -- cycle    ;
\draw  [fill={rgb, 255:red, 119; green, 221; blue, 197 }  ,fill opacity=1 ] (458.1,274) -- (435,274) -- (435,228) -- (468,228) -- (468,264.1) -- cycle -- (458.1,274) ; \draw   (468,264.1) -- (460.08,266.08) -- (458.1,274) ;
\draw    (520.41,218.31) .. controls (519.83,242.69) and (531.75,257.34) .. (468.93,246.34) ;
\draw [shift={(467,246)}, rotate = 370.15] [fill={rgb, 255:red, 0; green, 0; blue, 0 }  ][line width=0.75]  [draw opacity=0] (8.93,-4.29) -- (0,0) -- (8.93,4.29) -- cycle    ;
\draw [shift={(520.5,216)}, rotate = 93.3] [fill={rgb, 255:red, 0; green, 0; blue, 0 }  ][line width=0.75]  [draw opacity=0] (8.93,-4.29) -- (0,0) -- (8.93,4.29) -- cycle    ;

\draw    (479,109) .. controls (527.02,122.72) and (471.32,147) .. (487.89,170.56) ;
\draw [shift={(489,172)}, rotate = 230.19] [fill={rgb, 255:red, 0; green, 0; blue, 0 }  ][line width=0.75]  [draw opacity=0] (8.93,-4.29) -- (0,0) -- (8.93,4.29) -- cycle    ;

\draw    (298.85,186.06) .. controls (285.58,145.79) and (303.42,154.29) .. (319.73,153.16) ;
\draw [shift={(321.5,153)}, rotate = 533.29] [fill={rgb, 255:red, 0; green, 0; blue, 0 }  ][line width=0.75]  [draw opacity=0] (8.93,-4.29) -- (0,0) -- (8.93,4.29) -- cycle    ;
\draw [shift={(299.5,188)}, rotate = 251.18] [fill={rgb, 255:red, 0; green, 0; blue, 0 }  ][line width=0.75]  [draw opacity=0] (8.93,-4.29) -- (0,0) -- (8.93,4.29) -- cycle    ;

\draw (131,104) node  [scale=1.3] [align=center] {instrument};
\draw (40,9) node [scale=1.5] [align=left] {config.yaml};
\draw (318,62) node [scale=1.5] [align=center] {nodes};
\draw (296,249) node [scale=1.5] [align=left] {monitor.py};
\draw (385,22) node  [scale=1.5] [align=left] {ROS};
\draw (462,74) node [scale=1.5] [align=left] {log.txt};
\draw (520.5,192.5) node  [scale=1.3] [align=center] {oracle};
\draw (450,215) node [scale=1.5] [align=left] {spec};
\draw (442,177) node [scale=1.5] [align=left] {online};
\draw (515,151) node [scale=1.5] [align=left] {offline};

\end{tikzpicture}

}
\caption{High-level overview of ROSMonitoring~\cite{DBLP:conf/taros/FerrandoC0AFM20}.}
\label{fig:rosmon-pipeline}
\end{figure*}
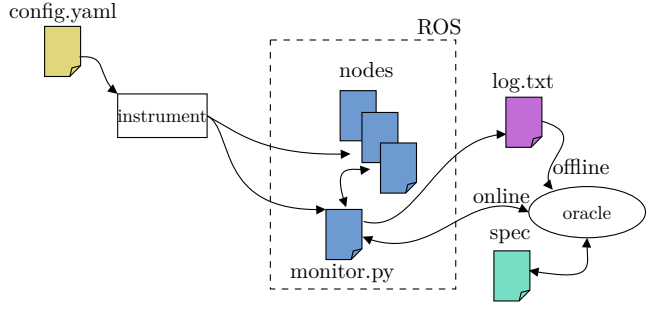

This work uses ROSMonitoring 1.0~\cite{DBLP:conf/taros/FerrandoC0AFM20}, which monitors \gls{ros} programs only through the topics they communicate over. As mentioned in \S\ref{sec:dsl}, the impact of this limitation is that the \gls{rcl} contracts can only state what topics their inputs and outputs correspond to, because this information is used during the synthesis of ROSMonitoring monitors. However, this means that \gls{rcl} contracts cannot link their inputs and outputs to services, though services are often used in \gls{ros} programs. 
An updated version of ROSMonitoring (2.0)~\cite{rosmon2} introduces support for monitoring services, {providing a route to enabling our approach to do the same. Updating \gls{rcl} and \Vanda to cater to \gls{ros} systems using services and actions is left as future work, and discussed in \S\ref{sec:conclusion}.}

\subsubsection*{From RCL topics to ROSMonitoring configuration file}
 
ROSMonitoring takes a configuration file as input (`config.yaml' in Fig.~\ref{fig:rosmon-pipeline}) where the user specifies the nodes and the topics that will be analysed by the monitors. Our tool, \textsc{Vanda}, automatically generates a configuration file from an \gls{rcl} specification. Generating the configuration file only requires the name of the topics to monitor and their message types. An example of a configuration file is shown later when we describe the case study (\S\ref{sec:case1step5}). 

Given a configuration file, ROSMonitoring generates one (or multiple) runtime monitor(s) and changes the \gls{ros} system nodes to allow the monitors to intercept the message exchanges needed for the verification. The remaining step is to produce the \emph{Oracle}, the component that performs the formal verification.

\subsubsection*{From RCL guarantees to an RML Oracle}

{In an assume-guarantee contract, the guarantee of a component is conditional on its assumptions being satisfied. However, for connected software nodes, our calculus checks that the assumptions of a receiving node follow from the guarantees of the producing node(s). In such cases, monitoring the receiving node’s assumption in addition to the producing node's guarantee would duplicate runtime checks and increase overhead. We therefore focus on guarantee monitoring, which provides a low-overhead safety net for detecting when software components fail to produce outputs that satisfy their contracts.

This choice has implications for debugging and failure diagnosis. A guarantee violation identifies where the monitored contract is violated, but the root cause may be an unmet assumption. When the unmet assumption is due to another software node producing incorrect information, our approach still captures this, since the producing node's guarantee is the source of the failure. However, assumptions that correspond directly to environmental conditions may require explicit monitoring to distinguish environmental failures from software failures. Extending Vanda with assumption monitors and richer verdicts that distinguish assumption violations from guarantee violations is therefore a useful direction for future work.}

The \texttt{Oracle} (shown in Fig.~\ref{fig:rosmon-pipeline}) verifies that a system's trace obeys the formal specification, producing a \textit{verdict} as output. In ROSMonitoring, the Oracle is decoupled from the monitoring framework and can be specified with any formalism, here we have chosen the \gls{rml}~\cite{ANCONA2021102610} because the translation from \gls{rcl} to \gls{rml} was the most intuitive and direct.

An \gls{rml} property is a tuple $\langle t,ETs \rangle$, with a term $t$ , and a set of event types $ETs=\{ET_1,\ldots,ET_n\}$. An event type $ET$ is represented as a set of pairs $\{k_1:v_1,\ldots,k_n:v_n\}$, where each pair identifies a specific piece of information ($k_i$) and its value ($v_i$). Given an event type $ET$, an event $Ev$, denoted as a set of pairs $\{k_1':v_1',\ldots,k_m':v_m'\}$ as well, matches $ET$ if $ET \subseteq Ev$, which means $\forall (k_i:v_i) \in ET \cdot \exists (k_j:v_j) \in Ev \cdot k_i = k_j \land v_i = v_j$. In practice, an event type $ET$ specifies the requirements that an event $Ev$ must satisfy to be considered valid. 
For instance, an event type $ET$ could be $\{pos:(waypoint1)\}$, meaning that all events containing $pos$ with value $(waypoint1)$ are valid. An event $Ev$ generated by a moving robot could be $\{speed:1.6,pos:(waypoint1)\}$, meaning that the robot is moving at speed $1.6$ [m/s], and is currently at position $(waypoint1)$. 

An \gls{rml} term $t$ can be:
\begin{compactitem}
    \item $ET$, denoting a singleton set containing the events $Ev$ s.t. $ET \subseteq Ev$
    \item $t_1 \; t_2$, denoting the concatenation of two sets of traces 
    \item $t_1 \land t_2$, denoting the intersection of two sets of traces
    \item $t_1 \lor t_2$, denoting the union of two sets of traces
    \item $\{ let \; x; t' \}$, denoting the set of traces $t'$ where the variable $x$ can be used
    \item $t' *$, denoting a chain of concatenations of trace $t'$
\end{compactitem}
where $t_1$, $t_2$ and $t'$ are \gls{rml} terms.

Event types can be negated. Given an event type $ET$, the term $\lnot ET$ denotes its negation. Specifically, $\forall_{Ev}.ET \subseteq Ev \iff \lnot ET \nsubseteq Ev$. In the rest of the paper, we also apply the notion of negation to the other \gls{rml} terms. For instance, if the term is $ET_1 \land ET_2$, its negation is $\lnot ET_1 \lor \lnot ET_2$; and the same reasoning is applied for the other operators. 

Event types can contain variables. Considering the previous event type, we could have its parametric version as $ET(x)=\{pos:(x)\}$, where we do not force any value for the waypoint. 
This event type matches all events containing $pos$ with any $x$ value. When an event matches an event type with variables, the variables get the values from the event. 

An example of using \gls{rml} to define contracts, consider the term $ET_1 ET_2 \lor ET_3 ET_4$, with $ET_i$ being some event type describing a certain kind of event (as before). However, differently from before, here, we have a union of two concatenations; which means the language recognised by this term is 
$$\{ Ev \; Ev' \;|\; ET_1 \subseteq Ev, ET_2 \subseteq Ev' \} \cup \{ Ev \; Ev' \;|\; ET_3 \subseteq Ev, ET_4 \subseteq Ev' \}$$ 
Naturally, the same reasoning can be applied to the $\land$ operator; the only difference would be in using $\cap$ instead of $\cup$. 

\newcommand{\Rule}[4]{\scriptstyle{\textrm{({#1})}}{\displaystyle\frac{#2}{#3}}\ #4}
\newcommand{\trans}[1]{\stackrel{{#1}}{\rightarrow}}

\begin{figure}[t]
\centering
\small{
\hspace*{-0.5cm}
\begin{math}
\begin{array}{c}
\Rule{atom}
{}
{atom \trans{} \langle atom, \{\} \rangle}
{}
\qquad
\Rule{and/or}
{g_1 \trans{} \langle t_1,ET_1 \rangle \land g_2 \trans{} \langle t_2,ET_2 \rangle}
{g_1 op\: g_2 \trans{} \langle t_1 op\: t_2, ET_1 \cup ET_2 \rangle}
{op \in \{ \land,\lor \}}
\\[3ex]
\Rule{iff}
{g_1 \trans{} \langle t_1,ET_1 \rangle \land g_2 \trans{} \langle t_2,ET_2 \rangle}
{g_1 \iff g_2 \trans{} \langle (t_1 \land t_2) \lor (\lnot t_1 \land \lnot t_2), ET_1 \cup ET_2 \rangle}
{}
\\[3ex]
\Rule{implies}
{g_1 \trans{} \langle t_1,ET_1 \rangle \land g_2 \trans{} \langle t_2,ET_2 \rangle}
{g_1 \implies g_2 \trans{} \langle (\lnot t_1 \land t_2) \lor (\lnot t_1 \land \lnot t_2) \lor (t_1 \land t_2), ET_1 \cup ET_2 \rangle}
{}
\\[3ex]
\Rule{equals}
{}
{a_1 == a_2 \trans{} \langle ET, \{ \langle ET,(a_1:a_2) \rangle \} \rangle}
{}
\\
\Rule{not-equals}
{}
{a_1 != a_2 \trans{} \langle \lnot ET, \{ \langle ET,(a_1:a_2) \rangle \} \rangle}
{}
\\[3ex]
\Rule{exists}
{g \trans{} \langle t,ET \rangle}
{\exists x \cdot g \trans{} \langle \{ let\: x; t \},ET \rangle}
{}
\Rule{guarantee}
{g \trans{} \langle t,ET \rangle}
{G (g) \trans{} \langle (t)*,ET \rangle}
{}
\\[3ex]
\Rule{guarantees}
{g_1 \trans{} \langle t_1,ET_1 \rangle \land \ldots \land g_n \trans{} \langle t_n,ET_n \rangle}
{G(g_1), \ldots, G(g_n) \trans{} \langle t_1 \land \ldots \land t_n, ET_1 \cup \ldots \cup ET_n \rangle}
{}
\end{array}
\end{math}}

\caption{RML partial translation.}\label{trans-fig}
\end{figure}

As previously mentioned, \textsc{Vanda} synthesises \gls{rml} specifications from the \gls{rcl} contracts.
Here we describe how \textsc{Vanda} translates the \gls{rml} guarantees into \gls{rml}. 
In principle, both the assumptions \textit{and} guarantees can be translated into \gls{rml}, {since both are expressed as RCL formulae. In the current implementation, Vanda translates only guarantees. This keeps the runtime monitoring layer lightweight and avoids duplicating checks that are already represented compositionally, since the assumptions of many nodes are entailed by the guarantees of the nodes that provide their inputs. Translating assumptions would follow a similar process but would be most useful for assumptions that correspond to external environmental conditions or for applications that require finer-grained runtime diagnosis.}

Fig.~\ref{trans-fig} shows the operational semantics of the translation function from \gls{rcl} descriptions to \gls{rml} specifications that it implements. Here, each rule formalises a single-step translation.
For instance, the (and/or) rule denotes how to translate the conjunction/disjunction of two RCL guarantees. This can be done first by translating the two guarantees $g_1$ and $g_2$ into \gls{rml}; then, by combining the results obtained using the corresponding \gls{rml} operator.

For the (equals) rule, at the guarantee level we require that a certain variable in a topic has a specific value. This is mapped into an event type that requires the event to have the same value. Recall that an event matches an event type if the latter is included in the former. Since we want to guarantee that the information identified by $a_1$ has value $a_2$, the resulting event type has to check that this holds, which can be done by adding the pair $a_1:a_2$ in the event type. 

The (exists) rule tackles the use of variables. In \gls{rcl}, the contract says that there exists an $x$ for which the guarantee $g$ holds. This can be straightforwardly mapped into a parametric term in \gls{rml}, where the variable $x$ is used inside the term $t$ (the one derived by $g$). As usual, the (forall) rule can be obtained by negating the (exists) rule.
Finally, the (guarantee) and (guarantees) rules denote the translation step for single and multiple guarantees, respectively. 
To translate multiple guarantees, each guarantee $g_i$ is translated separately, and the resulting \gls{rml} terms $t_i$ are combined using the $\land$ \gls{rml} operator. In this way, all guarantees are required to be satisfied. 
To translate a guarantee $G(g)$, first $g$ is translated into the \gls{rml} term $t$. A $*$ post-fix operator 
is then added to $t$, meaning that $t$ can be repeated as many times as necessary. We need this operator because the guarantees need to be checked continuously, not just once.

\subsubsection*{From RCL to ROSMonitoring monitors}

The generation of monitors in ROSMonitoring from RCL guarantees requires one more additional step. The RCL specifications are based on an abstraction of the system, while the monitors in ROSMonitoring are deployed in the real system under execution. Therefore, to ensure the monitors will be able to capture all of the necessary events (execution traces) produced by the system, it is necessary to instrument the system with any missing events. For example, these missing events can be a mismatch of names used in the contracts versus what is used in the implemented system, different representations of data types, implicit information that is not directly being published in a ROS topic, etc.

\section{Case Study: Remote Inspection}
\label{sec:exampleSystem}

Our case study is a \emph{Jackal}\footnote{Jackal: \url{https://clearpathrobotics.com/jackal-small-unmanned-ground-vehicle/} Accessed: 03/11/2023} rover performing remote inspection in a nuclear waste store, in a 3D Gazebo\footnote{Gazebo Simulator: \url{http://gazebosim.org/} Accessed: 03/11/2023} simulation. The Jackal uses a (simulated) sensor to take radiation measurements at given waypoints.
We adapted this case study from the simulation described in~\cite{robotics10030086}, adding an autonomous agent that makes the high-level decisions that control the Jackal.

\begin{figure*}[t]
\centering
\includegraphics[width=\linewidth]{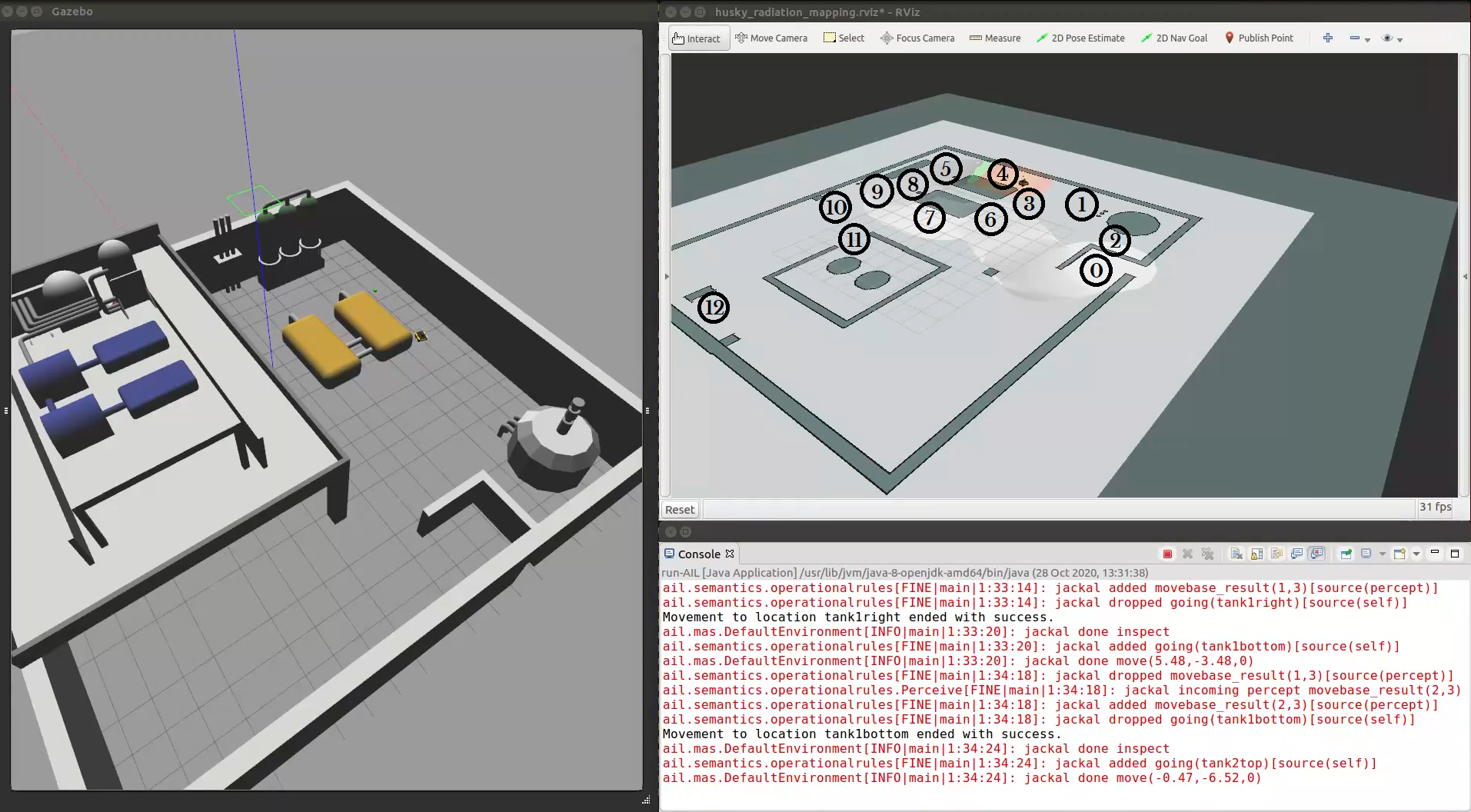}
\caption{Simulation environment of the remote inspection case study. On the left-hand side is the Gazebo 3D simulation window. On the top right-hand side is the RViz window, a visualisation interface containing the map that the rover uses for path planning, a visual representation of radiation levels (green, orange, red) around the robot, and the waypoints from 0 to 12 that the robot must inspect. On the bottom right-hand side is the autonomous agent console with the log of its execution.}
\label{fig:remoteinsp}
\end{figure*}

The rover's goal is to inspect 12 waypoints that are located inside the simulated nuclear waste store (see Fig.~\ref{fig:remoteinsp}). The autonomous agent decides which waypoint to inspect next, depending on the current radiation readings. It should always be aware of high radiation values, since this could put the robot in danger, as well as indicating a problem in the waste store that requires further investigation. The rover's battery is not simulated, so we assume that the rover has enough power for the duration of the scenario. We use the \gls{ros} \emph{move base} package for low-level path planning. 

At a high level the system should obey the following requirements.
\begin{description}
\item[REQ1:] The robot should inspect each waypoint as long as the radiation level at a waypoint is not high (``red").
\item [REQ2:] Each waypoint only needs to be inspected at least once.  
\item[REQ3:] If the radiation level at a waypoint is too high, then the robot will abandon the mission and return to the entry point.
\item [REQ4:] Eventually, all waypoints will have been inspected; or the mission will have  terminated in failure, due to high levels of radiation.   

\end{description}
In the rest of this section, we describe how to apply our approach to the remote inspection case study, from the manual specification of the contracts to the automatic generation of runtime monitors. The simulation environment, contracts, and verification assets for this chapter are available from the Zenodo repository that accompanies this paper: \url{https://doi.org/10.5281/zenodo.6941344}.

\subsection{Step 1: Creating a System Model}
\label{sec:case1step1}

\begin{figure*}[htbp]
\centering
\includegraphics[width=1.4\textwidth,angle=90]{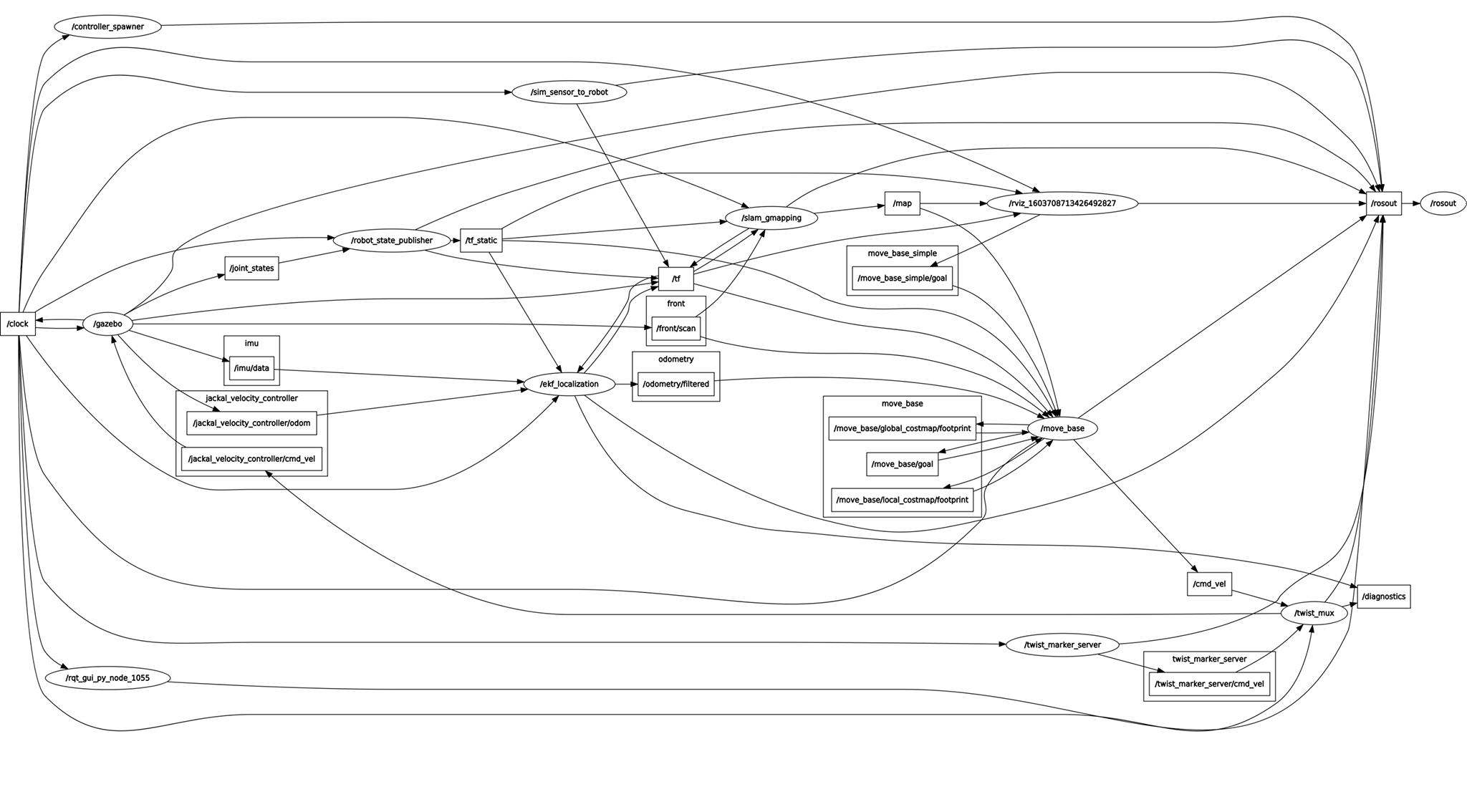}
\caption{ROS graph automatically generated for the remote inspection case study. The ovals represent ROS nodes and the rectangles represent topics. The arrows connecting nodes and topics show the direction of messages.  }
\label{fig:case1graph}
\end{figure*}

In this section we describe how we generated the system model for the remote inspection case study. Fig.~\ref{fig:case1graph} shows the \gls{ros} graph of the remote inspection robotic system (minus the autonomous agent, because it is not programmed using \gls{ros}).
As described in \S\ref{sec:specifyingRobots}, we often find it helpful to adapt the \gls{ros} graph to produce a system model that is more amenable to analysis and verification. 

\begin{figure}[t]
\centering
\scalebox{0.9}{
\hspace{-90pt}
\begin{tikzpicture}[node distance=1.5em]

\node (anchor) {}; 
\node (amcl) [process, right of = anchor, xshift = 20em] {\parbox[t][][t]{3cm}{\centering Localisation}};
\node(amclspec)[right of = amcl, xshift = 8em]{\parbox[t][][t]{3cm}{\centering \begin{tabular}{l}
		\textbf{$\overline{i_L}$:} sensors : SensorsType\\
		\textbf{$\overline{o_L}$:} position : PositionType
		\end{tabular}}};

\node (navigation) [process, below of = amcl, yshift = -4em] {\parbox[t][][t]{3cm}{\centering Navigation}};
\node(navigationspec)[right of = navigation, xshift = 8em]{\parbox[t][][t]{3cm}{\centering \begin{tabular}{l}
		\textbf{$\overline{i_N}$:} position : PositionType, \\ \quad command : CommandSet\\ 
		\textbf{$\overline{o_N}$:} at : AtType
		
		\end{tabular}}};

\node (agent) [process, below of = navigation, yshift = -4em] {\parbox[t][][t]{3cm}{\centering Agent}};
\node(agentspec)[right of = agent, xshift = 8em]{\parbox[t][][t]{3cm}{\centering \begin{tabular}{l}
		$\overline{i_A}$: wayP : WayP, at : AtType, \\ \quad radiationStatus : RadStat, \\ \quad inspected : InspectedType \\
		\textbf{$\overline{o_A}$:} command : CommandSet
		\end{tabular}}};

\node (radsensor) [process, below of = agent, yshift = -4em] {\parbox[t][][t]{3cm}{\centering Radiation Sensor}};
\node(radsensorspec)[right of = radsensor, xshift = 8em]{\parbox[t][][t]{3cm}{\centering \begin{tabular}{l}
		$\overline{i_R}$: r : REAL,\\ \quad command : CommandSet \\ 
		$\overline{o_R}$: radiationStatus : RadStat, \\ \quad inspected : InspectedType
        \end{tabular}}};

\draw [arrow] (amcl) to (navigation);
\draw [arrow] (agent) to [out=120,in=-120] node[right]{}(navigation);
\draw [arrow] (navigation) to [out=-30,in=30] node[right]{}(agent);
\draw [arrow] (radsensor) to [out=120,in=-120] node[right]{}(agent);
\draw [arrow] (agent) to [out=-30,in=30] node[right]{}(radsensor);

\node(types)[left of = amcl, yshift = -9em, xshift = -12em]{\parbox[t][][t]{4cm}{\centering \footnotesize \textbf{Context} \begin{tabular}{l}
	 $ WayP : \mathbb{R} \times \mathbb{R} \rightarrow \mathbb{N} $ \\
		 $ RadStat : \{red, orange, green\} $ \\
		 $ CommandSet : \{move(\mathbb{R}, \mathbb{R}), inspect(\mathbb{N})\} $ \\
		 $ PositionType : \mathbb{R} \times \mathbb{R} \rightarrow \mathbb{B} $ \\
		 $ AtType : \mathbb{R} \times \mathbb{R} \rightarrow \mathbb{B} $ \\
		 $ InspectedType : \mathbb{N} \rightarrow \mathbb{B} $ \\
		 $ SensorsType : \emptyset $
		\end{tabular}}};

\end{tikzpicture}}

\caption{System model for the remote inspection case study. Arrows indicate data flow between the nodes.}
\label{fig:contracts}
\end{figure}
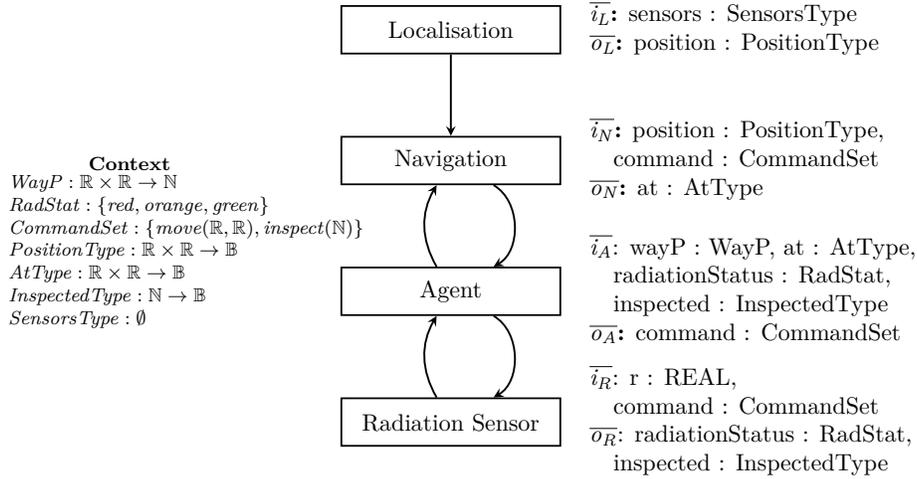

Here, most of the nodes are well-tested off-the-shelf libraries, such as the Jackal \emph{velocity controller} and the \emph{move base} library. The former relates to the Jackal's internal velocity controllers, while the latter provides a suite of path planners. Both relate to the robot's navigation, and thus we abstract them into a \emph{compound} node called \textbf{Navigation}. Similarly, we merge the nodes \emph{imu} and \emph{odometry},into the \textbf{Localisation} \emph{compound} node. We obtain the \textbf{Radiation Sensor} node in a similar way. Finally, because the autonomous agent is external to the \gls{ros} system, we create the \textbf{Agent} node as an abstraction of its implementation. Thus, the system model is composed of these four abstracted nodes, shown in Fig.~\ref{fig:contracts} and described below.\smallskip 

\noindent\textbf{Localisation:} abstracts the \gls{amcl} \gls{ros} package\footnote{Adaptive Monte Carlo Localisation package: \url{http://wiki.ros.org/amcl} Accessed: 03/11/2023}, which uses sensor input to perform localisation. This is a well established package, so we do not define a contract for its specifics. Instead, it is enough (for our purposes) to know that the node will output what it believes to be the robot's position. The \textbf{Localisation} node takes input from hardware (a variety of sensors), but our contracts focus on the links between the \textit{software} nodes; so it is enough for our contracts to specify its input as $\emptyset $ (defined as $SensorsType$ in the specification's \texttt{context} clause, because the inputs and outputs are name-type pairs), knowing that the node will be able to access the sensor information.
\smallskip 

\noindent\textbf{Navigation:} abstracts the well known \emph{move base} package, which performs low-level path planning. Its inputs are the estimated position from the \textbf{Localisation} node, and the command from the \textbf{Agent} node to move to a particular position. It outputs a function mapping coordinates to a boolean, which is \texttt{true} if the Robot is at that position.\smallskip 

\noindent\textbf{Agent:} the most critical node to verify; because it makes the high-level decisions for the robot, and its implementation is not based on preexisting \gls{ros} packages. The agent is implemented in the agent programming language \textsc{Gwendolen}~\cite{dennis17gwen}. Its inputs are the output of the \textbf{Navigation} and \textbf{Radiation Sensor} nodes. It outputs a high-level command (either move or inspect).\smallskip 

\noindent\textbf{Radiation Sensor:} takes as input a radiation measurement $r$ from its internal sensor and the command from the \textbf{Agent} to inspect a particular waypoint $i$. It outputs a confirmation that a particular waypoint has been inspected.\smallskip

\noindent Once the model has been distilled, the next step is to define contracts for the individual nodes.

\subsection{Step 2: Specifying Node Contracts}
\label{sec:case1step2}

This step builds the contracts used to guide the heterogeneous verification and to enable synthesis of the runtime monitors. Each contract is written in \gls{rcl} (described in \S\ref{sec:dsl}), and consists of a node's inputs, outputs, assumptions, and guarantees (which are all written in a plain-text encoding of \gls{fol}). A contract also links the inputs and outputs to the topics in the underlying \gls{ros} program.

As an example, Listing~\ref{list:contextRCL} shows the \texttt{context} clause that defines the types and constants used in the specification of our case study and Listing~\ref{list:CS1AgentRCL} shows the \gls{rcl} contract for the \textbf{Agent} node. In the \textbf{Agent}'s contract, Listing~\ref{list:CS1AgentRCL}, lines 3 and 5 identify the node's \textcolor{blue}{\texttt{inputs}} and \textcolor{blue}{\texttt{outputs}}, respectively. 

\begin{sloppypar}
Lines 7--11 show the \gls{ros} \texttt{topics} that the \textbf{Agent} uses and which input or output they correspond to (in the \textcolor{red}{\texttt{matches}} statement). For example, line 9 shows that the \texttt{at} field in the \texttt{gazebo\_radiation\_plugin/Snapshot} topic matches \texttt{in.at}. On line 14, the \textcolor{blue}{\texttt{assume}} statement contains the assumption of the \textbf{Agent}'s contract. 
From line 16 onward, the \textcolor{blue}{\texttt{guarantee}} statements contain a plain-text encoding of the \textbf{Agent}'s \gls{fol} guarantees. The links between the four nodes, and the inputs and outputs that link them, are shown in Fig.~\ref{fig:contracts}.
\end{sloppypar}

\begin{table}[t]
\begin{tabularx}{\textwidth}{l|l|X}

\multicolumn{3}{c}{\textbf{Localisation}}  \\ \hline \hline 

\multirow{2}{*}{Input} & \textbf{Text} & Input from hardware (a variety of sensors) \\ \cline{2-3}
& \textbf{FOL} & $sensors: SensorsType$ \\ \hline \hline 

\multirow{2}{*}{Output} & \textbf{Text} & The robot's estimated position \\ \cline{2-3}
& \textbf{FOL} & $ position : PositionType$ \\ \hline \hline 

\multirow{2}{*}{Assume} & \textbf{Text}  & N/A \\ \cline{2-3}
 & \textbf{FOL} & $TRUE$ \\ \hline \hline 
 
\multirow{2}{*}[-1em]{Guarantees} & \textbf{Text}  & The node outputs a unique $(x,y)$ coordinate that is the robot's estimated position \\ \cline{2-3}
 & \textbf{FOL} &  $\exists! x, y \in \mathbb{R} \bullet out.position(x, y)$ \\ \hline \hline 
\end{tabularx}

\caption{A summary of the \textbf{Localisation} node's inputs, outputs, assumptions, and guarantees. \label{tab:contracts_local}}

\end{table}

In \S\ref{sec:dsl} we describe the general rules that we uses to map \gls{ros} topics into \gls{rcl} contract variables. Conceptually, the values from each \gls{ros} topic populates the \gls{rcl} variable that it matches. However, it is important to note that the topics in the contracts are only used to synthesise the monitors, and there is no automatic connection between the contracts and the \gls{ros} program. We assume that there is a mapping that implements the rules, to link the more abstract specification in the \gls{rcl} contracts with the concrete program.

Using Listing~\ref{list:CS1AgentRCL} as an example, we show how the rules map between the \gls{ros} program and the \gls{rcl} contracts. The \inlineRCL{in.inspected} and \inlineRCL{in.at} variables are functions (as shown in Listing~\ref{list:contextRCL}) so we apply the \textbf{Function} rule, which maps the values in the \gls{ros} message to the parameters of the function. For \inlineRCL{in.inspected}, this maps an \texttt{int16} variable in the \gls{ros} message \texttt{gazebo\_radiation\_plugin/Snapshot inspected} to the \texttt{NATURAL} parameter of the function in \gls{rcl} using the \textbf{Scalar} rule. For \inlineRCL{in.at}, the \gls{ros} message contains two \texttt{float64} variables, so the \textbf{Scalar} rule is applied to each in turn, mapping them to the two \texttt{REAL} parameters of the \inlineRCL{in.at} function.

\begin{sloppypar}
The \inlineRCL{in.wayP} variable is also a function. Here the matching \gls{ros} topic (\texttt{gazebo\_radiation\_plugin/Snapshot wayPNow}) has three variables, but the \inlineRCL{in.wayP} function only takes two parameters. For this mapping, the \textbf{Function} rule takes only the variables that match the types of the two \inlineRCL{REAL} parameters of the function.
\end{sloppypar}

The \inlineRCL{in.radiationStatus} variable's type is the set \inlineRCL{\{red, orange, green\}} and the \gls{ros} topic contains only a single \texttt{string} variable. For this mapping we use the  \textbf{Collection} rule, which maps the \texttt{string} variable to one of the three elements of the set. Given that we know the program only communicates strings ``red'', ``orange'', or ``green'', the verification steps assume this to be true. As a safety net, the runtime monitors (\S\ref{sec:case1step5}) would treat strings outside this set in the same way as ``red'' or ``orange''.

\smallskip 
\begin{lstlisting} [language=RCL, caption={The \textbf{Context} clause for the RCL contracts. This defines global types and constants.},captionpos=b, label={list:contextRCL}, basicstyle=\scriptsize\ttfamily, aboveskip=0em, belowskip=0em]
context{
 WayP : REAL x REAL --> NATURAL ;
 RadStat : {red, orange, green} ;
 CommandSet : { move(REAL, REAL), inspect(NATURAL) };
 PositionType : REAL x REAL --> BOOL ;
 AtType: REAL x REAL --> BOOL;
 InspectedType :  NATURAL --> BOOL;
 SensorsType : {}; }
\end{lstlisting}

\smallskip 
\begin{lstlisting} [language=RCL, caption={The \textbf{Agent}'s RCL contract.},captionpos=b, label={list:CS1AgentRCL}, basicstyle=\scriptsize\ttfamily, aboveskip=0em, belowskip=0em]
node agent{

inputs( wayP : WayP, at : AtType, radiationStatus : RadStat, inspected : InspectedType )

outputs( command : CommandSet  )

topics( gazebo_radiation_plugin/Snapshot command matches(out.command),
gazebo_radiation_plugin/Snapshot inspected matches(in.inspected),
gazebo_radiation_plugin/Snapshot at matches(in.at),
gazebo_radiation_plugin/Snapshot wayPNow matches(in.wayP),
gazebo_radiation_plugin/Snapshot radiationStatus matches(in.radiationStatus) )

assume( in.radiationStatus in {red, orange, green} )

guarantee(forall(x', y' in REAL, i in NATURAL |
  in.at(x', y') == TRUE and in.wayP(x', y') == i and in.inspected(i) == TRUE and in.radiationStatus !in {red, orange} ->
  exists(x, y in REAL | (in.wayP(x, y) == i + 1  or 
    (forall( x'', y'' in REAL | in.wayP(x'', y'') != i+1 and in.wayP(x, y) == 0))) and out.command == move(x, y) )) )

guarantee(forall(x', y' in REAL, i in NATURAL |
  in.at(x', y') == TRUE and in.wayP(x', y') == i and in.inspected(i) == FALSE -> out.command == inspect(i) ) )

guarantee(forall(x', y' in REAL, i in NATURAL |
  in.radiationStatus in {red, orange} or not exists( x, y in REAL | in.wayP(x, y) == i + 1 )
  -> exists (x'', y'' in REAL | in.wayP(x'', y'') == 0 and out.command == move(x'', y'') ) ) )
}
\end{lstlisting}

\begin{sloppypar}
Finally, to handle the mapping between the \inlineRCL{out.command} variable and the \texttt{gazebo\_radiation\_plugin/Snapshot command} topic we add a bespoke rule, \textbf{CommandMapping}. The \texttt{command} topic contains a \texttt{string}, which is either ``move'' or ``inspect''; and an \texttt{int16}, which is the id of a waypoint. The \textbf{CommandMapping} rule checks the value of the \texttt{string}. If the \texttt{string == "move"} then it uses the inverse of \inlineRCL{WayP} (\inlineRCL{WayP^-1}) to obtain the coordinates of the of the waypoint that the rover should move to, then uses the \textbf{Function} rule to map these two coordinate values to the parameters of the \inlineRCL{move()} function. If the \texttt{string == "inspect"} then \textbf{CommandMapping} uses the \textbf{Scalar} rule to map the \texttt{int16} waypoint id to the \inlineRCL{NATURAL} parameter of the \inlineRCL{inspect()} function. 
\end{sloppypar}

Tables~\ref{tab:contracts_local}, \ref{tab:contracts_nav}, \ref{tab:contracts_agent}, and \ref{tab:contracts_rad} summarise the inputs, outputs, assumptions, and guarantees of each contract. The tables describe the statements in the contract in English, and then shows the relevant part of the contract. The \gls{fol} has been rendered with  mathematical symbols for the convenience of the reader. The full contracts, in \gls{rcl}, are shown in~\ref{app:contracts}.

The information contained in the contracts is useful for guiding both formal and non-formal verification of the nodes, as we describe later in \S\ref{sec:case1step4}. \Vanda{}, the \gls{rcl} tool, can also automatically synthesise runtime monitors from the contracts, which we describe in \S\ref{sec:case1step5}. \smallskip

\noindent In the next step, we use our calculus to derive system-level properties from the individual node contracts.

\subsection{Step 3: Deriving System-Level Properties}
\label{sec:case1step3}

In this section, we give an example of how our calculus (\S\ref{sec:combination}) works by presenting the manual derivation of one of the system-level properties from the contracts.

We start with the \textbf{Agent} node, which takes input from the \textbf{Navigation} and \textbf{Radiation Sensor}. Because \textbf{Agent} takes several inputs, we instantiate \textbf{R3}, the input union rule, to get D1:
$$
\small{
\begin{array}{l}
\forall \overline{i_N},\overline{o_N}\cdot \mathcal{A}_N(\overline{i_N})\ \Rightarrow\ \lozenge \mathcal{G}_N(\overline{o_N})\\
\forall\overline{i_R},\overline{o_R}\cdot\ \mathcal{A}_R(\overline{i_R})\ \Rightarrow\ \lozenge \mathcal{G}_R(\overline{o_R}) \hspace{100pt} \text{(D1)} \\
\forall\overline{i_A},\overline{o_A}\cdot\ \mathcal{A}_A(\overline{i_A})\ \Rightarrow\ \lozenge \mathcal{G}_A(\overline{o_A})\\
\overline{i_A} = \overline{o_N}\cup \overline{o_{R}}\\
\vdash\ \bigg( \forall\overline{i_A},\overline{o_N}, \overline{o_{R}}\cdot \mathcal{G}_N(\overline{o_N})\land \mathcal{G}_{R}(\overline{o_{R}}) \Rightarrow \mathcal{A}_A(\overline{i_A}) \bigg) \\ 
\hline 
\bigg( \forall\overline{i_N}, \overline{i_{R}},\overline{o_A}\cdot \mathcal{A}_N(\overline{i_N}) \land  \mathcal{A}_{R}(\overline{i_{R}})\Rightarrow \lozenge\mathcal{G}_A(\overline{o_A}) \bigg)
\end{array}}
$$
\smallskip

\begin{table}[t]

\begin{tabularx}{\textwidth}{l|l|X}

\multicolumn{3}{c}{\textbf{Navigation}}  \\ \hline \hline

\multirow{2}{*}{Input} & \textbf{Text} & Robot's estimated position from the \textbf{Localisation} node, and the command from the \textbf{Agent} node \\ \cline{2-3}
 & \textbf{FOL} & $ position : PositionType, \quad command : CommandSet$ \\  \hline \hline 
 
\multirow{2}{*}{Output} & \textbf{Text} &  Robot's position \\ \cline{2-3}
& \textbf{FOL} &   $ at : AtType$ \\  \hline \hline 

\multirow{2}{*}{Assume} & \textbf{Text}  & There is a unique $(x,y)$ coordinate position that represents the Robot's estimated position \\ \cline{2-3}
 & \textbf{FOL}  &  $\exists!  x, y \in \mathbb{R} \bullet in.position(x, y) = TRUE $ \\  \hline \hline 

\multirow{2}{*}[-2em]{Guarantees} & \textbf{Text}  &  If the Robot is told to move to $(x,y)$ and the \textbf{Localisation} node says that the Robot is at $(x,y)$; then \textbf{Navigation} will conclude that it is at $(x,y)$\\ \cline{2-3}

 & \textbf{FOL}  & $
\begin{array}{l}
\forall x, y \in \mathbb{R} \bullet (\ (in.command = move(x, y)) \land \\ \hspace{0.1cm} (in.position(x, y) = TRUE))\iff out.at(x, y) = TRUE 
\end{array}
$  \\  \hline \hline 

\end{tabularx}
\caption{A summary of the \textbf{Navigation} node's inputs, outputs, assumptions, and guarantees. \label{tab:contracts_nav}}
\end{table}

\noindent To apply rule \textbf{R3}, we must show that we can deduce that the guarantees of the \textbf{Navigation} and \textbf{Radiation Sensor} nodes imply the assumption of the \textbf{Agent} node: {$\forall\overline{i_A},\overline{o_N}, \overline{o_{R}}\cdot \mathcal{G}_N(\overline{o_N})\land \mathcal{G}_{R}(\overline{o_{R}}) \Rightarrow \mathcal{A}_A(\overline{i_A})$}. For these three nodes, this is instantiated as:
$$
\small{
\begin{array}{l}
\forall\overline{i_A},\overline{o_N}, \overline{o_{R}}\cdot [\forall x, y \in \mathbb{R} \cdot\\ \quad in.command = move(x,y)  \land in.position(x, y) = true \Leftrightarrow  out.at(x,y) = true]\\ \t1 \land [ \forall i \in \mathbb{N} \cdot in.command = inspect(i)\Rightarrow (out.inspected(i) = true  \land\\
\quad\land (0 \leq in.r < 120 \Rightarrow out.radiationStatus = green)\\ 
\quad\land (120 \leq in.r < 250 \Rightarrow out.radiationStatus = orange)\\
\quad\land (250 \leq in.r \Rightarrow out.radiationStatus = red))] \\ 
\quad \Rightarrow out.radiationStatus \in \{red, orange, green\}
\end{array}}
$$
\noindent which is true by virtue of our definition of $radiationStatus$. 

Applying \textbf{R3} here here allows us to conclude that if the nodes are correctly linked (by showing that the property after the $\vdash$ in D1 holds) then we can say that the correct input to the \textbf{Navigation} and \textbf{Radiation Sensor} nodes will result in the guarantee of the \textbf{Agent} being preserved. 
Thus, the system-level property that we derive is: \\
\centerline{$ \forall\overline{i_N}, \overline{i_{R}},\overline{o_A}\cdot (\exists ! x,y \in \mathbb{R} \cdot position(x,y) = true) \land  (0 \leq r) \Rightarrow \lozenge\mathcal{G}_A(\overline{o_A})$}

\noindent which tells us that when the robot is in a valid, unique position; and the observed radiation level is valid; then eventually the guarantee of the \textbf{Agent} will hold. This demonstrates that the \textbf{Agent}'s guarantee is dependent on the correct input and functioning of the \textbf{Navigation} and \textbf{Radiation Sensor} nodes.

The \textbf{Agent}'s guarantees (see Table~\ref{tab:contracts_agent}) support the four requirements described in the introduction to this section. The \textbf{Agent} guarantees that that each waypoint will be inspected, unless dangerous levels of radiation are detected by the \textbf{Radiation Sensor} (REQ1 and REQ4); that a waypoint is only inspected if it hasn't been inspected yet (REQ2); that if the radiation level at a waypoint is too high, then it moves to the original waypoint (REQ3 and REQ4); and that when there are no more waypoints to inspect, the robot will return to the original waypoint (REQ1 and REQ4).

\FloatBarrier
\begin{table}[!t]

\begin{tabularx}{\textwidth}{l|l|X}

\multicolumn{3}{c}{\textbf{Agent}} \\ \hline \hline 

\multirow{2}{*}{Input} & \textbf{Text} & Map of $x,y$ pairs to waypoint ids, function that is \texttt{true} if the robot is at that position, and Radiation Status and inspected values from \textbf{Radiation Sensor} \\ \cline{2-3}
 & \textbf{FOL} &   $ wayP:WayP,\quad at:AtType,\quad radiationStatus:RadStat,\quad inspected:InspectedType$ \\ \hline \hline

\multirow{2}{*}{Output} & \textbf{Text} &  The command (move or inspect)\\ \cline{2-3}
 & \textbf{FOL} &  $ command : CommandSet$ \\ \hline \hline

\multirow{2}{*}{Assume} & \textbf{Text}  & The Radiation Status that it receives is either red, orange, or  green\\ \cline{2-3}
&  \textbf{FOL} & $in.radiationStatus \in \{red, orange, green\}$ \\ \hline \hline

\multirow{6}{*}[-6em]{Guarantees} & \textbf{Text}  & 
If the Robot is at, and has inspected waypoint $i$, and the radiation level was not dangerous, then it will move to the next waypoint or to waypoint 0 (the exit) if there is no next waypoint. \\ \cline{2-3}
 & \textbf{FOL} &$\begin{array}{l}\forall x', y' \in \mathbb{R}, i \in \mathbb{N} \bullet(\ (in.at(x', y') = TRUE)\\ \quad \land (in.wayP(x', y') = i) \land (in.inspected(i) = TRUE) \\\quad \land (in.radiationStatus \notin \{red, orange\})\ )\\ \quad\implies (\exists x, y \in \mathbb{R} \bullet ((in.wayP(x, y) = i+1 \\ \qquad \lor (\forall x'', y'' \in \mathbb{R} \bullet  in.wayP(x'', y'') \neq i + 1 \\\qquad \quad  \land in.wayP(x, y) = 0))\\\qquad \quad \land out.command = move(x, y) ) )\end{array}$ \\ \hhline{~==}

& \textbf{Text}  & And, if the Robot is at, and has not inspected, waypoint $i$, then inspect it. \\ \cline{2-3}
 & \textbf{FOL} & $ \begin{array}{l} \forall x', y' \in \mathbb{R}, i \in \mathbb{N} \bullet ((in.at(x', y') = TRUE)\\ \qquad \land (in.wayP(x', y') = i) \\ \qquad\land (in.inspected(i) = FALSE))\\ \quad\implies out.command = inspect(i) \end{array} $  \\ \hhline{~==}

& \textbf{Text}  & And, if the Robot inspects a waypoint and the radiation level is dangerous, or there are no more waypoints, then return to the original waypoint (the exit) \\ \cline{2-3}
 & \textbf{FOL}   & $\begin{array}{l} \forall x', y' \in \mathbb{R}, i \in \mathbb{N} \bullet\\ \quad( (in.radiationStatus \in \{red, orange\})\\ \qquad\lor (\neg \exists x, y \in \mathbb{R} \bullet in.wayP(x, y) = i+1)\ )\\ \t1\implies (\exists x'', y'' \in \mathbb{R} \bullet (in.wayP(x'', y'') = 0)\\ \qquad \land (out.command = move(x'', y'')\ )\ ) \end{array}$
\\  \hline \hline
\end{tabularx}
\caption{A summary of the \textbf{Agent} node's inputs, outputs, assumptions, and guarantees. \label{tab:contracts_agent}}
\end{table}

\subsection{Step 4a: Heterogeneous Verification}
\label{sec:case1step4}

This section describes our verification of the four nodes in our remote inspection case study. Each node presents its own verification challenges, and we have chosen a suitable approach to verify that each node obeys its contract. As mentioned in \S\ref{sec:guidingHV}, the link between the contracts and the verification steps is \emph{not} formal. The information in the contracts informs the verification steps, enables the use of either formal or non-formal methods as best suits the node being verified. 

\begin{table}[!t]

\begin{tabularx}{\textwidth}{l|l|X}

\multicolumn{3}{c}{\textbf{Radiation Sensor}}  \\ \hline \hline 

\multirow{2}{*}{Input} & \textbf{Text} &The radiation value from the sensors, and the command from the \textbf{Agent}  \\ \cline{2-3}
 & \textbf{FOL}  & $r : \mathbb{R}, command : CommandSet$ \\ \hline  \hline

\multirow{2}{*}{Output} & \textbf{Text} &  Radiation Status and if this waypoint has been inspected or not  \\ \cline{2-3}
 & \textbf{FOL}  & $radiationStatus : RadStat, \quad inspected : InspectedType$ \\ \hline  \hline

\multirow{2}{*}{Assume} & \textbf{Text}  &The radiation reading is less than or equal to 0 \\ \cline{2-3}
 & \textbf{FOL}  & $ 0 \leq in.r $ \\ \hline  \hline

\multirow{2}{*}[-3em]{Guarantees} & \textbf{Text}  &  If the command received was to inspect waypoint $i$, then eventually waypoint $i$ will be inspected; and the radiation at waypoint $i$ will be categorised as either green, orange, or red \\ \cline{2-3}
 & \textbf{FOL}  &$\begin{array}{l} \forall i \in \mathbb{N} \bullet (in.command = inspect(i))\\ \t1\implies (\ (out.inspected(i) = TRUE)\\ \t2\land (0 \leq in.r \land in.r < 120 \implies\\ \t3 out.radiationStatus = green)\\ \t2\land (120 \leq in.r \land in.r < 250 \implies\\ \t3 out.radiationStatus = orange)\\ \t2\land (250 \leq in.r \implies\\ \t3 out.radiationStatus = red)\ )\end{array} $ \\ \hline  \hline 
\end{tabularx}
\caption{A summary of the \textbf{Radiation Sensor} node's inputs, outputs, assumptions, and guarantees. \label{tab:contracts_rad}}
\end{table}

\subsubsection{Localisation Node}
\label{sedc:verifyLocal}

The \textbf{Localisation} node's contract specifies that it should output a unique position ($\exists!~ x, y \in \mathbb{R} \ \cdot Position(x, y) $). Using code review, we checked that noise from the sensors did not change the position estimate if the robot has not moved. In practical terms, it might be necessary to allow a short time-window in which the node could obtain several sensor readings before converging on a single position estimate.

It would have been temping to verify this property experimentally, as we do for the \textbf{Navigation} node (see \S\ref{sec:verifyNav}). However, the 3D Gazebo simulation of the nuclear waste store does not simulate noisy sensors (\textit{i.e.}, in any given position the value returned by the simulated sensors was deterministic). Therefore, the position estimates do not change and while we could have verified this, it would not necessarily have told us anything useful.  

Code review was chosen because of the previously mentioned limitations of the simulation and because the \textbf{Localisation} node relies on well-used \gls{ros} libraries. We inspected the code in the \textbf{Localisation} node and observed that the message specification for the node could only return one position at a time. 

While a code review was enough for the purposes of this paper, stronger demonstrations may be needed for (\textit{e.g.}) regulatory approval. For example, the performance of the \textbf{Localisation} node could be tested on a physical robot operating in a safe, test environment.  We could specify a time window, after which a position estimate should have converged to a single value (or a set of values all within acceptable bounds of each other) and then test, using a number of routes around the mapped space, that this behaviour occurred.

\subsubsection{Navigation Node}
\label{sec:verifyNav}

The \textbf{Navigation} node's contract specifies that if there was a command to move to a position $(x,y)$ and the estimated current position of the rover is $(x,y)$, then the rover has successfully arrived at its destination. We verified this experimentally in simulations.

Our verification used the 3D Gazebo simulation of the nuclear waste store as our test environment, and we created an autonomous agent that 
would take a random location within the waste store, navigate to that location and stop. We were then able to compare the goal location with the robot's actual position within the store, as reported by Gazebo.  

The random location was generated as an action available to the agent in the Java environment that linked it to \gls{ros}.  This used Java's built-in random number generation to select an $x$ and a $y$ coordinate within the the waste store map --- random coordinates were regenerated if the original pair were in some inaccessible location, such as one of the tanks or pillars. 

In total, we ran 47 experiments. On average the final location of the agent was 21.4cm away from the goal location with a standard deviation of 4mm.  The worst result was a final position 43cm away from the goal location and the best result was 8cm from the goal location.  The small standard deviation here implies that the simulated Jackal nearly always ends up approximately 21cm away from its goal location.  From a verification perspective, for this component to meet its contract we must decide whether a 21cm error counts as having ``arrived at its destination''.  If this is within tolerance then the component has met its contract, if it is not then the component has not.  In our case we had specifically configured \texttt{movebase} to have a 25cm tolerance for the controller in the $x$ and $y$ coordinates when achieving a goal, and so we can confirm that the node meets its contract.

\subsubsection*{Other Verification Approaches to Localisation, Mapping and Navigation}

Robotic systems often contain a module for \gls{slam}, which can be targeted by specific verification techniques~\cite{carlone2015duality,carlone2015lagrangian}. Given an estimate $x$ (\textit{i.e.}, a solution returned by a state-of-the art iterative solver), the verification approach evaluates whether $x$ corresponds to a global optimum of a cost function $f(x)$. If the answer is positive, then the estimate can be trusted; if the answer is negative, then some recovery technique needs to be performed, because the estimate is not accurate and it is not safe to use. Moreover, this verification technique can be integrated seamlessly in standard \gls{slam} pipelines, and provides a sanity check for the solution returned by standard iterative solvers. In~\cite{briales2016fast} an improved extension builds upon the work from~\cite{carlone2015duality,carlone2015lagrangian} and introduces a novel formulation leading to a higher efficiency, reducing verification times by up to 50 times. \gls{slam} relies on nonlinear iterative optimisation methods that in practice perform both accurately and efficiently. However, due to the non-convexity of the problem, the obtained solutions come with no guarantee of global optimality and may get stuck in local minima.

Other approaches targeting \gls{slam} systems include~\cite{hasegawa2016experimental,duchovn2019verification}, which are more experimental works that focus more on testing rather than verification of \gls{slam}. Also, in~\cite{sivy2016verification}, a mathematical model is developed for \gls{slam} verification and for physical model operation in the environment.

Our Localisation and Navigation nodes combine a SLAM capability with a navigation capability, without an explicit \gls{slam} module so these approaches were not appropriate here.

\subsubsection{Agent Node}
\label{sec:verifyAnget}

The \textbf{Agent} node makes the high-level decisions for the robot. It is implemented in the \Gwen{}~\cite{dennis17gwen} agent programming language and we use the agent-program model checker \gls{ajpf}~\cite{Dennis2012,dennis18:mcapl} to verify its decisions. \gls{ajpf} is an extension of \gls{jpf}~\cite{Visser2003} that enables formal verification of \gls{bdi}~\cite{rao:95b} agent programs by providing a property specification language that extends \gls{ltl} with \gls{bdi} constructs.

In the \gls{bdi} model, agents use their \textit{beliefs} (information that the agent believes about the world) and \textit{desires} (a goal state that the agent wants to achieve) to select an \textit{intention} for execution. To verify that the \textbf{Agent} node obeys its contract, we encode its guarantees into the property specification language for \gls{ajpf} and check that the agent program meets these specifications.

The main parts of the \Gwen{} code for the \textbf{Agent}\footnote{The specification is available in the Zenodo repository: \url{https://doi.org/10.5281/zenodo.6941344} Accessed: 16/01/2026} are shown in Listing~\ref{lst:case1gwen}. The agent starts with a list of static beliefs, such as \emph{location\_coordinate} that takes as parameters a location's numerical identifier (\texttt{Location} variable), name, and its map coordinates; and \emph{next\_location}, with two location identifiers as parameters, where the second parameter is the next location to visit after the first parameter. 

\begin{figure*}[t]
\begin{lstlisting}[caption={Partial code of the \Gwen{} agent.}, captionpos=b, label={lst:case1gwen}, basicstyle=\scriptsize\ttfamily]
+!inspect(Location) : { ~B danger_red, ~B danger_orange, ~B going(0), 
                        B location_coordinate(Location,LocationName,X,Y,Z) } 
	<- +going(Location), move(X,Y,Z);

+movebase_result(Id,3) : { B going(L1), B next_location(L1, L2) }
	<- -going(L1), inspect, +inspected(L1), +!inspect(L2);
+movebase_result(Id,3) : { B going(0) } 
    <- print("Decontamination"), do_nothing;
+movebase_result(Id,2) : { B going(Location) } 
    <- print("Failure");

+danger_red : { ~B going(0), B location_coordinate(0,door,X,Y,Z) } 
    <- +going(0), move(X,Y,Z);
+danger_orange : { ~B going(0), B location_coordinate(0,door,X,Y,Z) } 
    <- +going(0), move(X,Y,Z); 
\end{lstlisting}
\end{figure*}

\begin{sloppypar}
The \texttt{inspect(Location)} plan (line 1) is triggered by the addition of the goal \texttt{inspect}. When the system starts, the \textbf{Agent} begins with a goal to inspect location \texttt{1}. The guard (\textit{e.g.}, context or pre-conditions) of the plan is expressed inside the curly brackets. Here, the guard is that the \textbf{Agent} does not have the beliefs \texttt{danger\_red}, \texttt{danger\_orange}, or \texttt{going(0)}; where the first two beliefs indicate that there is radiation in the current location, and the latter is a \textit{bookkeeping} belief used to track the location that the robot is moving towards (\textit{0} indicates the entrance of the nuclear waste store, which is also the decontamination zone). The last belief, \texttt{location\_coordinate(Location, LocationName, X, Y, Z)}, is a query to the belief base which will use the \texttt{Location} value (obtained from calling the plan, \textit{e.g.}, \texttt{1} when the system starts) to match with its respective belief in the belief base and in turn unify the remaining open variables (\textit{e.g.}, search for a \texttt{location\_coordinate} belief where the \texttt{Location} term is 1, and then unify the remaining open variables with the values from the matched belief). If the guard test is successful, then the plan body (preceded by \lstinline{<-}) is selected for execution. The plan body is executed sequentially, in this case first the bookkeeping belief \texttt{going} is added (\texttt{+}) and then the action \texttt{move} is called with the coordinates of the desired location.
\end{sloppypar}

The second plan, \texttt{movebase\_result(Id, Result)} on line 5, has three variations. All of them are triggered by adding the \texttt{movebase\_result} belief, which reports that the robot's movement (controlled by the move base library) is complete, with the second parameter indicating either success (value $3$) or failure (value $2$). The first variation (lines 5--6) is the main plan, which tests where the rover was going before the action succeeded (\texttt{going(L1)}) and where the next location is, and then removes the outdated bookkeeping belief, performs the \texttt{inspect} action, add a belief that the location $L1$ has been inspected, and then adds the goal to inspect the next location. The second variation (line 7) is for when the robot is moving to the decontamination and should thus not take any additional actions. The third variation (line 9) is for when a \texttt{move} action fails, in which case we simply log that it has failed. 

Finally, the third (line 12) and fourth plans (line 14) are triggered by the addition of the beliefs \texttt{danger\_red} or \texttt{danger\_orange}, respectively. In both cases we test that we are not yet heading to the entrance (\texttt{going(0)}) and get its coordinates from the belief base. Then, if the test succeeds, we add the bookkeeping belief and send the \texttt{move} action for execution (this is published in a \gls{ros} topic that is subscribed to by a move base node).

Clearly the Gwendolen agent uses a different set of abstractions to the \gls{rcl} contracts.  The \gls{ajpf} property specification language has constructs for referring to the beliefs and actions of an agent -- $\lbelief{ag}{\phi}$ means that $\phi$ appears in the agent $ag$'s belief base and $D_{ag}{\phi}$ means that the last action performed by agent $ag$ was $\phi$.  The expressions, $\phi$, map naturally to expressions in the agent program.
Table~\ref{tab:gwen_rcl} presents a mapping from the properties in the \gls{rcl} contracts to properties about this Gwendolen program expressed in the \gls{ajpf} property specification language.

\begin{table}[!t]
\begin{tabular}{c|p{5cm}} 
FOL Predicate & Gwendolen \\ \hline \hline
$at(x, y) \land wayP(x, y) = i$ & $\lbelief{\texttt{jackal}}{going(i)} \land \lbelief{\texttt{jackal}}{movebase\_result(\_, 3)}$\\ \hline

$inspected(i)$ & $\lbelief{\texttt{jackal}}{inspected(i)}$ \\ \hline

$radiationStatus \not\in \{red, orange\}$ & $\neg \lbelief{\texttt{jackal}}{danger\_red} \land \neg \lbelief{\texttt{jackal}}{danger\_orange}$
 \\ \hline
 
$radiationStatus \in \{red, orange\}$ & $ \lbelief{\texttt{jackal}}{danger\_red} \lor  \lbelief{\texttt{jackal}}{danger\_orange}$  \\ \hline
 
$\forall x, y \in \mathbb{R} \bullet  wayP(x, y) \neq i + 1 $ & $\lbelief{\texttt{jackal}}{going(12)} \land \lbelief{\texttt{jackal}}{movebase\_result(\_, 3)}$ \\ \hline 

$wayP(x, y) = i \land command = move(x, y)$ & $\lbelief{\texttt{jackal}}{going(i)}$\\
$command = inspect(i)$ & $D_{\texttt{jackal}}{inspect}$ \\ \hline
\end{tabular}
\caption{A mapping of RCL contract predicates to the AJPF property specification language}
\label{tab:gwen_rcl}
\end{table}

Some of the mappings in Table~\ref{tab:gwen_rcl} are straightforward, but a couple are less so.  While the Gwendolen code utilises location coordinates, the \gls{ajpf} property specification language does not handle floats well so this information is abstracted to reasoning at the level of waypoints -- for instance the property of being at coordinates $(x, y)$ which are the coordinates of waypoint $i$ (the first property in table~\ref{tab:gwen_rcl}) becomes a belief that the Jackal rover was going to waypoint $i$ ($\lbelief{\texttt{jackal}}{going(i)}$) and that the move action has completed successfully ($\lbelief{\texttt{jackal}}{movebase\_result(\_,~ 3)}$ -- here we leverage a detail of the \gls{ajpf} property specification language where we may use $\_$ to stand in for any result, in this case the ID number returned by movebase). Because model-checking is finite state, and we are performing program model checking on a program working with a specific scenario the property that there is no waypoint $i + 1$ (property 5 in table~\ref{tab:gwen_rcl}) becomes the property that the agent is at waypoint 12.

The \gls{ajpf} property specification language is \gls{ltl} while the \gls{rcl} contracts are expressed in FOL.  In general the properties in the \gls{rcl} contracts are of the form $H \Rightarrow C$ where $H$ expresses some condition on the inputs to the node and $C$ expresses some property of the outputs.  These become properties of the form $\always (H' \Rightarrow \eventually C')$ (where $H'$ is the form of $H$ as mapped into the \gls{ajpf} property specification language and $C'$ is the form of $C$).  Since we have no quantifiers in LTL, and model checking is finite state, a general property about \emph{all waypoints} in \gls{rcl} becomes $n$ properties about each of the $n$ waypoints for \gls{ajpf}.

In all there are three guarantees specified in the \gls{rcl} contract for this node.  We consider each of these in turn.

\paragraph{Guarantee 1 (lines 15--18 in Listing~\ref{list:CS1AgentRCL})}  This guarantee states that 
``If the Robot is at, and has inspected waypoint $i$, and the radiation level was not dangerous, then it will move to the next waypoint or to waypoint 0 (the exit) if there is no next waypoint. ''

We express this in the \gls{ajpf} property specification language as 12 properties of the form:
$$
\begin{array}{l}
\varphi_1^i = \always ((\lbelief{\texttt{jackal}}{going(i)} \land \lbelief{\texttt{jackal}}{movebase\_result(\_, 3)} \land \\
\qquad \lnot \lbelief{\texttt{jackal}}{danger\_red} \land  \lnot \lbelief{\texttt{jackal}}{danger\_orange} \land \lnot \lbelief{\texttt{jackal}}{going(0)}) \\
\qquad \Rightarrow \eventually B_{\texttt{jackal}}going(i + 1)) \\
\end{array}
$$
where $i$ is instantiated with the waypoint numbers from 0 to 11 -- e.g., as 
$$
\begin{array}{l}
\varphi_1^4 = \always ((\lbelief{\texttt{jackal}}{going(4)} \land \lbelief{\texttt{jackal}}{movebase\_result(\_, 3)} \land \\
\qquad \lnot \lbelief{\texttt{jackal}}{danger\_red} \land  \lnot \lbelief{\texttt{jackal}}{danger\_orange} \land \lnot \lbelief{\texttt{jackal}}{going(0)}) \\
\qquad \Rightarrow \eventually B_{\texttt{jackal}}going(5)) \\

\end{array}
$$
Plus a 12th property:
$$\begin{array}{l}
\varphi_1^12 = \always (\lbelief{\texttt{jackal}}{going(12)} \land \lbelief{\texttt{jackal}}{movebase\_result(\_, 3)} \\
\qquad \Rightarrow \eventually B_{\texttt{jackal}}going(0)) \\
\end{array}
$$
The RCL contract for this guarantee has an additional condition $inspected(i) = TRUE$ which we have omitted giving us a more general property.  We have also omitted the conditions that the radiation status is not orange or red from $\varphi_1^12$ again to give us a more general property (in this case it remains true no matter what the radiation status since the robot also goes to waypoint 0 in the case of dangerous radiation).

\paragraph{Guarantee 2 (Lines 20--21 in Listing~\ref{list:CS1AgentRCL})} This guarantee states that ``if the Robot is at, and has not inspected, waypoint $i$, then inspect it.''  We verified two variants of this guarantee in \gls{ajpf}.  The first, $\varphi_2^i$, are $n$ properties (with $i$ instantiated from 1 to 12) that follow directly the mapping of the RCL contract predicates into the \gls{ajpf} property specification language as presented in Table~\ref{tab:gwen_rcl}.
$$
\begin{array}{l}

\varphi_2^i = \always ((\lbelief{\texttt{jackal}}{going(i)} \land \lbelief{\texttt{jackal}}{movebase\_result(\_, 3)} \land \\
\qquad \lnot \lbelief{\texttt{jackal}}{danger\_red} \land \lnot \lbelief{\texttt{jackal}}{danger\_orange} \land \lnot \lbelief{\texttt{jackal}}{going(0)}) \\
\qquad \Rightarrow \eventually D_{\texttt{jackal}}inspect) 

\end{array}
$$
However, since the $inspect$ action doesn't actually specify the waypoint inspected, it was possible for the Gwendolen program to satisfy this property while missing a waypoint (e.g., it could fail to inspect waypoint 4, but the property would become true when waypoint 5 was inspected).  Therefore we also verified a variant that checked that the agent \emph{believed} it had checked all waypoints - utilising the fact that the agent adds this belief after each waypoint is inspected.  Our variant $\varphi_2^{i'}$ properties were:
$$
\begin{array}{l}

\varphi_2^{i'} = \always ((\lbelief{\texttt{jackal}}{going(i)} \land \lbelief{\texttt{jackal}}{movebase\_result(\_, 3)} \land \\
\qquad \lnot \lbelief{\texttt{jackal}}{danger\_red} \land \lnot \lbelief{\texttt{jackal}}{danger\_orange} \land \lnot \lbelief{\texttt{jackal}}{going(0)}) \\
\qquad \Rightarrow \eventually \lbelief{\texttt{jackal}}{inspect}) 

\end{array}
$$
While these properties are not such a close mapping from the \gls{rcl} contract, we believe they better capture the intent behind the guarantee.

\paragraph{Guarantee 3 (Lines 23--25 in Listing~\ref{list:CS1AgentRCL})} This guarantee states that ``if the Robot inspects a waypoint and the radiation level is dangerous, or there are no more waypoints, then return to the original waypoint (the exit)''.
This became two properties (one for a radiation status of red and one for orange).  The final part of the guarantee (about the behaviour when there are no more waypoints, was effectively covered by $\varphi_1^12$)

$$
\begin{array}{l}

\varphi_3^a = \always (\lbelief{\texttt{jackal}}{danger\_red} \Rightarrow \eventually \lbelief{\texttt{jackal}}{going(0)})\\

\end{array}
$$

$$
\begin{array}{l}

\varphi_3^b = \always (\lbelief{\texttt{jackal}}{danger\_orange} \Rightarrow \eventually \lbelief{\texttt{jackal}}{going(0)})\\

\end{array}
$$

These \gls{ltl} properties were only used for verification of the agent node at design time, using the \gls{ajpf} model checker and were not used elsewhere (e.g., for the runtime monitors).  A discussion of \gls{ajpf} and its efficiency can be found in~\cite{dennisfisher23}.

\subsubsection{Radiation Sensor Node}
\label{sec:verifyRad}

The \textbf{Radiation Sensor} node interprets information from the (simulated) radiation sensor, categorising the values into either \texttt{green} (low), \texttt{orange} (medium), or \texttt{red} (high). We modelled the behaviour of the \textbf{Radiation Sensor} as a simple program in a Hoare Logic-style language~\cite{hoare1969axiomatic} and proved properties corresponding to the node's guarantee by hand.

\begin{figure*}[t]
\begin{lstlisting}[caption={Program for the radiation sensor node.}, captionpos=b, label={lst:radsensor}]
input := radiation_at(i)
IF (input < 120) THEN 
   output := green 
ELSE
   IF (input < 250) THEN 
      output := orange 
   ELSE
      output := red
\end{lstlisting}
\end{figure*}

The \textbf{Radiation Sensor} node's contract assumes that the measured radiation is $0$ or positive.
Its guarantee specifies that if the node is asked to inspect waypoint $i$ (\texttt{command = inspect(i)}) then:
\begin{enumerate}
    \item the proposition $inspected(i)$ becomes true, $inspected(i) = TRUE$;
    \item low-level radiation is categorised as \texttt{green}, $0 <= r < 120 \Rightarrow\\ radiationStatus = green$;
    \item medium-level radiation is categorised as  \texttt{orange}, $120 <= r < 250 \\ \Rightarrow radiationStatus = orange$; and,
    \item high-level radiation is categorised as \texttt{red}, $250 <= r \Rightarrow radiationStatus \\= red$.
\end{enumerate}
We proved these four properties by hand using Hoare Logic.  If \texttt{P} is our program from Listing~\ref{lst:radsensor} then: 

\begin{enumerate}
    \item $\{\top\}$ \texttt{P}$\{input = radiation\_at(i)\}$ -- The input to the node is always the radiation level at $i$. This corresponds to the requirement that $inspected(i) = TRUE$ in the contract.
    \item $\{radiation\_at(i) < 120\}$ \texttt{P}$\{output = green\}$ -- if the radiation level at $i$ is less than 120 then, after execution of the program, $output$ is green.  
    \item $\{120 \leq radiation\_at(i) < 250\}$ \texttt{P}$\{output = orange\}$ -- if the radiation level at $i$ is between 120 and 250 then, after execution of the program, $output$ is orange.
    \item $\{radiation\_at(i) \geq 250\}$ \texttt{P}$\{output = red\}$ -- if the radiation level at $i$ is greater than 250 then, after execution of the program, $output$ is red.
\end{enumerate}
We show the proof for property (2)  in Fig.~\ref{fig:HoareProof}. The other proofs follow a similar pattern, and can be found in~\ref{app:proofs}.

\begin{figure*}[t]
\begin{lstlisting}[caption={Hoare Logic Proof that $\{radiation\_at(i) < 120\}$\textbf{P}$\{output = green\}$. The program is denoted by \textbf{bold} text, and the proof steps are indicated by $\{braces\}$ .},captionpos=b, label={fig:HoareProof}, aboveskip=0em, belowskip=0em]
%*$\{radiation\_at(i) < 120\}$*)
%*\textbf{input := radiation\_at(i)}*)
%*$\{input < 120\}$*) - Assignment Axiom
%*  \textbf{IF (input < 120) THEN}*) 
   %*$\{input\ <\ 120 \&\ input\ <\ 120\}$*) 
   %*$\{input\ <\ 120\ \&\ green\ =\ green\}$*) - strengthening
   %*  \textbf{output\ :=\ green} *) 
   %*$\{input\ <\ 120\ \&\ output\ =\ green\}$*) - Assignment Axiom
   %*$\{output\ =\ green\}$*) - weakening
%*  \textbf{ELSE}*)
   %*$\{input\ <\ 120\ \&\ input\ >= 120\}$*)
   %*$\{FALSE\}$*) - strengthening
%*  \textbf{IF (input < 250) THEN} *)
      %*$\{FALSE\ \& \ input < 250\}$*)
      %*$\{FALSE\}$*) - strengthening
      %*  \textbf{output := orange}*) 
      %*$\{FALSE\}$*) - Assignment axiom
    %*  \textbf{ELSE}*)
       %*$\{FALSE\ \&\ input\ >=\ 250\}$*)
       %*$\{FALSE\}$*) - strengthening
       %*  \textbf{output := red}*)
       %*$\{FALSE\}$*) - Assignment axiom
   %*$\{FALSE\}$*) - Conditional Rule
   %*$\{output\ =\ green\}$*) - weakening
%*$\{output\ =\ green\}$*) - Conditional Rule
\end{lstlisting}
\end{figure*}

\subsection{Step 4b: Automatic Synthesis of Runtime Monitors}
\label{sec:case1step5}

This step takes the \gls{rcl} contracts from Step 2 (\S\ref{sec:case1step2}) and synthesises \gls{rml}  monitors that are compatible with the ROSMonitoring framework~\cite{DBLP:conf/taros/FerrandoC0AFM20}. As mentioned in \S\ref{sec:dsl}, our tool \textsc{Vanda} parses the \gls{rcl} contracts, and then automatically generates the monitors and ROSMonitoring configuration files. \textsc{Vanda} uses a \textit{contract translator} to produce the configuration file and structure of each monitor, and calls a \textit{\gls{fol} translator} to translate each guarantee.

Listing~\ref{list:agentRML} shows the \gls{rml} that was derived from the \gls{rcl} contract in Listing~\ref{list:CS1AgentRCL}. Lines 1--6 contain the event types corresponding to the contract's topics, while on lines 8--12 we have the \gls{rml}  terms for the contract's guarantees. These terms are obtained through the process shown in \S\ref{sec:rv}.
\gls{rml}  offers a Prolog-like notation, where variables that are not of interest can be replaced with the \texttt{\_} symbol. This is interpreted by \gls{rml}  as a wildcard variable, which can be assigned to any value with no restrictions.

To aid the reader's understanding of this translation step, we show in more detail how a specific part of this \gls{rml} specification has been generated. The approach is described in \S~\ref{sec:rv}, and the same reasoning is applied to the translation of the other contracts.

\begin{figure*}[t]
\begin{lstlisting} [language=RCL, caption={A snippet of the \textbf{Agent}'s contract from Listing~\ref{list:CS1AgentRCL}.},captionpos=b, label={list:CS1AgentRCLSnippet}, basicstyle=\scriptsize\ttfamily, aboveskip=0em, belowskip=0em]
node agent{
...
topics( 
gazebo_radiation_plugin/Snapshot command matches(out.command),
gazebo_radiation_plugin/Snapshot inspected matches(in.inspected),
gazebo_radiation_plugin/Snapshot at matches(in.at),
gazebo_radiation_plugin/Snapshot wayPNow matches(in.wayP),
gazebo_radiation_plugin/Snapshot radiationStatus matches(in.radiationStatus)
)
...
guarantee(forall(x', y' in REAL, i in NATURAL | in.at(x', y') == TRUE and in.wayP(x', y') == i and in.inspected(i) == FALSE -> out.command == inspect(i) ))
...
}
\end{lstlisting}
\end{figure*}

Listing~\ref{list:CS1AgentRCLSnippet} shows a snippet of Listing~\ref{list:CS1AgentRCL} that contains only one of the \textbf{Agent} node's guarantees. The \textcolor{blue}{\texttt{guarantee}} in Listing~\ref{list:CS1AgentRCLSnippet} is translated into the \gls{rml} term \texttt{t2} in Listing~\ref{list:agentRML} using the $(guarantee)$ rule in Figure~\ref{trans-fig}. 

The body of the \textcolor{blue}{\texttt{guarantee}} consists of a universal quantifier (\texttt{forall}) that contains an implication (\lstinline{->}). First, we use the $(implies)$ rule in Figure~\ref{trans-fig} to translate the implication; $(implies)$ negates the left operand and puts it in disjunction with the right operand. The left operand is: \texttt{in.at(x', y') == TRUE and in.wayP(x', y') == i and in.inspected(i) == FALSE}, which produces the following \gls{rml}  specification:\\
\centerline{\texttt{($\lnot$at(x, y) $\lor$ $\lnot$wayP(x, y, i) $\lor$ inspected(i))}}.
\noindent Note, that the \gls{rml}  is negated because \texttt{(a -> b) = ($\lnot$a or b)}. Then, the right operand \texttt{in.command\,==\,inspect(i)} is directly mapped into RML as: \texttt{command(inspect, i)}. This follows from the $(equals)$ rule in Figure~\ref{trans-fig}, which maps the command into its corresponding event type and adds it to the set of event types. 

\lstset{
	morekeywords={not,none,any,empty,matches,all,let,if,else,silent,id,topics,node,log,nodes,monitor,monitors},
	keywordstyle=\color{blue},
	morestring=[b]',
	stringstyle=\color{red},
	morecomment=[l]{//},
	commentstyle=\color{olive},
	mathescape=true,
	basicstyle=\small\ttfamily,
	captionpos=b,
	tabsize=4,
	breaklines,
	breakatwhitespace,
	showstringspaces=false,
	keepspaces
}

\begin{figure*}[t]
\begin{lstlisting} [caption={The \textbf{Agent}'s \gls{rml}  derived from the RCL contract shown in Listing~\ref{list:CS1AgentRCL}.},captionpos=b, label={list:agentRML}, basicstyle=\scriptsize\ttfamily,escapeinside={(*}{*)}]
at(x, y) matches { topic: 'gazebo_radiation_plugin/Snapshot', at: {posX: x, posY:y} };
wayP(x, y, i) matches { topic: 'gazebo_radiation_plugin/Snapshot', wayPNow: {posX: x, posY: y, id: i} };
radiationStatus(Lvl) matches { topic: 'gazebo_radiation_plugin/Snapshot', radiationStatus: { level: Lvl } };
command(Cmd, x, y) matches { topic:'gazebo_radiation_plugin/Snapshot', command: {command: Cmd, posX: x, posY: y } };
command(Cmd, i) matches { topic:'gazebo_radiation_plugin/Snapshot', command: {command: Cmd, id: i} };
inspected(i) matches { topic: 'gazebo_radiation_plugin/Snapshot', inspected{ id: i } };

t1 = {let x1, y1, i; (((*$\lnot$*)at(x1,y1) (*$\lor$*) (*$\lnot$*)wayP(x1, y1, _) (*$\lor$*) (*$\lnot$*)radiationStatus(green))) (*$\lor$*) {let x2,y2; (wayP(x2, y2, i+1) (*$\land$*) command(move, x2, y2))}}*;

t2 = {let x, y, i; ((*$\lnot$*)at(x, y) (*$\lor$*) (*$\lnot$*)wayP(x, y, i) (*$\lor$*) inspected(i)) (*$\lor$*) (command(inspect, i))}*;

t3 = {(radiationStatus(green) (*$\lor$*) {let x, y; wayP(x, y, 0) (*$\land$*) command(move, x, y))})}*;

\end{lstlisting}
\end{figure*}

Next, the variables derived from the universal quantifier are added to the \gls{rml}  specification. Thus, we obtain \texttt{\{let x, y, i; t\}}, where \texttt{t} is the previously created \gls{rml} specification (the one denoting the implication). Through this quantification, in \gls{rml} we can generalise \texttt{t} over the set of variables \texttt{\{ x,y,i \}}. Finally, since the \textcolor{blue}{\texttt{guarantee}} ranges over all possible instantiations for the variables, an \texttt{*} is added at the end of \gls{rml}  specification (as specified in the $(guarantee)$ rule in Fig.~\ref{trans-fig}). This operator, as in regular expressions, requires the \gls{rml}  specification to be matched multiple times. As previously mentioned, the same reasoning is applied to the other guarantees in Listing~\ref{list:CS1AgentRCL}.

Once the \gls{rml}  specification (Listing~\ref{list:agentRML}) and the configuration file (Listing~\ref{list:agentConfig}) have been generated, the \gls{rv}  step can be applied, as previously presented in Fig.~\ref{fig:rosmon-pipeline}. Hence, the \gls{rml}  specification can be used to synthesise an oracle, which will then be queried at runtime by the corresponding \gls{ros}  monitors (automatically synthesised from configuration files).

\begin{figure*}[t]
\begin{lstlisting}[caption={The monitor configuration file for the \textbf{Agent} node.},captionpos=b, label={list:agentConfig}, basicstyle=\scriptsize\ttfamily,escapeinside={(*}{*)}]
monitors:
- monitor:
    id: monitor_agent
    log: ./agent_log.txt
    topics:
     - {action: (*log*), name: gazebo_radiation_plugin/Snapshot, type: gazebo_radiation_plugins.msg.Snapshot}
\end{lstlisting}
\end{figure*}

ROSMonitoring  uses the configuration file, shown in Listing~\ref{list:agentConfig}, to generate the \textbf{Agent} node's monitor. Lines 3 \& 4 show the agent monitor's identifier and the location where its log files should be stored. The monitor also requires the list of topics to subscribe to (lines 5--6) to observe the events at runtime. These topics are obtained from lines 7--12 in Listing~\ref{list:CS1AgentRCL}. Note that, in this specific scenario, we only have one single topic, called \texttt{Snapshot}, that contains the information we need.

 As previously mentioned, ROSMonitoring automatically synthesises runtime monitors as additional nodes in the \gls{ros}  program. The monitor nodes that ROSMonitoring synthesises subscribe to the topics of interest listed in the configuration file (derived from the \gls{rcl} specification). After that, they collect the messages that are published on those topics and perform the runtime analysis of the \gls{rml}  specification. If a monitor observes a violation of an \gls{rml}  specification, it reports the violation to the entire system by publishing a specific error message. This information can be used by the system to promptly react and possibly recover from the erroneous behaviour.

An additional step of instrumenting the implementation of the case study was necessary to fully capture the execution traces that the ROSMonitoring monitors needed to observe. For example, this instrumentation step included publishing information that was kept within the \textbf{Agent}, such as beliefs about which waypoint the robot is at, to a ROS topic. This allowed the ROSMonitoring monitors to observe this information and give their verdict. Note that the instrumentation step is only necessary to track the behaviour of components, such as the \textbf{Agent}, which otherwise would not be observed; it does not change the behaviour of the system components.

\begin{figure}
    \centering
    \begin{subfigure}[b]{0.45\textwidth}
        \includegraphics[width=\textwidth]{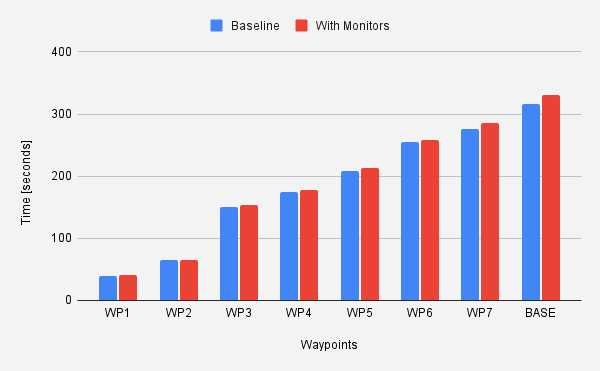}
        \caption{The y-axis indicates the time at which the rover completes its visit.}
        \label{fig:overhead1}
    \end{subfigure}
    \hfill
    \begin{subfigure}[b]{0.45\textwidth}
        \centering
        \includegraphics[width=\textwidth]{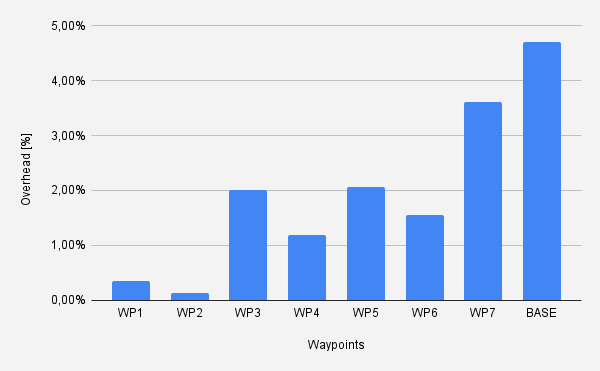}
        \caption{The y-axis indicates the percentage increase with respect to the baseline.}
        \label{fig:overhead2}
    \end{subfigure}
    \caption{Overhead experiment results in the remote inspection case study. The x-axis represents the visited waypoints. BASE is WP0, the door location. When reaching WP7 the rover detects dangerous radiation and then decides to move to the door for decontamination.}
\end{figure}

Using \gls{rml} monitors for \gls{rv} of our contracts only causes slight overheads. We conducted an analysis of the overhead introduced by the presence of monitors in our case study. Following a similar approach to~\cite{DBLP:conf/taros/FerrandoC0AFM20}, first we timed our system patrolling its simulated environment without monitors to establish a baseline, then timed the system with the monitors. Fig.~\ref{fig:overhead1} shows the time required for the robot to visit all the waypoints and return to the exit, averaged over 30 runs. 
The baseline, without monitors, is shown in blue, while the scenario with ROSMonitoring's monitors performing runtime verification is shown in red. As it can be observed, the overhead introduced by the monitors is negligible. Fig.~\ref{fig:overhead2} provides a more detailed view, showing the percentage overhead at waypoint (relative to the baseline execution without the monitors).
Further experiments on the  overhead of ROSMonitoring are presented in~\cite{DBLP:conf/taros/FerrandoC0AFM20}, where the overhead of \gls{rml} monitors in ROSMonitoring was analysed. This analysis empirically demonstrated that \gls{rml} monitors were introducing almost no overhead to the robotic system, as long as ROSMonitoring was used to detect failures and not to enforce correctness. 

Finally, thanks to the structure of the \gls{rml} terms that we synthesise, the \gls{rv} monitors operate in linear time (with respect to the length of the analysed trace). For more information about the time complexity of \gls{rml} monitoring, the reader can find additional details in~\cite{ANCONA2021102610}.

\section{Beyond the Case Study}
\label{sec:discussion}

This section goes beyond our case study, to demonstrate other features of our framework,
In \S\ref{sec:agentDafny}, we provide a worked example of how our framework enables the implementation of one node to be swapped with another, as long as the replacement node satisfies the specification. 
\S\ref{sec:towards} explores how our approach to composing the node specifications can be applied to systems that are not implemented in \gls{ros}. But first, we discuss the general utility of our approach, beyond what is presented in \S\ref{sec:exampleSystem}. The analysis of our approaches wider applicability, and the two formal models in this chapter are available from the Zenodo repository that accompanies this paper: \url{https://doi.org/10.5281/zenodo.6941344}.

While we have shown the applicability of the approach to our case study, there remains a question of whether it is useful in general for the verification of robotic systems. As a step towards assessing this, we examined fifty-six papers containing verifications of autonomous systems identified in our previous survey paper~\cite{Luckcuck2019}. For each paper, we assessed if our approach would be applicable to the system it describes. We identified 20 papers with a case study to which our general approach would be applicable.  Three of these, specifically involved the verification of \gls{ros} nodes.  Of these 20 papers, nine involved the verification of a component within a larger autonomous system and 10 involved the system-level verification of a modular system (the remaining one applied runtime verification to a modular system).

Of the systems where we deemed our approach to be inapplicable, the majority (20 in total) involved the verification of distributed or swarm systems where an unspecified number of similar components act to produce system level behaviour and, at present, our formal system is not equipped to work with such architectures.  Four examples involved the verification of monolithic systems.  In general, the behaviour of these monolithic systems was comparatively simple when compared to modular systems (typically these were hybrid control systems where the verification focused on obstacle avoidance behaviour).
We identified nine verification attempts as borderline.  These involved fixed-size teams of independent systems -- often individual robots but sometimes independent, communicating systems within a smart home or similar environment.  Our approach should be extensible to such systems by viewing sub-systems as components in a hierarchical fashion.  One paper contained an analysis of a failed verification attempt -- we also classified this as borderline.

\subsection{Compositionality enables Interchangeability}

\label{sec:agentDafny}

One of the advantages of our approach is that we can swap a node's implementation with another (similar) implementation, which may be modelled and verified using different specification languages and verification tools to the original node, as long as the new implementation conforms to the \gls{fol} contract. For example, instead of using a Gwendolen \gls{bdi} agent (as we do in \S\ref{sec:exampleSystem}) we might choose a simpler way of making the systems' executive decisions that still satisfies $\mathcal{G}_A(\overline{o_A})$. 

To demonstrate the modularity of our approach, this section describes a re-specification and verification of the \textbf{Agent} (\S\ref{sec:verifyAnget}) node in Dafny~\cite{leino2010dafny}. That is, we replace the Gwendolen BDI implementation and verification with this new Dafny implementation. Dafny is a programming language enriched with specification constructs -- for example: pre-/post-conditions and loop invariants\footnote{Note that our Dafny model uses \texttt{function method} constructs from Dafny 3.x which are deprecated in the latest version (4.9.1). However, the proof obligations are still discharged using normal functions in Dafny version 4.9.1.}. This enables the functional correctness of Dafny programs to be statically verified, by translating them into the intermediate verification language Boogie~\cite{barnett2005boogie} and using the theorem prover Z3~\cite{de2008z3} to automatically discharge the proof obligations for the specification statements in the program.

The idea here is that instead of using a Gwendolen \gls{bdi} program, the decision-making algorithm(s) will be implemented in a general-purpose programming language, for example C++ or Python for compatibility with \gls{ros}. So we use Dafny to implement and verify the algorithms that the new decision-making component will execute.
Our prior work demonstrates that the correspondence between Dafny and general-purpose languages, such as Python, makes it relatively straightforward to translate between formal models and implemented code \cite{farrell2021formal}.

Listing~\ref{list:dafnyagent} shows our Dafny model, in which the \texttt{Agent} method (line 5) provides an alternative implementation of the agent's behaviour to the Gwendolen \gls{bdi} version that was presented in \S\ref{sec:verifyAnget}. Importantly, the Dafny implementation follows the specification in its contract (Table~\ref{tab:contracts_agent}).

In Dafny, pre-conditions (assumptions) are indicated by the \texttt{requires} keyword. Lines 6--9 of Listing~\ref{list:dafnyagent} show the \texttt{Agent} method's pre-conditions, taken from its contract. One small change is that where the \textbf{Agent}'s contract has an assumption that the radiation status will be either red, orange, or green; our Dafny version encodes this constraint as the \texttt{RadiationLevel} datatype (line 3) using the radiation values, and using this in the parameters (for \texttt{radstat}) of the \texttt{Agent} method. 
Importantly, if the \texttt{Agent} method is called by another method, then the calling method's guarantees must not violate the assumptions of the \texttt{Agent} method.

\begin{figure}[!t]
\begin{dafny}[numbers=left, stepnumber=1, frame=single, caption={Dafny Agent}, label={list:dafnyagent}]
datatype Action = Inspect | MoveNext
datatype Command = Command(a:Action, w:int) 
datatype RadiationLevel = red | orange | green 

method Agent(waypoints: seq<int>, currentpos:int, radstat:RadiationLevel, wheelsready:bool) returns (actions: seq<Command>)
requires currentpos >= 0 && currentpos <= 12; //13 waypoints to patrol
requires |waypoints| == 13;
requires forall i:int :: 0<=i<|waypoints| ==> waypoints[i] == i;
requires forall i:int :: 0<=i<|waypoints| ==> i in waypoints;
ensures (wheelsready) == false ==> actions ==[];//safety check: if the hardware is not ready then do nothing.
ensures forall i:int :: 0<=i<|waypoints| ==> (wheelsready && at(i) && inspected(i) && radstat != red && radstat != orange  ==> exists next:int :: 0<=next<|waypoints| && Command(MoveNext, next) in actions);
ensures forall i:int :: 0<=i<|waypoints| ==> (wheelsready && at(i) && (! inspected(i)) ==> Command(Inspect, i) in actions);
ensures wheelsready && (radstat == red || radstat == orange) ==> Command(MoveNext, 0) in actions;
{
    var time, next := 0, 0; 
    var current:=currentpos;
    actions := [];

    if(wheelsready){
        while(wheelsready && time <200)   
        invariant next in waypoints;
        invariant current in waypoints;
        invariant time > 0 && (radstat == red || radstat == orange) ==> Command(MoveNext, 0) in actions && next == 0;
        invariant time > 0 && at(current) && (! inspected(current)) ==> Command(Inspect, current) in actions;
        invariant time > 0 && at(old(current)) && inspected(old(current)) && radstat != red && radstat != orange ==> Command(MoveNext, next) in actions && current == old(next);
        {
            if(at(current) && inspected(current) && radstat != red && radstat != orange){
                next := getnextwaypoint(current);
                actions := actions + [Command(MoveNext, next)];
            }
            if(at(current) && (! inspected(current))){
                actions := actions + [Command(Inspect, current)];
            }
            if(radstat == red || radstat == orange){
                next := 0;
                actions := actions + [Command(MoveNext, next)];
            }
            time := time +20;
            current := next;
        }
    }
}
\end{dafny}

\end{figure}





Post-conditions (guarantees) in Dafny are indicated by the \texttt{ensures} keyword.
Lines 10--13 of Listing~\ref{list:dafnyagent} show the \texttt{Agent} method's post-conditions, which are more detailed than the \textbf{Agent}'s contract, because they also check the status of the software controlling the rover's wheels before instructing it to carry out any commands. The post-condition on line 10 specifies that no actions are assigned when the wheels are not ready.
The post-conditions on lines 11--13 correspond to the \textbf{Agent}'s guarantees from its contract (Table~\ref{tab:contracts_agent}). For verification in Dafny, we must include loop invariants (lines 21--25), which help the verification tool (Z3) to prove the post-conditions.  We automatically discharged the associated proofs in Dafny 3.x with Z3 using Visual Studio Code on Ubuntu.

Another difference between our Dafny and Gwendolen implementations is that we had to add a notion of \texttt{time} to the Dafny model (the \texttt{time} variable is declared on line 15 and updated on line 38) to be able to prove loop termination. Dafny is primarily concerned with program safety, so termination is necessary for complete verification of Dafny programs.

This Dafny program verifies the decision-making algorithm, using a language that is closer to those in which \gls{ros} nodes can be implemented. Once the Dafny program has been verified against the \textbf{Agent} contract, it can be carefully reimplemented as a \gls{ros} node. 
The Dafny \textbf{Agent} can also be used with our calculus (\S\ref{sec:combination}) because it preserves the contract's assumptions and guarantees. Although the RCL and Dafny contracts use slightly different datatypes, we can see the correspondence between them easily. Specifically, the Dafny model returns a sequence of (action, waypoint) pairs that represents the \texttt{CommandSet} of the RCL contract which contains move and inspect commands. The \texttt{CommandSet} can thus be reconstructed from the Dafny output and vice versa if needed. Similarly, the belief base of the Gwendolen agent can be checked and shown to contain the same information as the \texttt{CommandSet}.

Not only does this allow the user to develop multiple models (and potentially implementations) of the same node, it also facilitates the use of predefined or library functions that have been verified previously and also meet the required contracts. This can potentially streamline the verification step by supporting the use of previously verified robotic system nodes.

The Dafny and Gwendolen agents described in this work provide potential alternatives to illustrate our approach. A developer should choose the formalism that most suitably matches their needs. Users of our framework should ensure that their FOL contracts are captured correctly in the formalism that they choose. To support this, systematic mappings between FOL and other logics/formalisms can be used but such mappings are out of scope for this paper.

Both the Dafny and Gwendolen agents are designed to visit a specific number of waypoints. This is necessary in Gwendolen since model-checking is used for verification. However, our Dafny model could be more liberal here and, since all sequences in Dafny are finite, omit the specific sequence size. The main modification would be to add a precondition: \texttt{requires currentpos in waypoints} to make sure that the input is valid. The \texttt{getNextWaypoint} function must also consider the size of the waypoints sequence when producing the next waypoint. In comparison with Dafny, Gwendolen supports temporal logic specifications which may be useful for a developer wanting to also verify additional temporal properties about their agent. Both Dafny and Gwendolen can be executed but Gwendolen, via ROSbridge, can be used directly on the robot. No such support currently exists for Dafny. Both Dafny and Gwendolen allow users to specify properties that were not specified in our FOL contracts including, for example, frame conditions in Dafny and temporal logic properties in Gwendolen. Dafny's verifier is a first-order SMT solver which makes it natural to express FOL contracts in Dafny. 

For simplicity, the Dafny and Gwendolen agents that we present in this paper are very similar in terms of functionality but, in practice, they could be slightly different so long as they obey the same FOL contract. For example, one of the agents could incorporate other behaviours than just inspection. It could both inspect and take samples from the locations that it visits. It may even stop to do self-maintenance before proceeding to the next waypoint.

\begin{figure}[t]
    \centering
    \includegraphics[width=\linewidth]{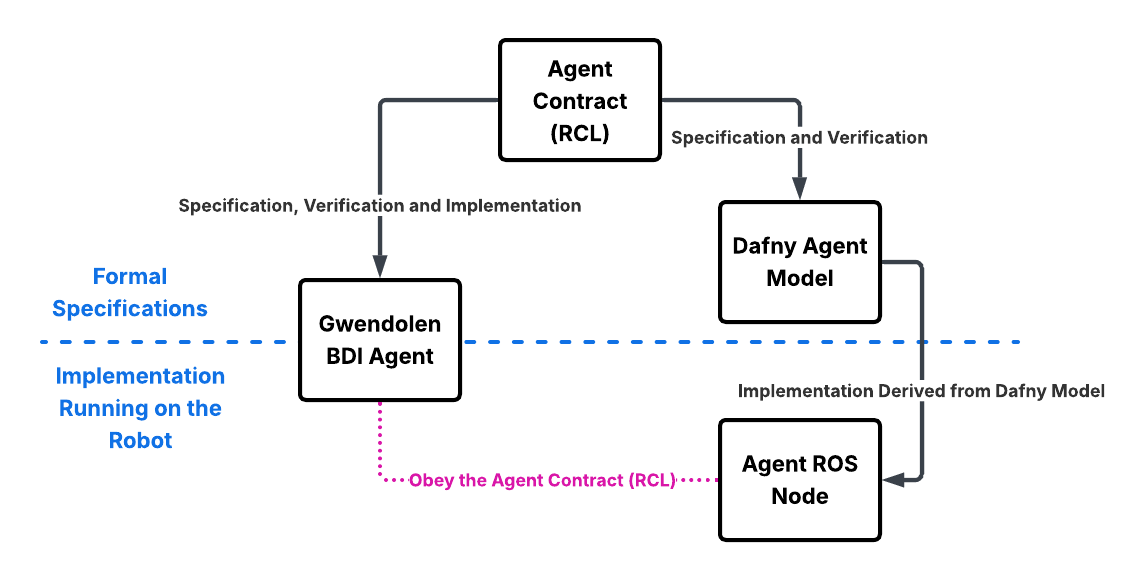}
    \caption{Summary of the different agent models, showing the differences and similarities. Both the Gwendolen BDI and Dafny agent models are specified by the \gls{rcl} contract, but the Dafny model must be implemented in a ROS node to be used as the \textbf{Agent}.}
    \label{fig:agentRelations}
\end{figure}

We illustrate the relationships between the different agent models in Fig.~\ref{fig:agentRelations}. Here, we can see that the \gls{rcl} contract is a specification for both the Dafny program and Gwendolen BDI agents, and that they are verified using formal techniques (model-checking for the Gwendolen BDI agent and Theorem Proving for the Dafny agent model). However, one of the main differences between these two representations is that the Gwendolen BDI agent is both a formal specification and an implementation that can be used directly in \gls{ros}, whereas the Dafny agent model needs to be manually implemented in a ROS node. Providing that the \gls{ros} implementation of the Dafny agent model is correct, both the Dafny and Gwendolen BDI agents obey the original \gls{rcl} contract. Though their actual behaviours may differ as described above, depending on the choice and needs of the developer.

\FloatBarrier

\subsection{Towards Applying our Approach to Non-ROS Systems}
\label{sec:non-robotic}

\label{sec:towards}

So far, we have focussed on \gls{ros} systems; this section explores how our approach can be adapted to robotic systems that do not use \gls{ros}. 
We use previous work~\cite{bourbouh2021integrating}~as our example, in which heterogeneous verification methods are used to verify an autonomous rover that is modelled in Simulink. The rover's mission is to autonomously navigate around a grid-world of known size, visiting waypoints to collect data, recharging as necessary. This example has similar functionality to our case study system in \S\ref{sec:exampleSystem}, but does not use \gls{ros}. 

\begin{figure}[ht]
    \centering
    \includegraphics[width =\textwidth]{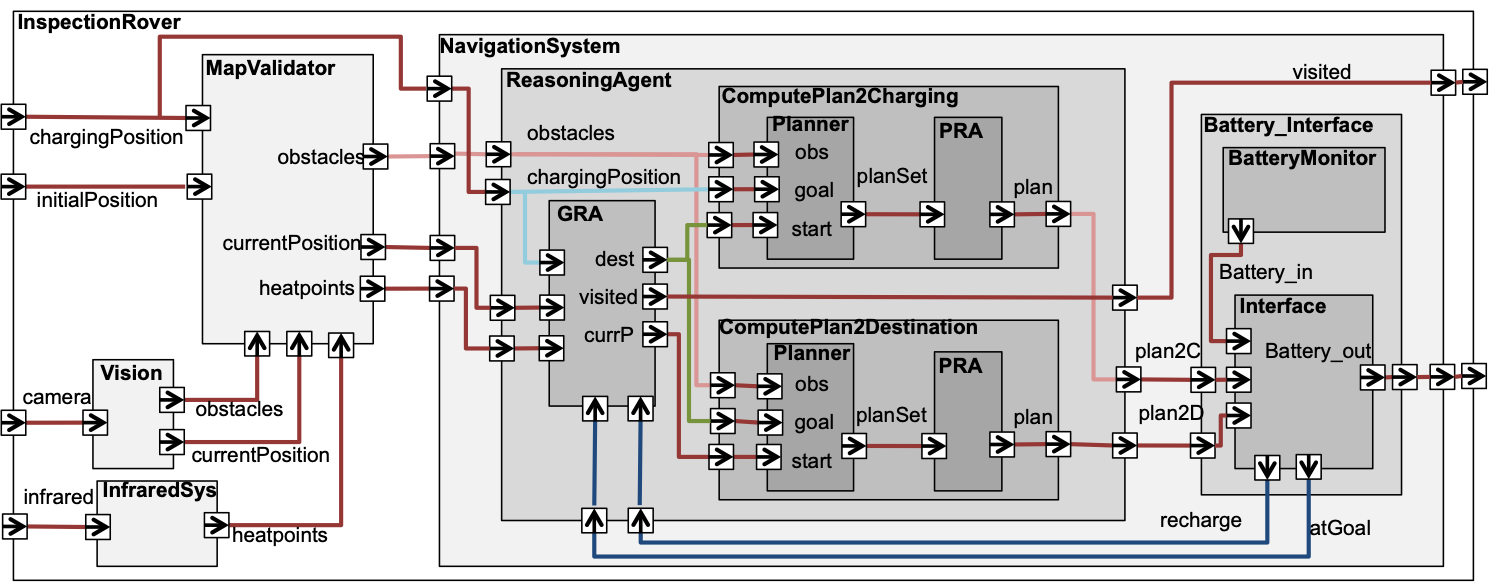}
    \caption{Architecture of autonomous rover from \cite{bourbouh2021integrating}.}
    \label{fig:nasarover} 
\end{figure}

Fig.~\ref{fig:nasarover}. shows the rover's architecture in \gls{aadl}.
This rover contains multiple connected modules, some of which provide functionality that is more critical than others -- for example the \textbf{ReasoningAgent}, which is the core decision-making component of the system. To account for this mixed-criticality, the work in \cite{bourbouh2021integrating} begins by eliciting the system's requirements using an approach driven by a detailed hazard analysis. 

One of the system's most important requirements is that \textit{the rover shall not run out of battery}. This particular requirement can be viewed as a system-level contract that relies on the behaviour of multiple system components. For example, the \textbf{Battery\_Interface} must function correctly, ensuring that the rover's goal location is set to the charge station when needed; and the \textbf{ReasoningAgent} must correctly select the shortest path, to conserve battery power.

The work in~\cite{bourbouh2021integrating} uses CoCoSim to enable the compositional verification of the Simulink model of the system.
This approach works by attaching contracts to nodes in the system and then defining a top node where the system-level contracts are specified. Here, compositional verification involves verifying (by model-checking) that the node-level contracts imply the system-level contracts. 

A key point of our approach is that it can derive system-level properties from the node-level contracts.
In \cite{bourbouh2021integrating}, both component-level contracts and system-level contracts are derived from the requirements and the component-level contracts are attached to individual system components. CoCoSim was then used to verify that the system-level properties could be deduced from the component-level contracts. 
The difference between this approach and ours is that CoCoSim requires a system-level property against which to verify systems, whereas we support deriving properties about a system's behaviour for systems built from (potentially independently developed) components. 

\begin{sloppypar}
To compare with this work and investigate how our approach performs on this non-\gls{ros} system, we examine its component-level contracts to see if we can derive similar system-level properties to those used in \cite{bourbouh2021integrating}. For this example, we focus our attention specifically on the \textbf{NavigationSystem} shown in Fig. \ref{fig:nasarover}, which is  composed of a \textbf{ReasoningAgent} and a \textbf{Battery\_Interface}. The \textbf{ReasoningAgent} contains a Goal Reasoning Agent (\textbf{GRA} in Fig.~\ref{fig:nasarover}), to select the rover's goal location; the \textbf{ComputePlan2Charging} component, which generates a plan to the charging point; and the \textbf{ComputePlan2Destination} component, which generates a plan to other destinations. The \textbf{Battery\_Interface} contains a \textbf{BatteryMonitor}  component, which regularly checks the battery level; and the \textbf{Interface}, which updates the \textbf{GRA} when the rover needs to recharge.
\end{sloppypar}

Despite this system not being implemented in \gls{ros}, we are able to apply our calculus to it because it is built from components that act similarly to a \gls{ros} nodes. We take Fig.~\ref{fig:nasarover} as our system model (which maps to Step 1 of our approach in \S\ref{sec:specifyingRobots}) and apply our calculus to the \textbf{ReasoningAgent} and \textbf{Battery\_Interface}.

First, we apply the calculus to the subcomponents of the \textbf{ReasoningAgent}. Because the output of the \textbf{GRA} is the input of both \textbf{ComputePlan2Charging} and the \textbf{ComputePlan2Destination}, we use the branching output rule, \textbf{R2}: 

$$
\small{
\begin{array}{l}
\forall \overline{i_{GRA}},\overline{o_{GRA}}\cdot \mathcal{A}_{GRA}(\overline{i_{GRA}})\ \Rightarrow\ \lozenge \mathcal{G}_{GRA}(\overline{o_{GRA}})\\
\forall\overline{i_{CPC}},\overline{o_{CPC}}\cdot \mathcal{A}_{CPC}(\overline{i_{CPC}})\ \Rightarrow\ \lozenge \mathcal{G}_{CPC}(\overline{o_{CPC}})\\
\forall\overline{i_{CPD}},\overline{o_{CPD}}\cdot \mathcal{A}_{CPD}(\overline{i_{CPD}})\ \Rightarrow\ \lozenge \mathcal{G}_{CPD}(\overline{o_{CPD}})\\
\overline{o_{GRA}} = \overline{i_{CPC}} \cup \overline{i_{CPD}}\\
\vdash\ \left( 
\begin{array}{l}
\forall\overline{o_{GRA}},\overline{i_{CPC}}\cdot \mathcal{G}_{GRA}(\overline{o_{GRA}})\ \Rightarrow\ \mathcal{A}_{CPC}(\overline{i_{CPC}}) \land \\ \forall\overline{o_{GRA}},\overline{i_{CPD}}\cdot \mathcal{G}_{GRA}(\overline{o_{GRA}})\ \Rightarrow\ \mathcal{A}_{CPD}(\overline{i_{CPD}})
\end{array}  \right)\\ 
\hline 
\bigg( \forall\overline{i_{GRA}},\overline{o_{CPC}}, \overline{o_{CPD}}\cdot \mathcal{A}_{GRA}(\overline{i_{GRA}})\ \Rightarrow\ \lozenge\mathcal{G}_{CPC}(\overline{o_{CPC}}) \land \lozenge\mathcal{G}_{CPD}(\overline{o_{CPD}}) \bigg)
\end{array}}
$$

Intuitively this means that, when the assumptions of the \textbf{GRA} component hold, then eventually the guarantees of the 
\textbf{ComputePlan2Charging} ($CPC$) and the \textbf{ComputePlan2Destination} ($CPD$) hold.
This also captures the requirement for this system that all plans are valid and free from obstacles \cite{bourbouh2021integrating}.

Similarly we can use the union of inputs rule, \textbf{R3}, to link the three components that provide input to the \textbf{Interface} ($I$): the \textbf{ComputePlan2Charging} ($CPC$), the \textbf{ComputePlan2Destination} ($CPD$), and the \textbf{BatteryMonitor} ($BM$). 

Using \textbf{R3}, we can derive the following: 
$$
\small{
\begin{array}{l}
\forall\overline{i_{CPC}},\overline{o_{CPC}}\cdot \mathcal{A}_{CPC}(\overline{i_{CPC}})\ \Rightarrow\ \lozenge \mathcal{G}_{CPC}(\overline{o_{CPC}})\\
\forall\overline{i_{CPD}},\overline{o_{CPD}}\cdot \mathcal{A}_{CPD}(\overline{i_{CPD}})\ \Rightarrow\ \lozenge \mathcal{G}_{CPD}(\overline{o_{CPD}})\\
\forall\overline{i_{BM}},\overline{o_{BM}}\cdot \mathcal{A}_{BM}(\overline{i_{BM}})\ \Rightarrow\ \lozenge \mathcal{G}_{BM}(\overline{o_{BM}})\\
\forall\overline{i_{I}},\overline{o_{I}}\cdot \mathcal{A}_{I}(\overline{i_{I}})\ \Rightarrow\ \lozenge \mathcal{G}_{I}(\overline{o_{I}})\\
\overline{i_{I}} = \overline{o_{CPC}} \cup \overline{o_{CPD}} \cup \overline{o_{BM}} \\
\vdash\  \forall\overline{i_{I}},\overline{o_{CPC}}, \overline{o_{CPD}}, \overline{o_{BM}}\cdot \mathcal{G}_{CPC}(\overline{o_{CPC}})\land \mathcal{G}_{CPD}(\overline{o_{CPD}}) \land \mathcal{G}_{BM}(\overline{o_{BM}})  \Rightarrow\ \mathcal{A}_{I}(\overline{i_{I}})\\ 
\hline 
\bigg( \forall\overline{i_{CPC}},\overline{i_{CPD}, \overline{i_{BM}}, \overline{o_{I}}},\cdot \mathcal{A}_{CPC}(\overline{i_{CPC}}) \land \mathcal{A}_{CPD}(\overline{i_{CPD}})\land \mathcal{A}_{BM}(\overline{i_{BM}}) \Rightarrow\ \lozenge\mathcal{G}_{I}(\overline{o_{I}}) \bigg)
\end{array}}
$$

\begin{sloppypar}
Thus we can deduce that when the plan-computing components (\textbf{ComputePlan2Charging} and \textbf{ComputePlan2Destination}) and the \textbf{BatteryMonitor} are functioning correctly, then eventually the \textbf{Interface}'s guarantee is preserved. This means that battery usage is computed correctly and that the rover stays in the charger location until it has fully recharged. 
\end{sloppypar}

Both, this previous approach and the derivations above are able to derive that the rover produces obstacle-free plans. The approach in \cite{bourbouh2021integrating} was able to verify that the rover never runs out of battery but our rules above only allowed us to show that the rover recharges fully when required. These properties are fairly closely related. The reason that we could deduce the property that the rover never runs out of battery was down to the way that the approach using CoCoSim works. There, we had to specify top-level requirements including that the rover never runs out of battery and essentially show that the component-level contracts imply this. Our current approach is more flexible and does not require system-level properties in the same way, rather it derives the properties for the system from the component-level contracts. This has benefits, especially in the domain of autonomous systems where
there might be, previously unknown, emergent properties.

This subsection has illustrated how our contract-based compositional verification approach can be applied in the case of non-ROS systems to derive system-level contracts. We used the rover's \gls{aadl} model (Fig.~\ref{fig:nasarover}) as our system model, which maps to Step 1 in our approach (\S\ref{sec:spec}). Here, we did not need to abstract the model to make it more manageable, but this may be needed for more complicated models.
Then we use \gls{fol} to write and reason about contracts, which maps to Steps 2 and 3 of our approach. Heterogeneous verification (mapping to Step 4a) is still possible here, guided by the \gls{fol} contracts.
Automatically generating runtime monitors (Step 4b) is not possible here, because that part of our approach relies on information in the \gls{rcl} contracts that is specific to \gls{ros}. However, generating runtime monitors in a suitable generic framework that can be applicable to non-ROS systems is a useful avenue of future work.

\FloatBarrier

\section{Conclusions and Future Work}
\label{sec:conclusion}
\glsresetall

This paper contributes a compositional approach to the development of verifiable modular robotic systems, which focusses on systems that use the \gls{ros} -- though parts of it are applicable to non-\gls{ros} systems.
Each module (or \gls{ros} node) is specified using an assume--guarantee contract written in \gls{fol}, that guide its verification. We also present a calculus for composing these contracts, which caters for sequences, joins, and branches in the system's architecture. Each node can be verified using the most suitable method; some may be amenable to formal verification, while others may not. The verification is driven by the contracts, which specify the minimal set of properties that the verification must be able to show hold for that node (to an appropriate level of confidence for the node, and the system's regulatory environment). 

As a safety net, we automatically synthesise formal monitors to verify the contracts' guarantee(s) at runtime. The runtime verification is handled by ROSMonitoring~\cite{DBLP:conf/taros/FerrandoC0AFM20}, an existing tool for runtime verification of \gls{ros} systems.
Supporting this approach is the \gls{rcl}, a novel \gls{dsl} for writing \gls{fol} contracts for \gls{ros} systems; and \textsc{Vanda}, which is a novel prototype tool that parses \gls{rcl} and synthesises the runtime monitors.

In our remote inspection case study (\S\ref{sec:exampleSystem}) we specify contracts for an extant system, then use formal and non-formal verification techniques.  Our case study is designed to simply illustrate the core concepts of our approach: \gls{fol} contracts, combined with a calculus, guiding the most suitable verification method for each node. 

The benefit of letting the contract \textit{guide} the verification is that it is effective even if the specification language used to verify a node does not directly implement \gls{fol}. 
For example, the verification of the \textbf{Agent} node (\S\ref{sec:verifyAnget}) used program model checking of properties written in \gls{ltl} (which is inherently similar to \gls{fol}) enriched with \gls{bdi} concepts~\cite{Dennis2012}.
However, the other nodes were verified using a variety of techniques, both formal and non-formal. The verification of the \textbf{Radiation Sensor} (\S\ref{sec:verifyRad}) uses Hoare Logic, so the contract's \gls{fol} properties are easy to represent. The final two nodes were verified using testing/experimental approaches, so the contracts are a guide for the properties to be checked.

Our approach to the case study was to write contracts for, and verify the nodes of, a pre-existing robotic system. However, if we reverse this workflow, the \gls{rcl} contracts could form a contract-based development approach for building new robotic systems. Used in this way, the contracts would link the system's requirements to its low-level design. \gls{rcl} could be used to specify the assumptions and guarantees of the proposed system's nodes, and how their inputs and outputs connect the nodes together. Once this version of the design was checked using the calculus, the \gls{ros} topic information could be added. These completed contracts could be used to drive a low-level software design, as the starting point for an implementation. Interesting future work could involve extending \textsc{Vanda} to produce message type definitions and skeleton code for \gls{ros} nodes, based on the specification in the \gls{rcl} contracts.

{
Our approach benefits from three main assumptions or restrictions that nonetheless limit the effectiveness of our current methodology. 
First, the expressiveness of the streams that link our contracts together is limited by the assumption that each contract consumes one data element from the input stream and produces (at most) one element on the output stream. This means they are limited in handling bursts of data, asynchronous communications, or compound inputs and outputs. We note that the approach can be generalised to improve its expressivity, for example by pre-processing the stream to match the contracts or by splitting the contracts to match the types of data on the streams. 
Second, our calculus for combining the contracts is currently manually applied and assumes that outputs persist once generated to ensure that all of a contract's inputs are available at the same time. Removing this assumption and mechanising the calculus is left as future work. 
Finally, we only synthesise \gls{rv} monitors for a contract's guarantees, not its assumptions. We use our calculus to check that the assumptions of a contract follow the guarantees of the contract(s) that provide its inputs, therefore we chose to monitor only the guarantees to reduce the \gls{rv} overhead. However, this means that our monitors cannot differentiate between the violation of a guarantee and the violation of an assumption. There is nothing methodologically preventing our tool from synthesising assumption monitors as well, and this extension would enable the \gls{rv} to detect (for example) where the environment is operating outside of the design-time expectations. 
}  

As future work, we have identified three improvements to \gls{rcl} and \textsc{Vanda}. First, we will investigate adding notation to \gls{rcl} for specifying that an output is triggered by an input, which would allow contracts to specify some ordering of events.
\gls{rcl} could also be updated to enable more sophisticated specifications, for example involving real-time constraints or uncertainty.
Second, we intend to add support for calculating a contract's assumptions when composing two contracts. For this we would draw inspiration from work by Cobleigh et al.~\cite{Cobleigh2003}.

{
The third improvement to \gls{rcl} and \textsc{Vanda} is extending them to handle \gls{ros} services and actions. We will update \gls{rcl} to provide support for linking an input or output to a service or action, and \textsc{Vanda} will be updated to enable services or actions to be monitored. One route to achieving this is to build on our existing approaches use of ROSMonitoring. ROSMontoring 2.0~\cite{rosmon2} can monitor services by interceding between the client and server. Given that \gls{ros} actions are built on topics and services, an extension that combines these two approaches seems feasible. 
Another route to monitoring is via introspection, which ROS2 provides for both services\footnote{ROS2 Service Introspection: \url{https://docs.ros.org/en/ros2_documentation/kilted/Tutorials/Demos/Service-Introspection.html}.} and actions\footnote{ROS2 Action Introspection: \url{https://docs.ros.org/en/ros2_documentation/kilted/Tutorials/Demos/Action-Introspection.html}.} This would allow the monitor to read the internal state of the service or action, though this would require the the client to be configured to allow introspection.
}

It would also be useful to investigate the scalability of our approach; so we intend to apply it to a larger and more complex system, expanding the calculus to cater to other arrangements of nodes where needed. Finally, we intend to explore the level of confidence we can have in a system that has been verified using a mixture of methods; particularly how confidence can be calculated for more complex systems with loops in the information flow.

\bibliographystyle{elsarticle-num}
\bibliography{fol}

\newpage
\appendix

\section{Remote Inspection Contracts}
\label{app:contracts}

This appendix contains the full listing of the \gls{rcl} context and the contracts of all four nodes from our case study in \S\ref{sec:exampleSystem}.

\textbf{Context}
\medskip

$
\begin{array}{rl}
 WayP :& \mathbb{R} \times \mathbb{R} \rightarrow \mathbb{N} \\
 RadStat :& \{red, orange, green\} \\
 CommandSet :& \{move(\mathbb{R}, \mathbb{R}), inspect(\mathbb{N})\} \\
 PositionType :& \mathbb{R} \times \mathbb{R} \rightarrow \mathbb{B} \\
 AtType :& \mathbb{R} \times \mathbb{R} \rightarrow \mathbb{B} \\
 InspectedType :& \mathbb{N} \rightarrow \mathbb{B} \\
 SensorsType :& \emptyset 
\end{array}
$

\vfill
\textbf{Agent}

$
\begin{array}{rl}
inputs &( wayP : WayP, at : AtType, radiationStatus : RadStat, \\& inspected : InspectedType )\\

outputs &( command : CommandSet )\\

topics &( gazebo\_radiation\_plugin/Snapshot~command\\ & \qquad matches:~out.command, \\
&gazebo\_radiation\_plugin/Snapshot~inspected\\ & \qquad matches:~in.inspected, \\
&gazebo\_radiation\_plugin/Snapshot~at~matches:~in.at, \\
&gazebo\_radiation\_plugin/Snapshot~wayPNow\\ & \qquad matches:~in.wayP, \\
&gazebo\_radiation\_plugin/Snapshot~radiationStatus\\ & \qquad matches:~in.radiationStatus ) \\

assume & (in.radiationStatus \in \{red, orange, green\}) \\

guarantee & ( \forall x', y' \in REAL, i \in NATURAL \cdot in.at(x', y') = TRUE \\& \land in.wayP(x', y') = i  \land in.inspected(i) = TRUE\\ & \land in.radiationStatus \notin \{red, orange\} \\ & \implies \exists x, y \in REAL \cdot ( in.wayP(x, y) = i+1 \\ &\lor ( \forall x'', y'' \in REAL \cdot in.wayP(x'', y'') \neq i+1\\ & \land in.wayP(x, y) = 0 ) ) \land out.command = move(x, y) ) \\

guarantee & ( \forall x', y' \in REAL, i \in NATURAL \cdot in.at(x', y') = TRUE\\ & \land in.wayP(x', y') = i \land in.inspected(i) = FALSE\\ & \implies out.command = inspect(i) )\\

guarantee & ( \forall x', y' \in REAL, i \in NATURAL \cdot\\ & in.radiationStatus \in \{red, orange\} 
\\ & \lor \neg \exists x, y \in REAL \cdot in.wayP(x, y) = i+1 \\ & \qquad \implies \exists x'', y'' \in REAL\\ 
& \cdot ~ in.wayP(x'', y'') = 0 \land out.command = move(x'', y'') )
\end{array}
$
\newpage

\textbf{Localisation}
\medskip

$
\begin{array}{rl}
inputs &( sensors : SensorsType ) \\
outputs &( position : PositionType ) \\
topics &( gazebo\_radiation\_plugin/Snapshot~position\\ & \qquad matches:~out.position  )\\
assume &(TRUE) \\
guarantee &( \exists!~ x, y \in REAL \cdot out.position(x, y) )
\end{array}
$
\vfill
\textbf{Navigation}
\medskip

$
\begin{array}{rl}
inputs &( position : PositionType, command : CommandSet )\\
outputs &( at : AtType )\\
topics &( gazebo\_radiation\_plugin/Snapshot~command\\& \qquad matches:~in.command, \\
&gazebo\_radiation\_plugin/Snapshot~currentLoc\\ & \qquad matches:~in.position, \\
&gazebo\_radiation\_plugin/Snapshot~at~matches:~out.at  ) \\
assume &(\exists!~ x, y \in REAL \cdot in.position(x, y) = TRUE)\\
guarantee &(\forall x, y \in REAL \cdot in.command = move(x, y) \\ & \land in.position(x, y) = TRUE \iff out.at(x, y) = TRUE)
\end{array}
$
\vfill

\textbf{RadiationSensor}
\medskip

$
\begin{array}{rl}
inputs &( r : REAL, command : CommandSet )\\
outputs &( radiationStatus : RadStat, inspected : InspectedType )\\
topics &( gazebo\_radiation\_plugin/Snapshot~r~matches:~r, \\
&gazebo\_radiation\_plugin/Snapshot~command\\ & \qquad matches:~in.command, \\
&gazebo\_radiation\_plugin/Snapshot~inspected\\ & \qquad matches:~inspected, \\
&gazebo\_radiation\_plugin/Snapshot~radiationStatus\\ & \qquad matches:~out.radiationStatus ) \\
assume &(0 \leq in.r )\\
guarantee & ( \forall i \in NATURAL \cdot in.command = inspect(i) \implies \\ & ( out.inspected(i) = TRUE \\ & \land 0 \leq in.r \land in.r < 120 \implies out.radiationStatus = green \\ & \land 120 \leq in.r \land in.r < 250 \implies out.radiationStatus = orange \\ & \land 250 \leq in.r \implies out.radiationStatus = red ) )
\end{array}
$

\newpage
\section{Hoare Proofs}
\label{app:proofs}

This appendix contains the full Hoare Logic Proofs used in \S\ref{sec:verifyRad}

\subsection{Proof 1}

\noindent $\{radiation\_at(i) < 10\}$
\begin{lstlisting}
input := radiation_at(i)
\end{lstlisting}
$\{input < 10\}$ - Assignment Axiom
\begin{lstlisting}
IF (input < 10) THEN 
\end{lstlisting}
$\quad \quad \{input < 10 \land input < 10\}$ 

\noindent $\quad \quad \{input < 10 \land green = green\}$ - strengthening
\begin{lstlisting}
   output := green 
\end{lstlisting}
$\quad \quad \{input < 10 \land output = green\}$ - Assignment Axiom

\noindent $\quad \quad \{output = green\}$ - weakening
\begin{lstlisting}
   ELSE
    \end{lstlisting}
    $\quad \quad \{input < 10 \land input \geq 10\}$
    
\noindent     $\quad \quad \{\bot\}$ - strengthening
    \begin{lstlisting}
    IF (input < 20) THEN 
\end{lstlisting}
    $\quad \quad \quad \quad \{\bot \land input < 20\}$
    
\noindent     $\quad \quad \quad \quad \{\bot\}$ - strengthening
    \begin{lstlisting}
      output := orange 
\end{lstlisting}
$\quad \quad \quad \quad \{\bot\}$ - Assignment axiom
    \begin{lstlisting}
    ELSE
 \end{lstlisting}
    $\quad \quad \quad \quad \{\bot \land input \geq 20\}$
    
\noindent     $\quad \quad \quad \quad \{\bot\}$ - strengthening
    \begin{lstlisting}
      output := red
\end{lstlisting}
$\quad \quad \quad \quad \{\bot\}$ - Assignment axiom
     
\noindent   $\quad \quad \quad \quad \{output = green \}$ - weakening
 
\noindent    $\quad \quad \{output = green\}$ -- Conditional Rule

\noindent $\{output = green\}$ -- Conditional Rule

\subsection{Proof 2}

\noindent $\{10 \leq radiation\_at(i) < 20\}$
\begin{lstlisting}
input := radiation_at(i)
\end{lstlisting}
$\{10 \leq input < 20\}$ - Assignment Axiom
\begin{lstlisting}
IF (input < 10) THEN 
\end{lstlisting}
$\quad \quad \{10 \leq input < 20\land input < 10\}$ 

\noindent $\quad \quad \{\bot\}$ - strengthening
\begin{lstlisting}
   output := green 
\end{lstlisting}
$\quad \quad \{\bot\}$ - Assignment Axiom

\noindent $\quad \quad \{output = orange\}$ - weakening
\begin{lstlisting}
   ELSE
    \end{lstlisting}
    $\quad \quad \{10 \leq input < 20 \land input \geq 10\}$
    
\noindent     $\quad \quad \{10 \leq input < 20\}$ - strengthening
    \begin{lstlisting}
    IF (input < 20) THEN 
\end{lstlisting}
    $\quad \quad \quad \quad \{10 \leq input < 20 \land input < 20\}$
    
\noindent     $\quad \quad \quad \quad \{10 \leq input < 20 \land orange = orange\}$ - strengthening
    \begin{lstlisting}
      output := orange 
\end{lstlisting}
$\quad \quad \quad \quad \{10 \leq input < 20 \land output = orange\}$ - Assignment axiom

\noindent $\quad \quad \quad \quad \{output = orange\}$ - weakening
    \begin{lstlisting}
    ELSE
 \end{lstlisting}
    $\quad \quad \quad \quad \{10 \leq input < 20 \land input \geq 20\}$
    
\noindent     $\quad \quad \quad \quad \{\bot\}$ - strengthening
    \begin{lstlisting}
      output := red
\end{lstlisting}
$\quad \quad \quad \quad \{\bot\}$ - Assignment axiom
     
\noindent   $\quad \quad \quad \quad \{output = orange \}$ - weakening
 
\noindent    $\quad \quad \{output = orange\}$ -- Conditional Rule

\noindent $\{output = orange\}$ -- Conditional Rule

\subsection{Proof 3}

\noindent $\{radiation\_at(i) \geq 20\}$
\begin{lstlisting}
input := radiation_at(i)
\end{lstlisting}
$\{input \geq 20\}$ - Assignment Axiom
\begin{lstlisting}
IF (input < 10) THEN 
\end{lstlisting}
$\quad \quad \{input \geq 20 \land input < 10\}$ 

\noindent $\quad \quad \{\bot\}$ - strengthening
\begin{lstlisting}
   output := green 
\end{lstlisting}
$\quad \quad \{\bot\}$ - Assignment Axiom

\noindent $\quad \quad \{output = red\}$ - weakening
\begin{lstlisting}
   ELSE
    \end{lstlisting}
    $\quad \quad \{input \geq 20 \land input \geq 10\}$
    
\noindent     $\quad \quad \{input \geq 20\}$ - strengthening
    \begin{lstlisting}
    IF (input < 20) THEN 
\end{lstlisting}
    $\quad \quad \quad \quad \{input \geq 20 \land input < 20\}$
    
\noindent     $\quad \quad \quad \quad \{\bot\}$ - strengthening
    \begin{lstlisting}
      output := orange 
\end{lstlisting}
$\quad \quad \quad \quad \{\bot\}$ - Assignment axiom

\noindent $\quad \quad \quad \quad \{output = red\}$ - weakening
    \begin{lstlisting}
    ELSE
 \end{lstlisting}
    $\quad \quad \quad \quad \{input \geq 20 \land input \geq 20\}$
    
\noindent     $\quad \quad \quad \quad \{input \geq 20 \land red = red\}$ - strengthening
    \begin{lstlisting}
      output := red
\end{lstlisting}
$\quad \quad \quad \quad \{input \geq 20 \land output = red\}$ - Assignment axiom
     
\noindent   $\quad \quad \quad \quad \{output = red \}$ - weakening
 
\noindent    $\quad \quad \{output = red\}$ -- Conditional Rule

\noindent $\{output = red\}$ -- Conditional Rule

\subsection{Proof 4}

\noindent $\{True\}$

\noindent $\{radiation\_at(i) = radiation\_at(i)\}$ - strengthening
\begin{lstlisting}
input := radiation_at(i)
\end{lstlisting}
$\{input = radiation\_at(i)\}$ - Assignment Axiom
\begin{lstlisting}
IF (input < 10) THEN 
\end{lstlisting}
$\quad \quad \{input = radiation\_at(i) \land input < 10\}$ 

\begin{lstlisting}
   output := green 
\end{lstlisting}
$\quad \quad \{input = radiation\_at(i) \land input < 10\}$ - Assignment Axiom

\noindent $\quad \quad \{input = radiation\_at(i)\}$ - weakening
\begin{lstlisting}
   ELSE
    \end{lstlisting}
    $\quad \quad \{input = radiation\_at(i) \land input \geq 10\}$
    
\noindent     $\quad \quad \{input = radiation\_at(i)\}$ - strengthening
    \begin{lstlisting}
    IF (input < 20) THEN 
\end{lstlisting}
    $\quad \quad \quad \quad \{input = radiation\_at(i) \land input < 20\}$
    \begin{lstlisting}
      output := orange 
\end{lstlisting}
$\quad \quad \quad \quad \{input = radiation\_at(i) \land input < 20\}$ - Assignment axiom

\noindent $\quad \quad \quad \quad \{input = radiation\_at(i) \}$ - weakening
    \begin{lstlisting}
    ELSE
 \end{lstlisting}
    $\quad \quad \quad \quad \{input = radiation\_at(i)  \land input \geq 20\}$
    
    \begin{lstlisting}
      output := red
\end{lstlisting}
$\quad \quad \quad \quad \{input = radiation\_at(i)  \land input \geq 20\}$ - Assignment axiom
     
\noindent   $\quad \quad \quad \quad \{input = radiation\_at(i) \}$ - weakening
 
\noindent    $\quad \quad \{input = radiation\_at(i)\}$ -- Conditional Rule

\noindent $\{input = radiation\_at(i)\}$ -- Conditional Rule

\end{document}